# Boosting Quantum Classifier Efficiency through Data Re-Uploading and Dual Cost Functions

**Sara Aminpour[1,2], Mike Banad[1], Sarah Sharif[1,2]***

[1]School of Electrical and Computer Engineering, University of Oklahoma, Norman, OK 73019, USA
[2]Center for Quantum and Technology, University of Oklahoma, Norman, OK 73019 USA
*Corresponding author: Sarah Sharif (email: s.sh@ou.edu)

## ABSTRACT

Quantum machine learning integrates quantum computing with classical machine learning techniques to enhance computational power and efficiency. A major challenge in Quantum machine learning is developing robust quantum classifiers capable of accurately processing and classifying complex datasets. In this work, we present an advanced approach leveraging data re-uploading, a strategy that cyclically encodes classical data into quantum states to improve classifier performance. We examine two cost functions—fidelity and trace distance—across various quantum classifier configurations, including single-qubit, two-qubit, and entangled two-qubit systems. Additionally, we evaluate four optimization techniques (L-BFGS-B, COBYLA, Nelder-Mead, and SLSQP) to determine their effectiveness in optimizing quantum circuits for both linear and non-linear classification tasks. Our results show that the choice of optimization method significantly impacts classifier performance, with L-BFGS-B and COBYLA often yielding superior accuracy. The two-qubit entangled classifier shows improved accuracy over its non-entangled counterpart, albeit with increased computational cost. Also the two-qubit entangled classifier are the best option for real word random dataset in order to accuracy and computational cost. Linear classification tasks generally exhibit more stable performance across optimization techniques compared to non-linear tasks. Our findings highlight the potential of data re-uploading in Quantum machine learning outperforming existing quantum classifier models in terms of accuracy and robustness. This work contributes to the growing field of Quantum machine learning by providing a comprehensive comparison of classification strategies and optimization techniques in quantum computing environments, offering a foundation for developing more efficient and accurate quantum classifiers.

## INTRODUCTION

Since its inception in 1959[1], machine learning (ML) has become one of the most transformative technologies of the modern era, revolutionizing how we classify, cluster, and recognize patterns in vast datasets. Today, ML is deeply integrated into various sectors of society, and even small advancements in the field can yield significant economic and technological benefits. In recent years, a natural extension of ML has emerged within the framework of quantum mechanics, leading to the rise of Quantum Machine Learning (QML). By the mid-2010s[2], QML began gaining momentum as researchers explored the potential of quantum computing to enhance classical ML techniques. Quantum computing leverages the principles of quantum mechanics, specifically entanglement, superposition, and interference, to execute computations[3]. Quantum information processing offers advantages in communication and computational tasks, such as solving algebraic problems, reducing sample complexity, and enhancing optimization processes. Notably, even simplified models of quantum computation can solve complex tasks, thereby holding promise for advancements in machine learning and artificial intelligence[4]. At the heart of contemporary QML practices is the training of quantum circuits, aimed at processing both classical and quantum[5-11].

In the emerging field of QML, quantum neural networks (QNNs) adapt this concept by leveraging quantum mechanics to process information[12]. These networks undergo a training process akin to their classical counterparts, where data is input into the quantum system, a cost function is computed based on the output, and the parameters of the QNN are iteratively adjusted through classical optimization techniques to minimize cost functions[13].

A notable breakthrough in the QML field is the concept of data re-uploading, which involves the cyclic encoding of classical information into a quantum system, allowing for the repeated integration of diverse datasets into the quantum processing workflow. Data re-uploading enables the construction of universal quantum classifiers[14], where a quantum circuit is meticulously organized into a series of stages dedicated to data integration and single-qubit operations[15]. This approach not only enhances the flexibility and adaptability of quantum classifiers but also significantly boosts their accuracy and efficiency in handling various classification tasks.

Several studies have explored various optimization techniques to enhance the performance of quantum classifiers. Lockwood[16] presents a comprehensive empirical review of optimization techniques for quantum variational circuits, comparing 46 different optimizer setups, including minimization methods such as L-BFGS-B, Nelder-Mead, and SLSQP, across different QML problems such as Variational Quantum Eigensolver[17], Quantum Approximate Optimization Algorithm[18], and Moon binary classification[19]. Similarly, Lee et al.[20] propose an iterative layerwise optimization strategy for the quantum approximate optimization algorithm to reduce optimization costs while maintaining high approximation ratios. Their numerical simulations compare the performance of L-BFGS-B and Nelder-Mead optimizers in conjunction with the proposed strategy on the Max-cut problem. Although these studies provide valuable insights, there is still a lack of research comparing different cost functions, minimization methods, and classification patterns in combination with data reuploading techniques. The impact of data reuploading on the performance of various minimization methods remains largely unexplored. Further investigation into the interplay between data reuploading and different optimization techniques could potentially lead to more efficient and effective QML algorithms.

In addition, there is a notable gap in the literature considering random datasets that closely mimic real-world scenarios[16,20,21]. By using random datasets that approximate actual data, we can assess how well QML algorithms perform under conditions that are more representative of real-world applications, identify potential weaknesses or limitations in current QML techniques when faced with diverse and unpredictable data patterns, and develop more resilient and adaptable QML algorithms that can handle a wider range of data types and structures. Our initial results indicate that the proposed methodology shows promise when applied to a randomized dataset[22]. This encouraging outcome suggests that further investigation is warranted to validate the effectiveness of the methodology across a broader range of random datasets, assess its generalizability to various types of real-world data and applications, and compare its performance against existing QML techniques.

To comprehensively evaluate our proposed model, we are considering applying it to random datasets that simulate real-world data for potential applications. Additionally, we plan to conduct comparative analyses between fixed datasets and random datasets across various situations. This comparison will help us identify any discrepancies in the model's performance between structured (fixed) and unstructured (random) data, assess the model's ability to generalize across different data distributions and patterns, and determine the robustness of the model when faced with unexpected or noisy data. This approach will contribute to the development of more efficient, reliable, and versatile QML techniques that can address a wide range of practical challenges.

Moreover, we introduce the trace distance cost function as an alternative to the fidelity cost function, highlighting its distinct advantages in quantum classification tasks for the first time[23]. Unlike fidelity, which measures the overlap between quantum states, trace distance directly quantifies how distinguishable two states are, ranging from 0 (indistinguishable) to 1 (perfectly distinguishable). This makes it particularly effective for applications where state differentiation is crucial. Moreover, the trace distance cost function helps address the barren plateau problem, where gradients of random parameterized quantum circuits vanish exponentially with the number of qubits and layers[24]. This issue is especially pronounced with global cost functions like fidelity. By employing trace distance, the classifier becomes less prone to this vanishing gradient effect, providing a more stable and scalable training process. Through this exploration, we aim to assess the classifier's adaptability and generalization potential under varying optimization criteria, offering valuable insights into its robustness and effectiveness across different conditions. Building on previous research that primarily examined the 'L-BFGS-B' method for fidelity cost function and fix data set[14], this paper significantly expands the scope by incorporating three additional minimization techniques: 'COBYLA,' 'Nelder-Mead,' and 'SLSQP' for trace distance cost function considering both fix and random datasets. In this study, we employed three distinct optimization methods - COBYLA, Nelder-Mead, and SLSQP – in addition to L-BFGS-B to explore a range of approaches suited to different problem characteristics. COBYLA was chosen for its ability to handle non-linear constraints without requiring derivative information, making it versatile for complex constraint landscapes[25-27]. Nelder-Mead, a derivative-free method, was selected for its effectiveness with potentially non-smooth functions and simplicity in low-dimensional spaces[16,28-30]. SLSQP was included for its efficiency in handling both constrained and unconstrained problems, particularly when gradient information is available[31]. This diverse selection allows us to compare the performance of gradient-based and derivative-free methods, as well as those specialized for constrained optimization, providing a more comprehensive understanding of our problem's optimization landscape than a single method like L-BFGS-B could offer. This dual evaluation of cost functions—fidelity and trace distance—allows for a more nuanced analysis of classifier behavior, revealing how different optimization methods interact with varied performance criteria.

Finally, we use linear classification patterns (LCP) and non-linear classification patterns (non-LCP) as a fundamental starting point for evaluating the performance of various optimization methods in quantum classifiers. This allows for a clear and controlled analysis of how quantum classifiers manage distinct types of data relationships.

While more complex patterns can be studied, we consider fundamental line patterns for more intricate linear relationships and the circle pattern for more advanced non-linear patterns.

The structure of the paper unfolds as follows: The results section presents a comprehensive analysis of our quantum classifier's performance across 52 various configurations. We evaluate single-qubit, two-qubit, and two-qubit entangled classifiers using four optimization methods (L-BFGS-B, COBYLA, Nelder-Mead, and SLSQP, two cost functions (fidelity and trace distance), two classification patterns (LCP and non-LCP) as well as two datasets (fix and random). We present findings on accuracy, computational efficiency, and the impact of increasing layers and training samples. We examine the trade-offs between accuracy and computational cost, the advantages of entanglement in quantum classifiers, and the relative performance of different optimization methods across various classification tasks. This section also discusses the potential applications of our findings and their contribution to the broader field of quantum computing. Finally, the methods section elucidates our experimental approach, detailing the implementation of data re-uploading strategies, the construction of quantum circuits for different classifier types, and the specifics of our optimization techniques. We also describe our data generation processes for both fixed and random datasets and explain our evaluation metrics and statistical analysis methods.

In addition to the main manuscript, we have released, supplemental documents providing preliminary analysis to identify an appropriate number of training samples and layers, more details about the result for each section, detailed exploration of the process of reuploading, including how it occurs and is handled within the quantum classifier framework, methods and methodology of modeling cost functions, as well as minimization methods. We also released our main code for public use.

## RESULTS

In this study, we explored a range of models and methodologies to assess the performance of quantum classifiers in binary classification tasks. Two main cost functions, Fidelity and Trace Distance, were examined alongside four minimization methods—L-BFGS-B, COBYLA, Nelder-Mead, and SLSQP—to provide a comprehensive evaluation. In addition, we studied quantum systems with 1-qubit, 2-qubit, and 2-qubit entangled configurations to capture the differences in performance across various quantum setups.

Both fixed and random datasets were considered to evaluate the robustness and adaptability of the classifiers. We generate a random dataset for non-LCP on a plane with coordinates $\vec{x}=\left(x_1, x_2\right)$ with $x_i \in [-1, 1]$ defined by $x_1^2 + x_2^2 < r^2$, aiming to classify these data based on whether they fall inside or outside a circle of radius $r=\sqrt{2/\pi}$. The radius is chosen in a way that ensures equal areas for the regions inside and outside the circle. This setup results in a balanced dataset, where randomly assigning labels to data points would yield a 50 percent accuracy rate by chance. To ensure uniformity across our experiments, a consistent seed was utilized for generating all data points when dataset is fixed. Conversely, for analyses involving random data, data points were generated entirely at random for each of the 20 iterations to ascertain the average accuracy.

The outcomes of these runs were averaged to present a more reliable and statistically significant assessment of the classifiers' performance. This approach enabled us to evaluate the quantum classifiers across diverse scenarios, capturing their true capabilities in both stable and unpredictable data environments.

We studied all models with a training sample size of up to 250, and we eliminated the results of overfitting, so the results presented vary from 50 to 250. This careful selection of data points ensured that the findings accurately reflected the performance of the classifiers without being skewed by overfitting. A conceptual overview of the studied models and methodologies is provided in Figure 1, illustrating the key components of this investigation.

Our next focus was on benchmarking the selection of classifiers across varying numbers of layers, with a particular emphasis on configurations comprising five layers. This emphasis was based on the hypothesis that a five-layer architecture could potentially achieve enhanced performance and accuracy.

The subsequent sections delve deeper into this exploration, providing detailed insights into the performances of specific algorithms when implemented using single-qubit and 2-qubit classifiers with the innovative technique of data re-uploading. This methodical approach not only enhances our understanding of the quantum classifier's potential but also sets the stage for future advancements in the field of QML, spotlighting the critical role of algorithmic diversity and adaptability in navigating the complexities of quantum data classification.

Before delving into the accuracy metrics of the 52 unique scenarios depicted in figure1, we embarked on a preliminary analysis to identify an appropriate number of training samples and layers. This preparatory step was crucial not only for establishing a consistent baseline for comparing training and test accuracies across various configurations but also for ensuring that our simulations remained feasible on our desktop computer with limited configurations. As illustrated in figures S1.1 and S1.2 in the supplementary note 1, we conducted a series of runs with our algorithm, varying the number of layers from 1 to 5 and using up to 250 training samples, to determine the conditions under which our algorithm would reach a test accuracy around 90%. This exploration led us to conclude selecting 5 layers of training for our study. To maintain a uniform evaluation framework, we subsequently used these values for all simulated cases.

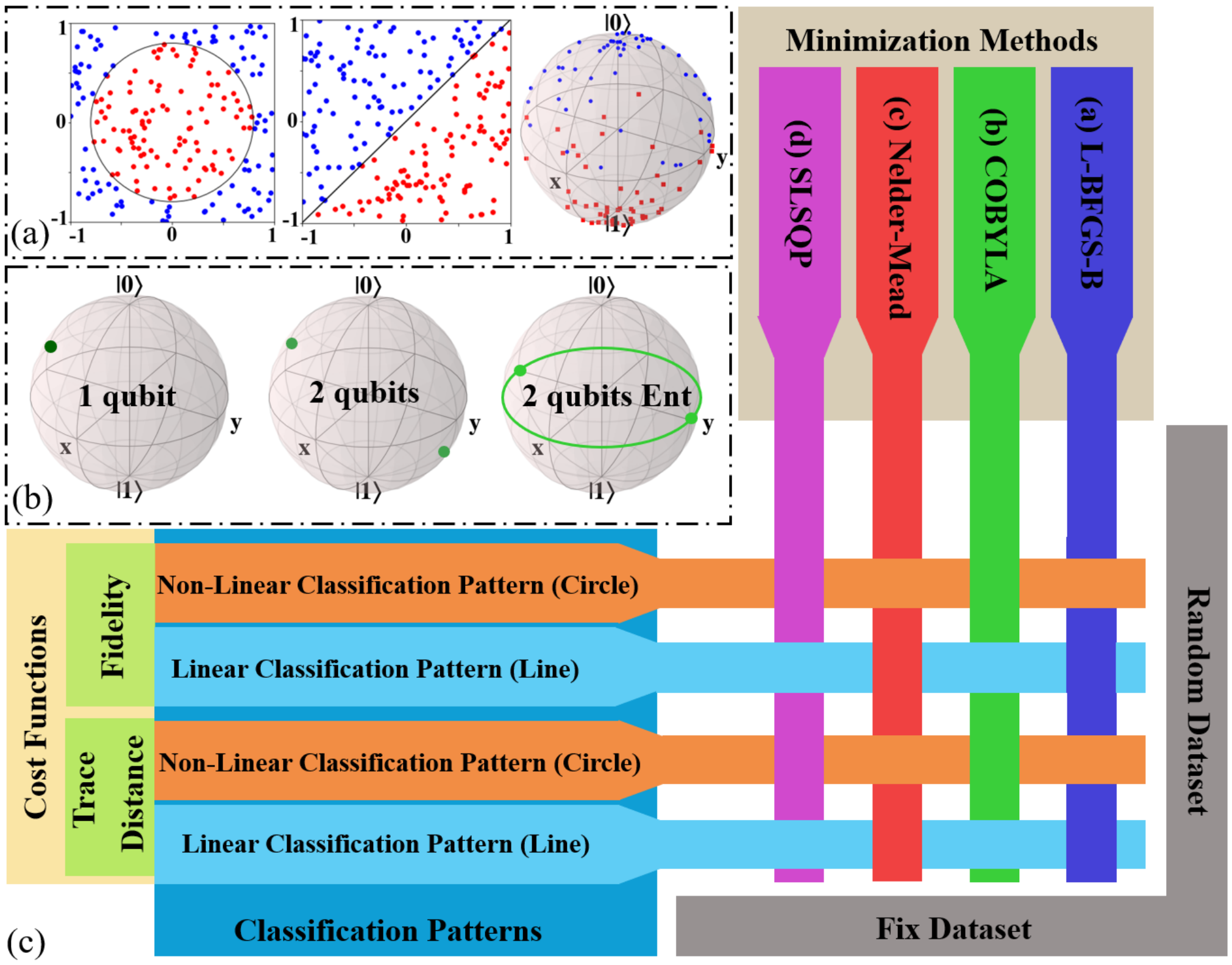

**Figure 1.** (a) Linear and non-linear classification pattern represented by a circle and line, and mapping of data points onto the Bloch sphere for quantum classification. Blue and red points represent different classes. (b) Illustrations of states on the Bloch sphere for single qubit, two qubits, and two qubits entangled. (c) Visualization of quantum classification concepts and schematics of 52 different cases studied in this paper. The figure outlines two cost functions, two classification patterns, and four minimization methods under two fix and random datasets.

## Evaluating linear and non-linear classification approaches for fidelity in fixed and random datasets for 1-qubit classifier for four different minimization methods

In our study, we devised a methodology to assess the performance of a single-qubit classifier across various conditions by constructing training datasets of varying sizes. For certain minimization methods, we employed different ranges of sample sizes to effectively control for overfitting, ensuring that the results accurately reflected the classifier's true performance.

The classifier's efficacy was then evaluated using a comprehensive test dataset consisting of 4000 data points.

This section is focused on the Fidelity cost function, and we presented the comparison of four optimization techniques (L-BFGS-B, COBYLA, Nelder-Mead, and SLSQP) for different datasets and classification patterns. Figures 2 and 3 show the result for the non-LCP for fixed and random datasets, respectively. Initially, all methods achieve perfect training accuracy, but L-BFGS-B shows greater resistance to overfitting, maintaining high training accuracy as sample size increases. In contrast, COBYLA, Nelder-Mead, and SLSQP display higher susceptibility to overfitting and performance fluctuations, especially in test accuracy. Notably, L-BFGS-B achieves peak accuracy with more training samples, while the others perform better with fewer samples, underscoring the importance of careful sample selection. These insights emphasize the need to balance the number of training samples and method choice to optimize accuracy and prevent overfitting.

Figures 4 and 5 compare four optimization techniques (L-BFGS-B, COBYLA, Nelder-Mead, and SLSQP) for classifying LCP using fidelity cost function, fixed and random datasets. In Figure 4, all methods achieve near-perfect training accuracy, with test accuracy improving as the number of training samples increases, peaking around 94%-98% with 125 samples. L-BFGS-B and COBYLA show a reduction in the training-test accuracy gap as sample size grows, while Nelder-Mead achieves precise classification with minimal gaps between train and test accuracies. Figure 5 demonstrates a similar trend on random datasets, with all methods surpassing 90% accuracy with 50 training samples. COBYLA and SLSQP exhibit superior generalization, while Nelder-Mead achieves the smallest gap between train and test accuracy for 50 number of training samples, underscoring its balance between training and generalization. Both figures highlight the importance of selecting an optimal number of training samples for effective performance and overfitting mitigation across methods.

A comparison of Figures 2 and 4 highlights that LCP exhibits more stable and consistent accuracy curves across all optimization techniques, while non-LCP shows greater variability and fluctuations. This difference may arise from several factors: (1) LCP likely aligns better with linear decision boundaries, making it easier for classifiers to generalize, whereas non-LCP involves more complex, non-linear patterns that challenge generalization and lead to overfitting or underfitting. (3) The algorithms' adaptability to specific classification tasks may also affect their performance. Similarly, figures 3 and 5 reveal that the classification problem's complexity impacts accuracy, with LCP requiring fewer samples for stable accuracy, while non-LCP fluctuates more, underscoring the importance of matching optimization methods to problem complexity. We explained more about the non-linear and linear classification approaches for fidelity in fixed and random datasets for 1-qubit classifier in supplementary note 2.

## Evaluating linear and non-linear classification approaches for Trace distance in fixed and random datasets for 1-qubit classifier for four different minimization methods

Figures 6 and 7 examine the use of the trace distance cost function for classifying non-LCP on fixed and random datasets, comparing four optimization techniques: L-BFGS-B, COBYLA, Nelder-Mead, and SLSQP. In Figure 6, all methods achieve perfect training accuracy with relatively few samples, but their test accuracy varies. COBYLA performs best, reaching 84.6% with 100 samples and demonstrating strong generalization. L-BFGS-B achieves 79.2%, while Nelder-Mead and SLSQP show more variability, with Nelder-Mead experiencing overfitting. Figure 7 reveals a similar trend on random datasets, with test accuracy improving as the number of training samples increases. L-BFGS-B reaches the highest test accuracy (77.8%) with 45 samples, followed by COBYLA at 72.9%, and SLSQP and Nelder-Mead showing moderate fluctuations. Overall, COBYLA excels in generalization for fixed datasets, while L-BFGS-B stands out for random datasets. Both figures highlight the importance of selecting appropriate optimization methods and training sample sizes based on the task complexity and dataset type.

Figures 8 and 9 compare the performance of four optimization methods—L-BFGS-B, COBYLA, Nelder-Mead, and SLSQP—for LCP using a trace distance cost function across both fixed and random datasets. In both figures, SLSQP stands out, achieving the highest test accuracy with minimal overfitting and consistent generalization, peaking at 93.3% in the fixed dataset and 88.3% in the random dataset. L-BFGS-B also shows strong performance, particularly with larger datasets, while COBYLA and Nelder-Mead exhibit fluctuations and signs of overfitting as the number of training samples increases for both datasets. Overall, SLSQP is the most robust method, handling both datasets effectively and maintaining accuracy without overfitting. For a more detailed analysis of the LCP and non-LCP approaches using trace distance cost function with fixed and random datasets for the 1-qubit classifier, please refer to Supplementary Note 3. Also, for a comprehensive performance comparison of 5-layer single-qubit quantum classifiers using fidelity and trace distance cost functions across various classification tasks and dataset types, please refer to supplementary note 4.

## Evaluating non-linear and linear classification approaches for fidelity in fixed and random datasets for 2-qubit and 2-qubit entangled classifiers

Building on our findings from Figure 4, where we analyzed four distinct minimization methods, we identified that the Nelder-Mead minimization method achieved the highest accuracy of 97.3% when assessed using both fidelity and trace distance cost functions for LCP. This observation prompted us to extend our investigation to 2-qubit and 2-qubit entangled systems, focusing specifically on the fidelity cost function in combination with the Nelder-Mead minimization method. To assess the efficiency of these classifiers for LCP, we analyzed the computational time required to achieve the highest accuracy using the fidelity cost function and the Nelder-Mead minimization method across 1-qubit, 2-qubit, and 2-qubit entangled classifiers. This approach allowed us to gain a comprehensive understanding of the performance and computational efficiency of the fidelity cost function paired with the Nelder-Mead minimization method across various quantum configurations. Figure 10 (a-c) presents a comparative analysis of accuracy in quantum classifiers for LCP using three different quantum systems: single-qubit, two-qubit, and two-qubit entangled configurations, while figure (d-f) represents the computational time vs number of training samples. Figure 10(a), the single-qubit system shows a steep learning curve, with accuracy rising from 51.6% to 92% after just 75 training samples, and stabilizing between 92% and 97.7% as the sample size increases to 250. To assess the computational efficiency of the quantum classifier, we extended the original implementation provided by Pérez-Salinas et al.[32] with datetime library in python. Our modified version incorporates time measurement commands to quantify the runtime of the classification algorithm across various problem instances and classifier configurations. This enhancement allows us to analyze the computational cost of the quantum approach. As shown in figure 10(d), its computational time reaches 62.15 seconds for 250 samples, making it both accurate and computationally efficient. In contrast, figure 10(b) presents the 2-qubit classifier, which starts with higher accuracy (73.2%) and gradually peaks at 95.7% with 175 samples but with a significantly higher computational time of 260 seconds for 250 samples, indicated in Figure 10(e). Figure 10(c) illustrates the performance of the 2-qubit entangled classifier, which, while achieving the highest peak accuracy (97.5%), also exhibits pronounced fluctuations in accuracy and matches the non-entangled system's computational cost of 260 seconds in Figure 10(f). This comparison highlights trade-offs: the single-qubit system offers stability and efficiency, the 2-qubit classifier provides robust initial accuracy with more resource demands, and the 2-qubit entangled system offers the highest peak accuracy but at the cost of increased computational complexity and performance variability. The choice of system depends on whether stability, efficiency, or peak performance is prioritized.

For the one-qubit case, our results indicate that the highest accuracy is achieved when the number of layers is set to 5. This configuration was therefore maintained as the initial condition for both the 2-qubit and 2-qubit entangled classifiers. We then explored the optimal number of training samples required to achieve the highest accuracy without inducing overfitting, identifying 175 samples as the threshold for fixed dataset. We then incrementally increased the number of layers from 1 to 20, using the optimal training sample size of 175, as illustrated in Figure 11. Figures 11(a-b) reveal that both classifiers improve in train and test accuracy as training samples increase, with the 2-qubit classifier showing higher initial test accuracy (73.5%) and more stable performance, while the 2-qubit entangled classifier starts lower (47.6%) but improves significantly with more samples. Figures 11(c- d) demonstrate that increasing the number of layers in the quantum circuit boosts accuracy for both classifiers, though the 2-qubit entangled classifier benefits more dramatically, especially early on. Both classifiers plateau in performance after 12-15 layers. Figures 11(e) and 11(f) show that computational time grows exponentially with the number of layers for both classifiers, reflecting a similar scaling trend regardless of entanglement use. The 2-qubit classifier generally achieves better accuracy with fewer training samples and maintains stable performance, while the 2-qubit entangled classifier, though more volatile, demonstrates greater potential for capturing complex patterns with more layers and samples. However, this advantage comes with higher computational costs and sensitivity to changes in training conditions.

The exploration of random dataset within the context of data reuploading has not been previously reported, prompting us to address this gap in the literature. To ensure a comprehensive analysis, we examined the fidelity cost function alongside four minimization methods: COBYLA, L-BFGS-B, Nelder-Mead, and SLSQP. This evaluation was conducted for both LCP and non-LCP scenarios, focusing on 2-qubit and 2-qubit entangled classifiers. To assess the effectiveness of these minimization methods, we also analyzed the associated computational time for each case. Figures 12 and 13 present a comparison of the train and test accuracy, as well as computational time for four optimization methods (COBYLA, L-BFGS-B, NELDER_MEAD, and SLSQP). Concluding the result from Figure 10 and 11, we kept the number of training samples and the number of layer constant of 250 and 5, respectively. Figure 12 shows the results when we applied the fidelity cost function to the LCP task and random dataset using 2-qubit and 2-qubit entangled classifiers. The results show, on average, the 2-qubit entangled classifier achieves approximately 2% higher test accuracy than the non-entangled classifier. In addition, L-BFGS-B not only has the highest test accuracy in comparison with 1-qubit (figure 5), but it also provides the highest accuracy for both 2-qubit and 2-qubit entangled for the amount of 96.3% and 97%, respectively.

**Fidelity** **Fixed Dataset**

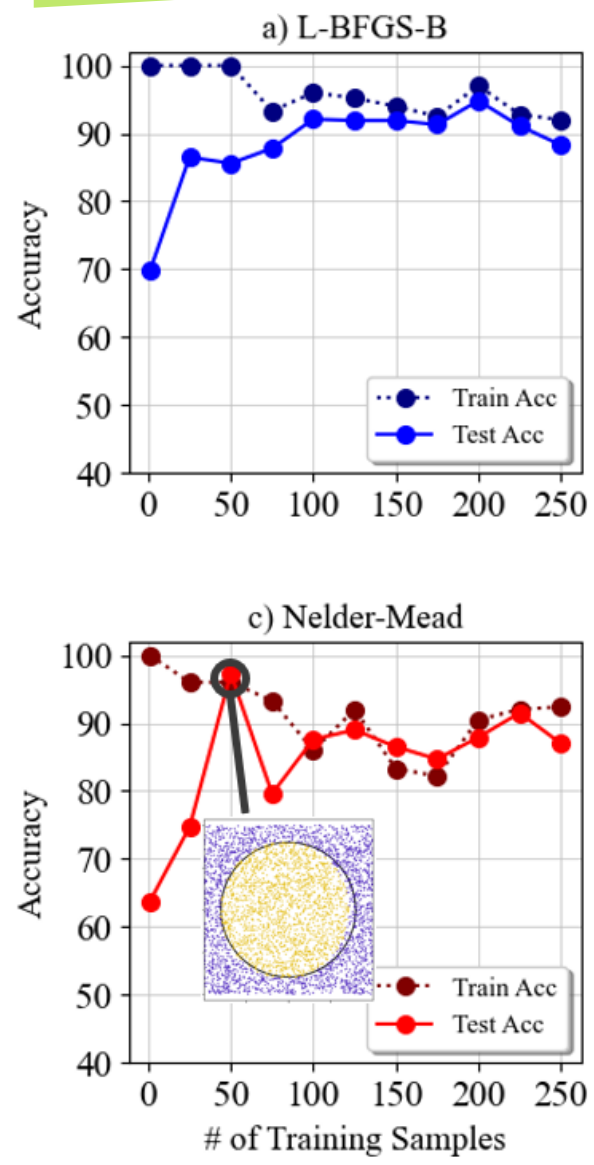

**Figure 2.** Train and test accuracy of fidelity for the 5-layer model of non-LCP and fix dataset for (a) L-BFGS-B, (b) COBYLA, (c) Nelder-Mead and (d) SLSQP minimization methods. The inset image in subplot (c) in the graph shows a visualization of a circle classification task with the highest accuracy of 97.3% in the Nelder-Mead minimization method.

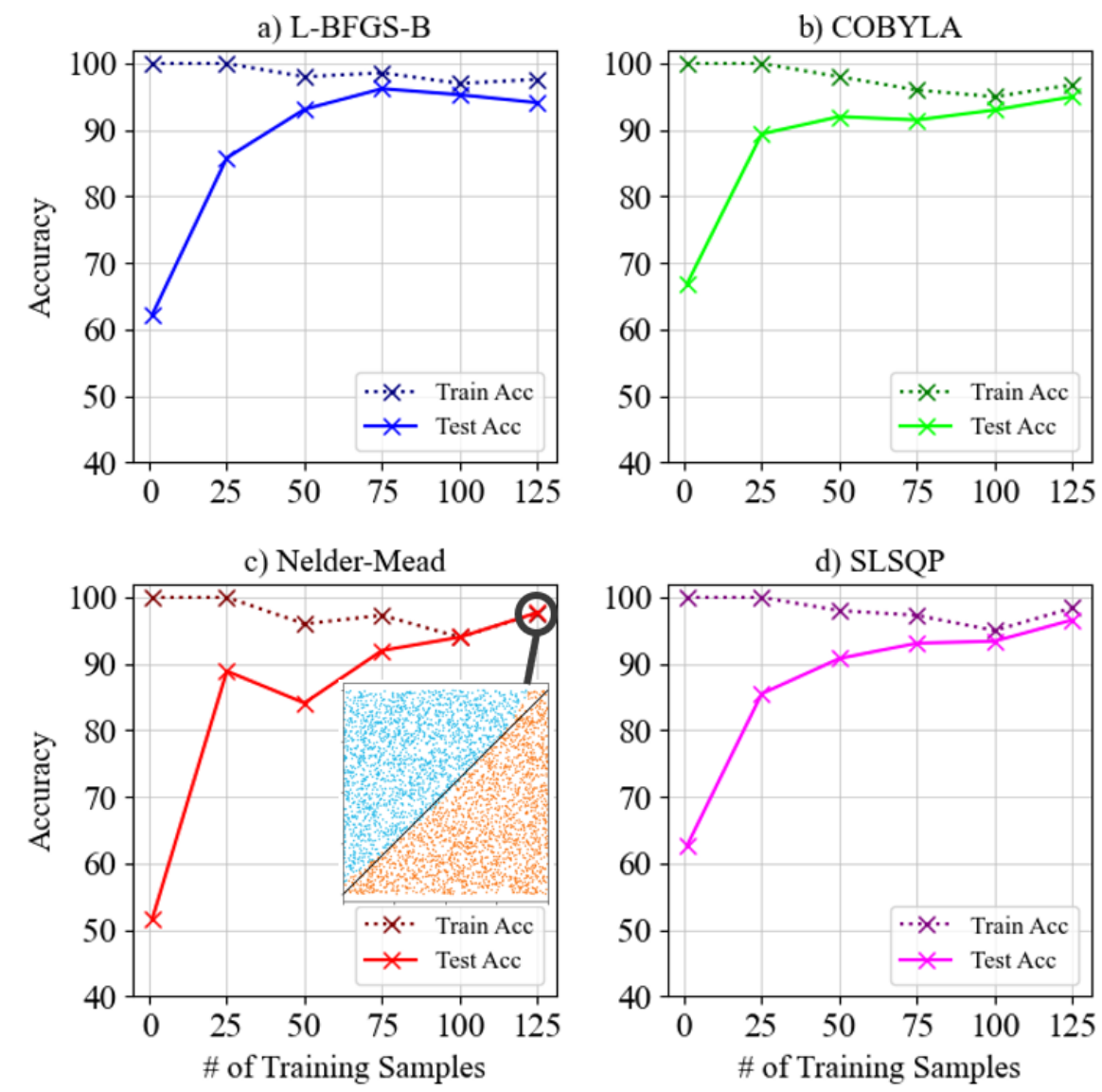

**Figure 4.** Train and test accuracy of fidelity for the 5-layer model of LCP and fix dataset for (a) L-BFGS-B, (b) COBYLA, (c) Nelder-Mead and (d) SLSQP minimization methods. The inset graph in subplot (c) shows the visualization of a line classification pattern with the highest accuracy of 97.7% in the Nelder-Mead minimization method.

**Fidelity** **Random Dataset**

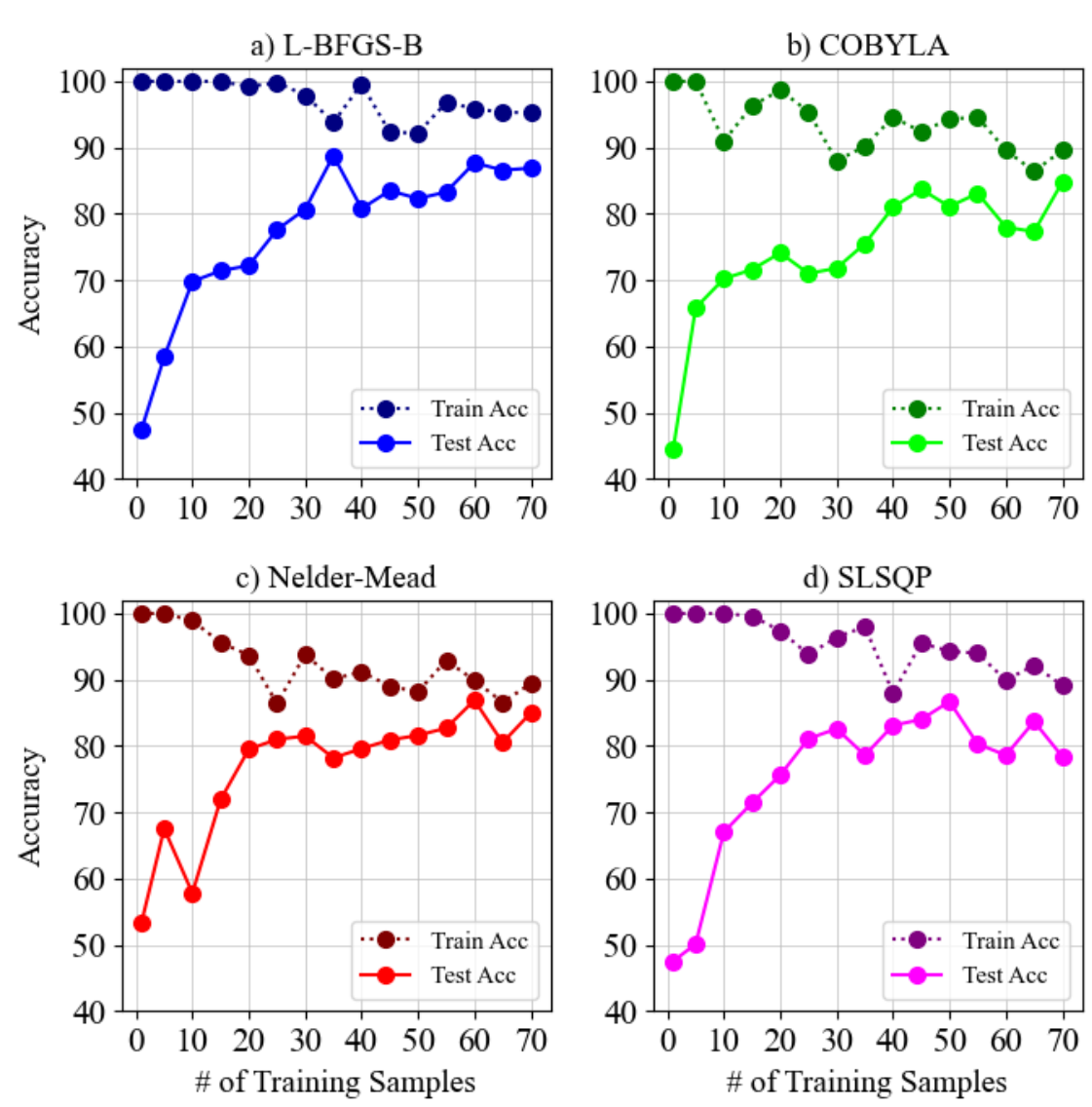

**Figure 3.** Train and test accuracy of fidelity for the 5-layer model of non-LCP and random dataset for (a) L-BFGS-B, (b) COBYLA, (c) Nelder-Mead and (d) SLSQP minimization methods.

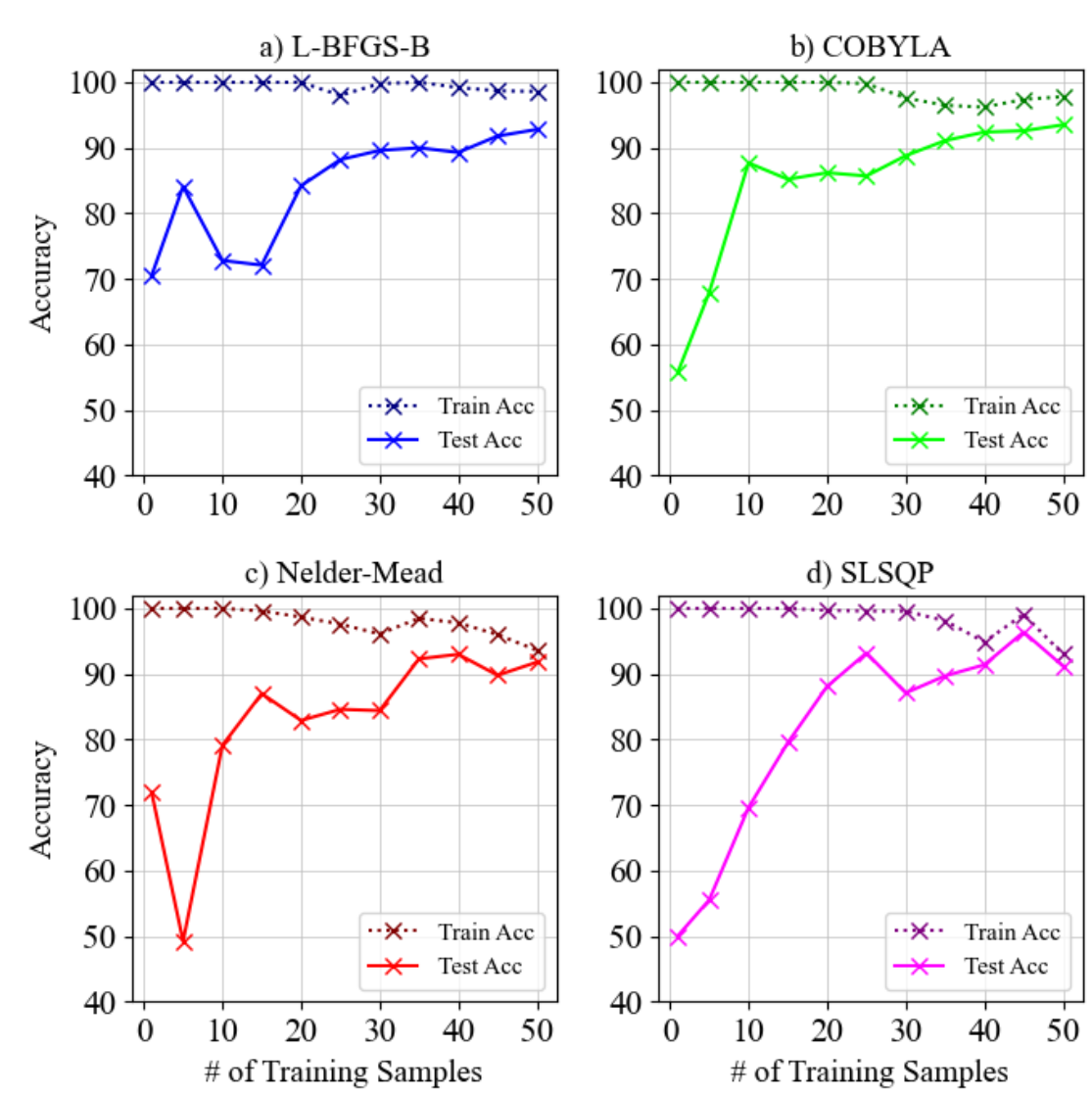

**Figure 5.** Train and test accuracy of fidelity for the 5-layer model of LCP and random dataset for (a) L-BFGS-B, (b) COBYLA, (c) Nelder-Mead and (d) SLSQP minimization methods.

**Trace Distance** **Fixed Dataset**

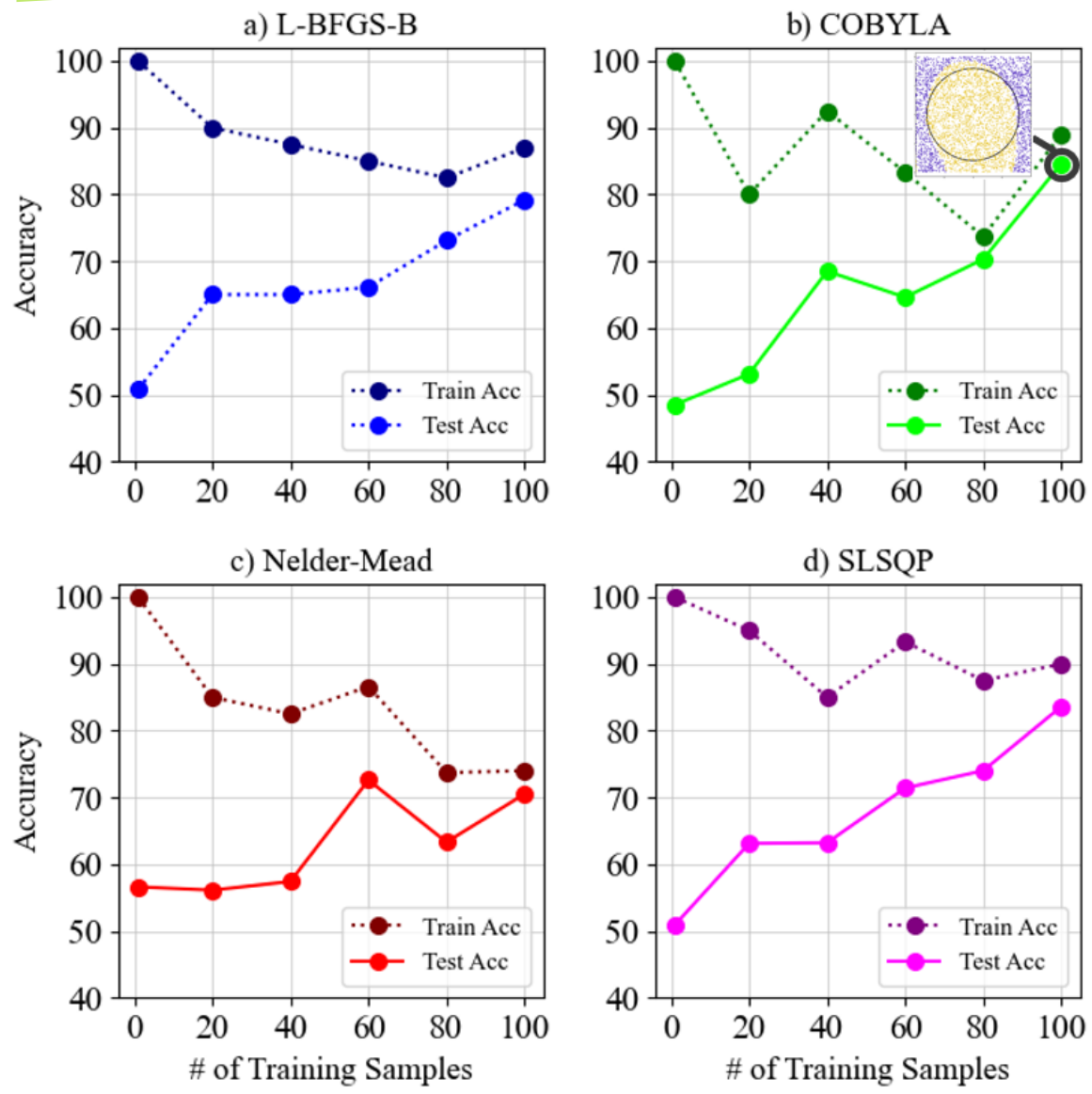

**Figure 6.** Train and test accuracy of trace distance for the 5-layer model of non-LCP and fixed dataset for (a) L-BFGS-B, (b) COBYLA, (c) Nelder-Mead and (d) SLSQP minimization methods. The inset graph in subplot (b) shows the visualization of a circle classification pattern with the highest accuracy of 84.6% in the COBYLA minimization method.

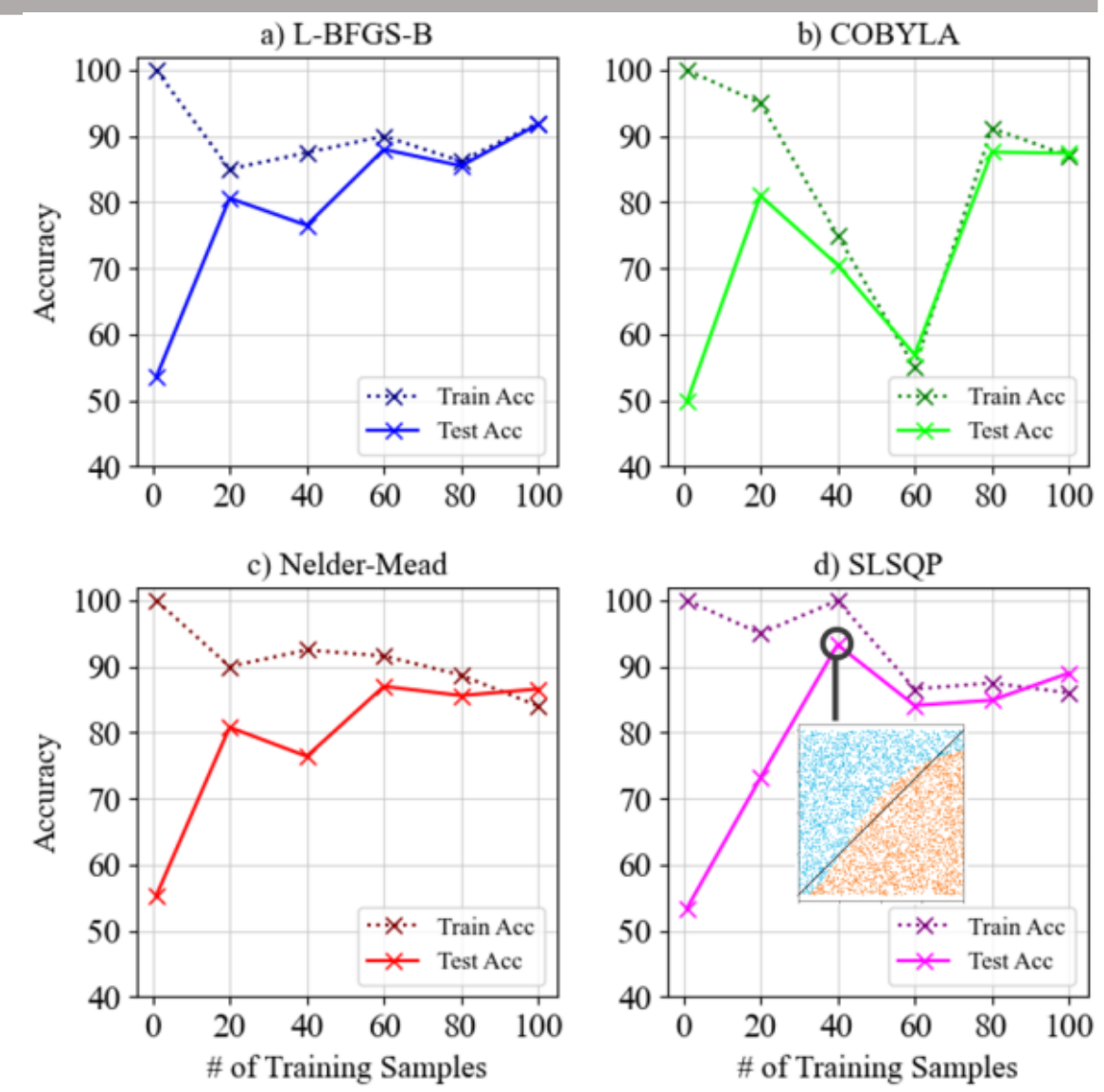

**Figure 8.** Train and test accuracy of trace distance for the 5-layer model of LCP and fix dataset for (a) L-BFGS-B, (b) COBYLA, (c) Nelder-Mead and (d) SLSQP minimization methods. The inset graph in subplot (c) shows the visualization of a line classification pattern with the highest accuracy of 93.3% in the SLSQP minimization method.

**Trace Distance** **Random Dataset**

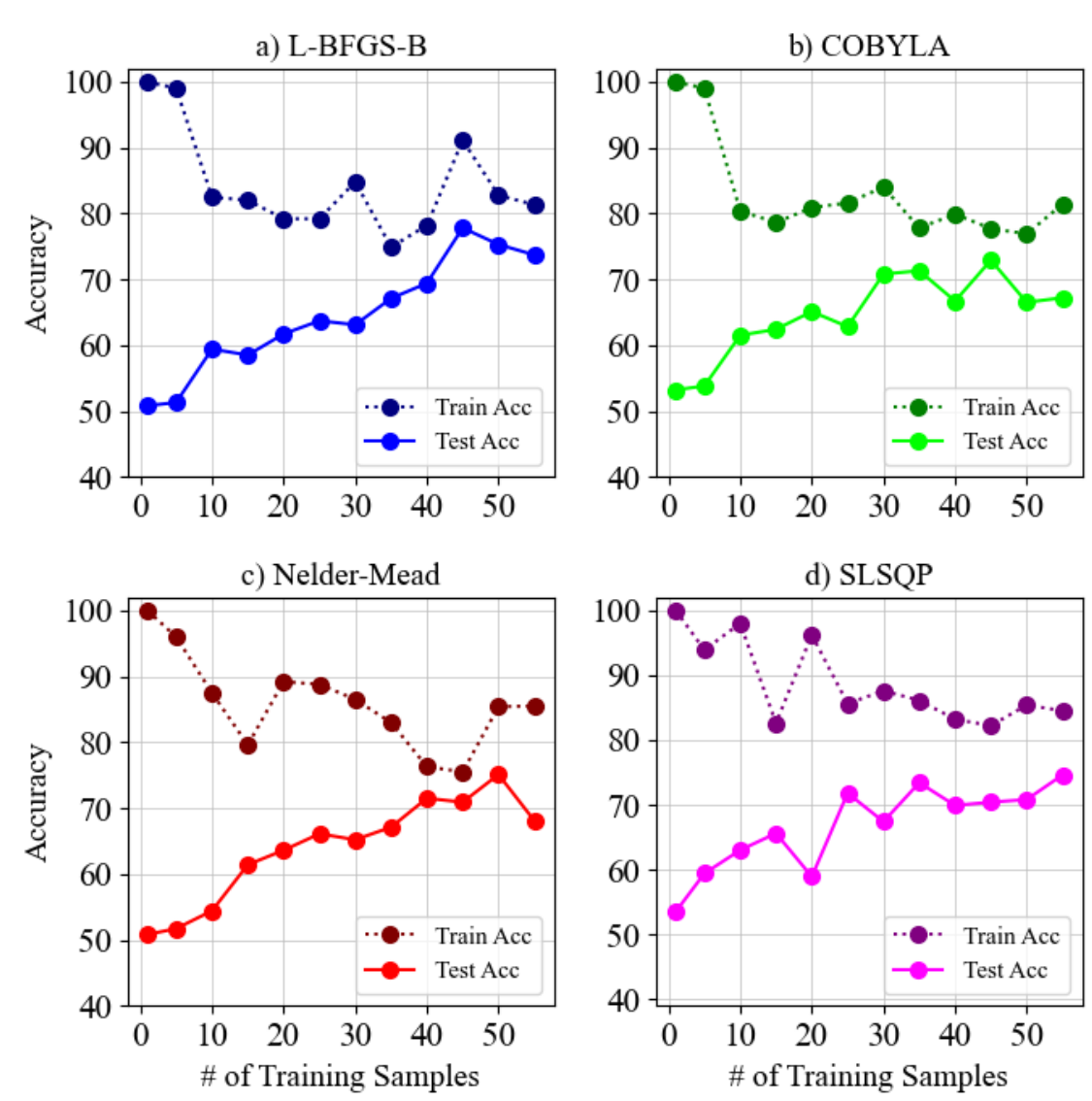

**Figure 7.** Train and test accuracy of trace distance for the 5-layer model of non-LCP and random dataset for (a) L-BFGS-B, (b) COBYLA, (c) Nelder-Mead and (d) SLSQP minimization methods.

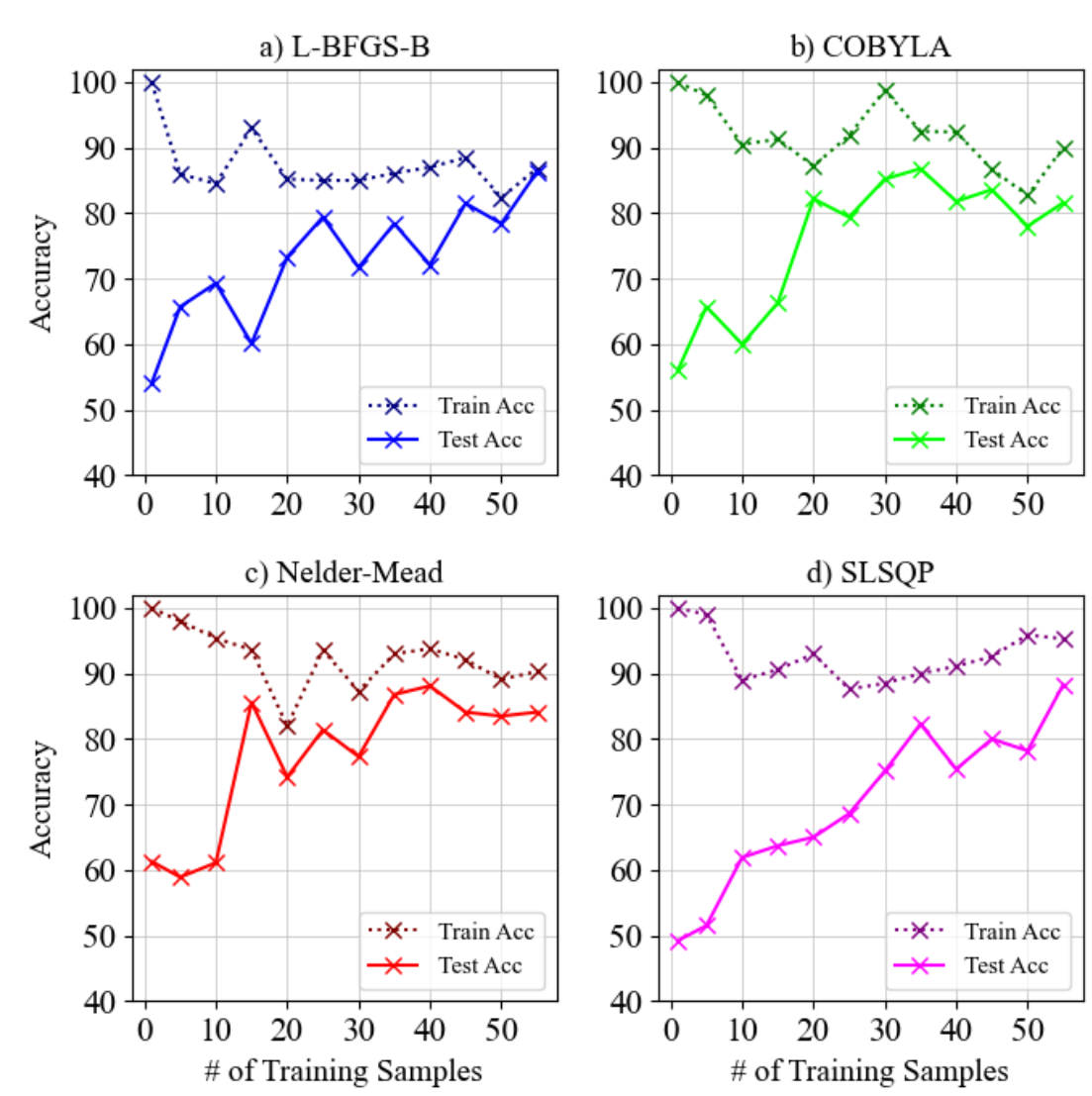

**Figure 9.** Train and test accuracy of trace distance for the 5-layer model of LCP and random dataset for (a) L-BFGS-B, (b) COBYLA, (c) Nelder-Mead and (d) SLSQP minimization methods.

Comparing the computational time, remarkably, the COBYLA minimization method completed the tasks for 2-qubit and 2-qubit entangled in just 9 minutes for both classifiers. While the COBYLA minimization method achieved the lowest test accuracy (94% for non-entangled and 95.3% for entangled), its computational time was approximately 10 times faster than the L-BFGS-B and Nelder-Mead methods, and 5 times faster than SLSQP method making it the fastest option for evaluating random datasets using the fidelity cost function and random dataset for data reuploading with 2 qubit classifier . In contrast, L-BFGS-B and NELDER MEAD take the most time, with 90 and 89 minutes for the non-entangled classifier, respectively, and reduced times of 71 and 87 minutes for the entangled version, respectively. This suggests that the 2-qubit entangled classifier offers better overall performance and generalization but requires more computational resources, highlighting the trade-off between accuracy and time in selecting the optimal minimization method in the present of real data analysis. Figure 13 presents a comparison of four optimization methods (COBYLA, L-BFGS-B, NELDER_MEAD, and SLSQP) for non-LCP using 2-qubit and 2-qubit entangled quantum classifiers, evaluating the accuracy and computational time with 250 training samples. In terms of accuracy, the 2-qubit classifier shows L-BFGS-B as the best performer, exceeding 90% in both train and test accuracies, while COBYLA has the lowest test accuracy at 76.7%. NELDER_MEAD and SLSQP offer intermediate results, with test accuracies between 82-87%. The 2-qubit entangled classifier shows improved accuracy overall, with L-BFGS-B still leading, and COBYLA notably improving to 85.4%, with smaller accuracy gaps between training and testing, indicating better generalization. On the computational side, COBYLA is the fastest for both classifiers, taking just 9 minutes, while L-BFGS-B is the slowest for the 2-qubit classifier at 130 minutes but reduces to 81 minutes in the entangled system. NELDER_MEAD and SLSQP show moderate times, with SLSQP remaining stable across both classifiers at around 42-45 minutes. The analysis highlights the trade-offs between accuracy and computational time, with the 2-qubit entangled classifier offering superior overall performance. L-BFGS-B provides the highest accuracy but with greater computational cost, while COBYLA is more efficient, making it a balanced choice for quantum classification tasks, especially in entangled systems. A detailed examination of LCP and non-LCP approaches for fidelity in fixed and random datasets for 2-qubit and 2-qubit entangled classifiers is provided in supplementary note 5.

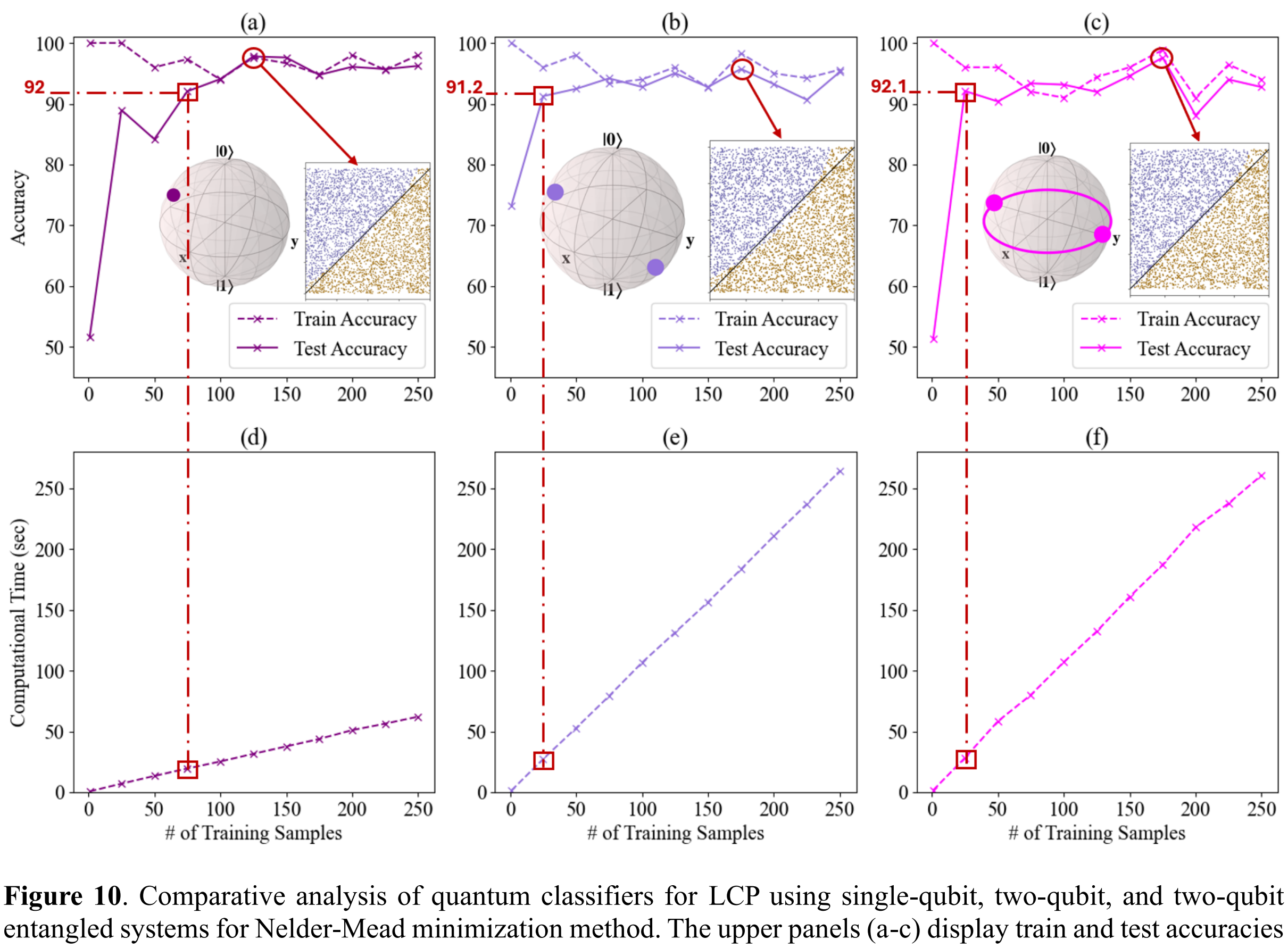

**Figure 10**. Comparative analysis of quantum classifiers for LCP using single-qubit, two-qubit, and two-qubit entangled systems for Nelder-Mead minimization method. The upper panels (a-c) display train and test accuracies as a function of training sample size, showcasing robust classification performance across all quantum configurations. The lower panels (d-f) present the relationship between computational time and the number of training samples, highlighting a substantial increase in computational complexity for two-qubit systems relative to the single-qubit implementation. This comprehensive evaluation elucidates the balance between classification accuracy and computational efficiency in quantum machine learning approaches.

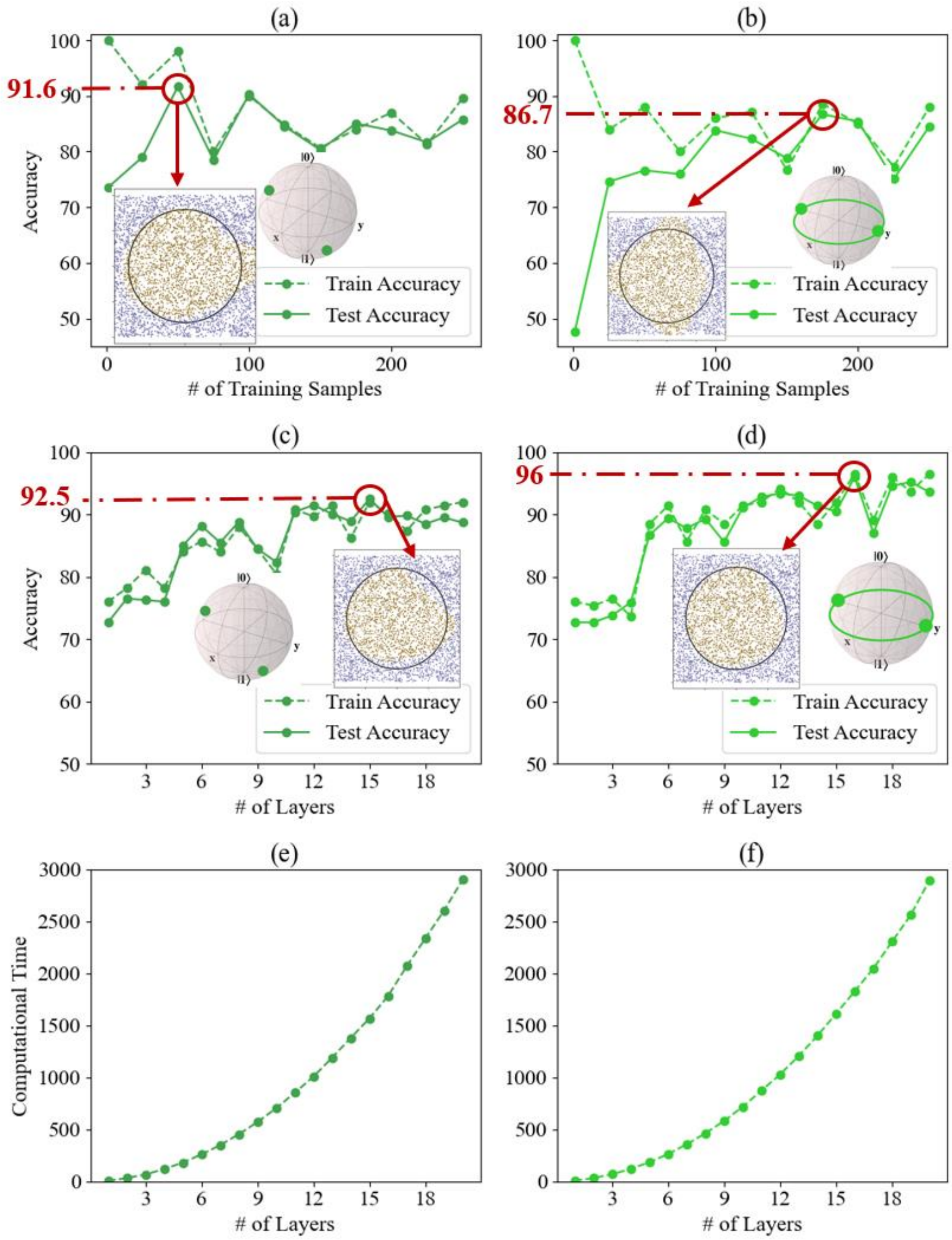

**Figure 11.** Performance analysis of 2-qubit and 2-qubit entangled classifiers for non-linear classification with Nelder-Mead minimization method for fixed dataset. (a-b) Accuracy vs. number of training samples. (c-d) Accuracy vs. number of layers. (e-f) Computational time vs. number of layers. Left column (a, c, e) shows results for the 2-qubit classifier, right column (b, d, f) for the 2-qubit entangled classifier. Subplots (c) to (f) use 175 training samples based on accuracy convergence observed in (a)and(b).

**Fidelity** **Random Data**

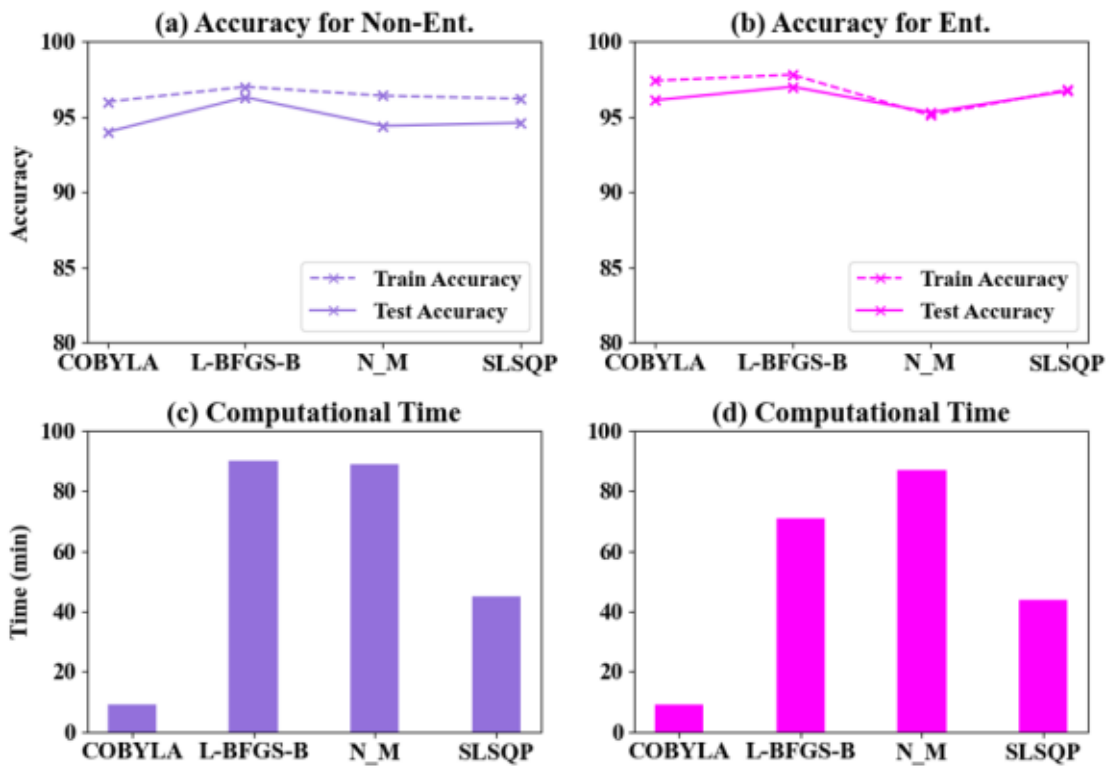

**Figure 12.** Performance analysis of 2-qubit quantum classifiers for linear classification using a random dataset with 250 training samples. (a) Accuracy for the 2-qubit classifier, (b) Accuracy for the 2-qubit entangled classifier, (c) Computational time for 2-qubit classifier, and (d) Computational time for 2-qubit entangled classifier. Results compare four optimization methods (COBYLA, L-BFGS-B, NELDER-MEAD, SLSQP) using a fidelity cost function, illustrating trade-offs between accuracy and computational efficiency.

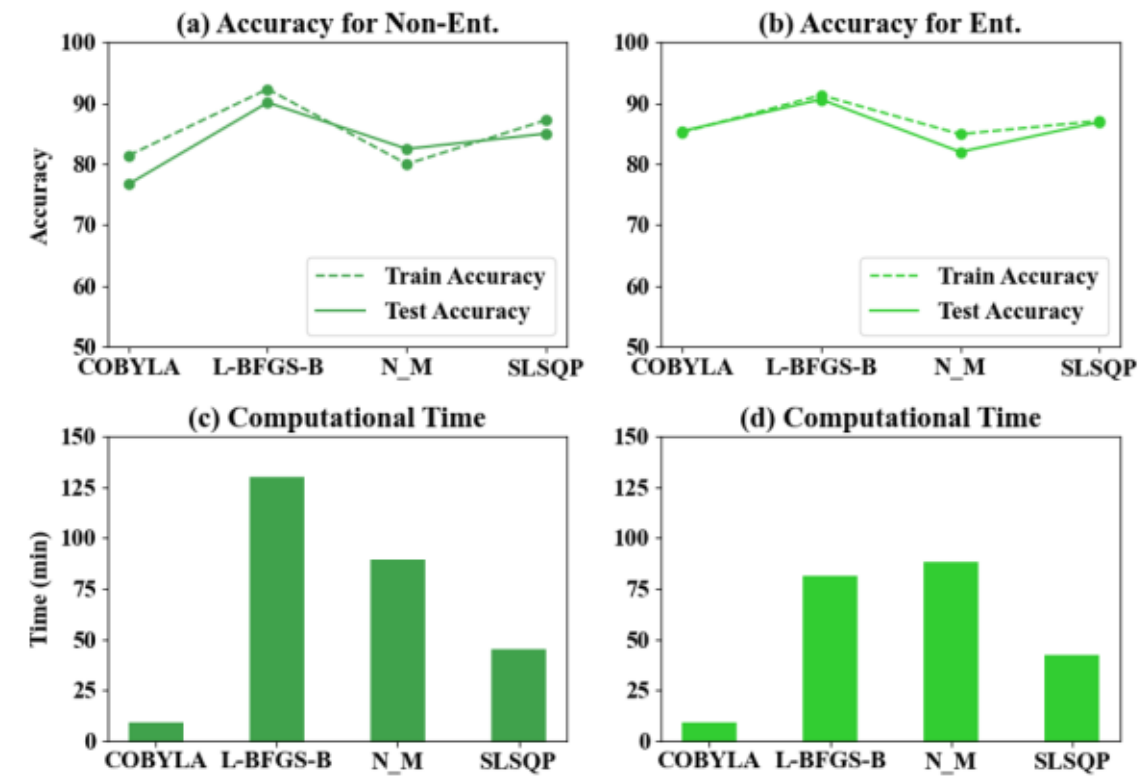

**Figure 13.** Performance comparison of 2-qubit quantum classifiers for non-linear classification using a random dataset with 250 training samples. (a) Accuracy for non-entangled classifier, (b) Accuracy for entangled classifier, (c) Computational time for non-entangled classifier, and (d) Computational time for entangled classifier. Results compare four optimization methods (COBYLA, L-BFGS-B, NELDER-MEAD, SLSQP) using a fidelity cost function, demonstrating the trade-offs between classification accuracy and computational efficiency for non-linear (circular) decision boundaries.

## DISCUSSION

This work presents a pioneering investigation into enhancing quantum classifier performance through strategic data re-uploading, exploring its impact across both linear and non-linear classification patterns. By integrating novel cost functions and employing various new optimization methods, we significantly advanced the accuracy and robustness of quantum classifiers. Our approach, which leverages the unique properties of quantum mechanics, demonstrates substantial improvements over traditional models, particularly in handling complex patterns within minimal datasets. Through comprehensive comparisons across diverse datasets and classification tasks, we underscore the adaptability of our methodology to different learning scenarios, thereby offering a versatile tool for QML applications.

Our findings contribute to the theoretical foundations of QML and provide practical insights into the design and optimization of quantum classifiers. Exploring different cost functions reveals their distinct impacts on model performance, highlighting the importance of careful selection based on the task at hand. Furthermore, our study illustrates the effectiveness of data re-uploading in enhancing model expressivity, a key factor in achieving high classification accuracy with fewer training samples, especially in the representation of real-world data sets.

Future work will focus on extending these methodologies to more complex quantum systems and exploring their application in broader quantum computing tasks. By continuing to unravel the capabilities of quantum classifiers and refining their design, we move closer to realizing the full potential of quantum computing in addressing some of the most challenging problems in machine learning and beyond. Our study also initialized the fastest method with respect to the minimization method, a number of qubit/s, and the present data sets, which show promising results using introduced methods for real-world data sets, which we will consider for our next research work.

This research not only paves the way for further advancements in QML but also highlights the transformative impact quantum computing can have across various scientific and technological domains.

## METHOD

The methodology employed in this study utilizes data re-uploading, a technique that enables sophisticated integration of data input and processing within a unified quantum circuit. The circuit's efficacy is enhanced through the optimization of rotational angles, which are governed by classical parameters. These parameters are refined iteratively by minimizing a specific cost function. This function quantifies the circuit's proficiency in categorizing data points into distinct regions on the Bloch sphere, with each region corresponding to a unique class. The final stage of the process involves quantum measurement, wherein the overlap between the resultant quantum state and predefined label states is determined, facilitating classification decisions. The architecture and operational principles of this quantum classifier can be elucidated through comparisons with classical neural networks, as illustrated in figure 14.

Figure 14(a) depicts a rudimentary two-input classical neural network, which finds its quantum counterpart in the single-qubit circuit shown in figure 14(b). This quantum circuit employs alternating data upload operations U(x) and trainable unitary operations U(ϕ). The classical network's single hidden layer corresponds to a quantum classifier utilizing a solitary qubit. The neurons in the classical hidden layer are analogous to the processing units or "layers" in the quantum classifier, denoted as "A", "B", through “N” in figure 14. In this quantum circuit, |0⟩ represents the initial qubit state, U(x) denotes data encoding, U(ϕ) represents trainable quantum operations, and the measurement symbols indicate the final readout process. The combination of U(x) and U(ϕ) constitutes a single layer in the quantum framework.

More intricate classical networks featuring additional layers correspond to multi-qubit quantum circuits implementing data re-uploading strategies. Advanced quantum circuits can incorporate entanglement between qubits, demonstrating the potential for enhanced quantum processing capabilities that surpass classical limitations.

Figure 14(c) illustrates a two-input classical neural network with two hidden layers, which is equivalent to a two-qubit quantum classifier shown in Figure 14(d). The number of hidden layers in the classical neural network corresponds to the number of processing units or layers in the quantum classifier. Similarly, the number of neurons in the classical hidden layers indicates the number of processing units or layers in the quantum classifier, represented as "A","B", through “N”, for 2-qubit as well as "A' ","B' ", through “N'” for 2-qubits entangled. In this quantum circuit, |0⟩ and |1⟩ represent the initial qubit states.

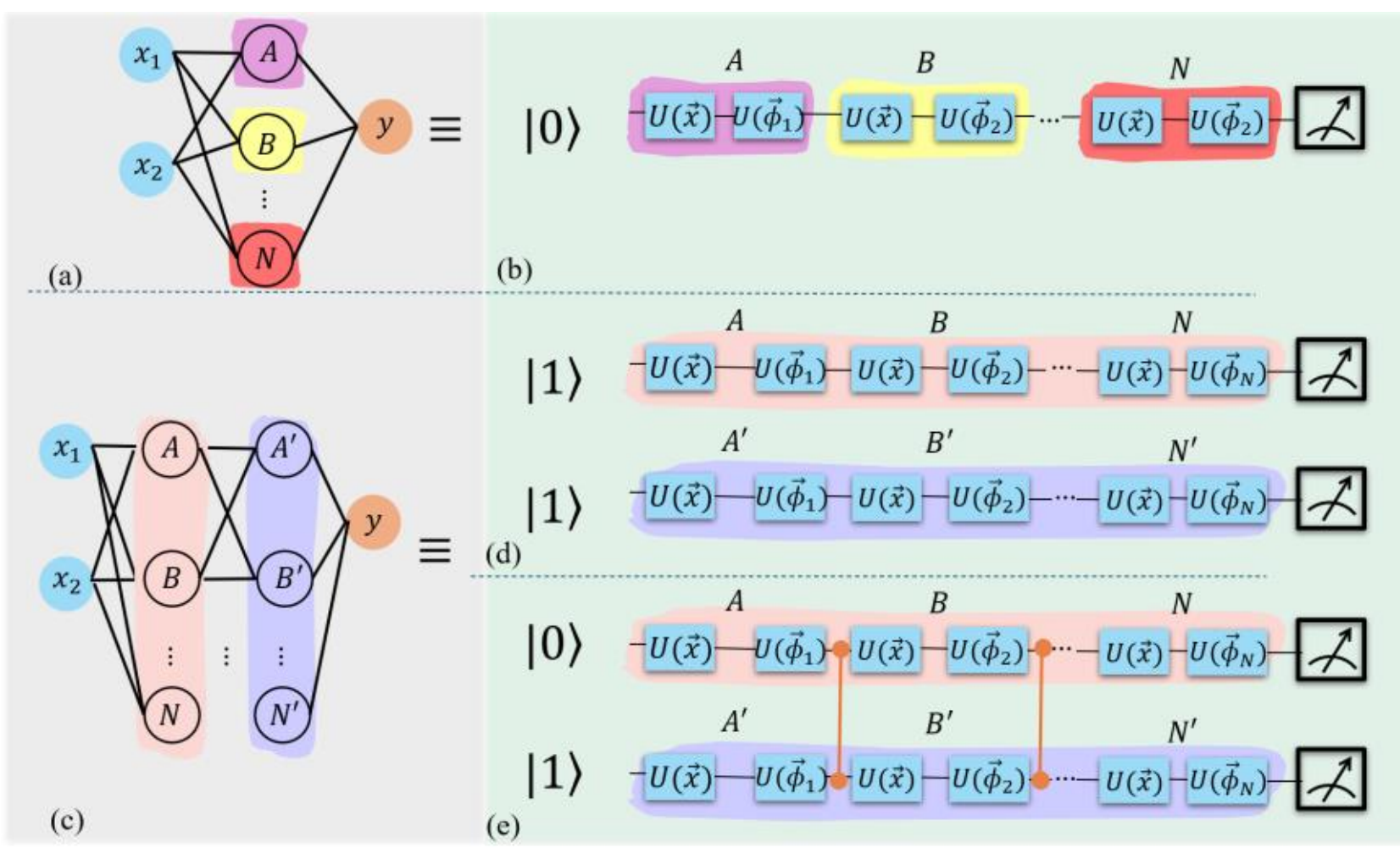

**Figure 14.** Comparison of classical neural networks and their quantum counterparts for classification tasks. (a) Classical neural network with one hidden layer (gray background) and (b) its single-qubit quantum circuit equivalent with 2 processing units (green background). The quantum circuit shows alternating data upload U(x) and learnable unitary operations U(ϕ). (c) Classical neural network with 2 hidden layers and (d) corresponding two-qubit quantum circuit implementation, demonstrating data re-uploading strategy. (e) Two-qubit entangled quantum circuit, showcasing potential for enhanced quantum processing. |0⟩ and |1⟩ represents the initial qubit states, and the measurement symbol indicates the final readout. U(x) denotes data encoding, while U(ϕ) represents trainable quantum operations.

Figure 14(e) depicts a two-qubit entangled classifier. This configuration is similar to the two-qubit classifier but incorporates a CZ gate after each processing unit or layer to induce entanglement between the two qubits, thereby enhancing the quantum circuit's computational capabilities. Supplementary Note 6 presents a thorough and detailed exploration of the process of reuploading, including how it occurs and is handled within the quantum classifier framework. It also delves into the application of various cost functions, explaining their role in optimizing the classifier's performance. Additionally, the note outlines the universality of the single-qubit classifier and transition from using a single-qubit classifier to a two-qubit classifier, discussing the steps involved, the challenges faced, and the improvements in performance that arise from the inclusion of an additional qubit. This discussion provides valuable insights into the evolution and scaling of quantum classifiers.

LCP and non-LCP approaches, and cost functions for the quantum classifier, please refer to supplementary note 6.

Optimizing a single-qubit classifier involves minimizing a function across a complex parameter space. This paper evaluates four distinct minimization techniques: L-BFGS-B, COBYLA, Nelder-Mead, and Sequential Least Squares Programming (SLSQP). L-BFGS-B, a quasi-Newton method, efficiently handles large-scale problems with linear memory usage. COBYLA, designed for constrained optimization, doesn't require derivative calculations. Nelder-Mead, a direct search method, is effective for problems lacking derivative information. SLSQP minimizes functions while adhering to specific constraints, using a quadratic approximation of the objective function. Each method has unique strengths and limitations, with their effectiveness varying based on the specific classification task and dataset characteristics. The choice of optimization method significantly impacts the classifier's performance, especially when dealing with smaller training sets and the inherent challenges of local minima in quantum circuits. For a detailed explanation of the minimization methods employed in this study, please refer to supplementary note 7. supplementary note 8 presents the comparison between references[14] and our development in detail.

## DATA AVAILABILITY

All data generated or analyzed during this study are included in this published article and its supplementary information files.

# Boosting Quantum Classifier Efficiency through Data Re-Uploading and Dual Cost Functions
# Supplementary Documentation

**Authors:** Sara Aminpour[1,2], Mike Banad[1], Sarah Sharif[1,2]*

**Author Affiliations:**
[1] School of Electrical and Computer Engineering, University of Oklahoma, Norman, OK 73019, USA
[2]Center for Quantum and Technology, University of Oklahoma, Norman, OK 73019 USA

*Corresponding author: Sarah Sharif (email: s.sh@ou.edu)

**Table of contents**

## Supplementary Note 1: Range of training samples and number of layers

Figure S1.1 illustrates the performance of a quantum classifier utilizing a fidelity cost function within a five-layer framework for circular pattern classification in a fixed dataset, employing the L-BFGS-B optimization method. The analysis encompasses training data up to 250 samples to benchmark our algorithm against the findings from the reference [1]. The diagram depicts training accuracy with a blue dashed line and test accuracy with a solid blue line, underscoring the algorithm's efficacy. A red dot highlights a notable benchmark from the reference, showing an 89% accuracy with 200 training samples, demonstrating parity with this published result. The inset provides a visual representation of the classification process. Notably, test accuracy begins at approximately 70%, rising impressively to 96% for a slightly expanded dataset of 210 samples. Remarkably, with as few as 60 training samples, the model achieves a test accuracy of 91.8%, and the discrepancy between training and test accuracy diminishes with the inclusion of 90 samples. This observation underscores the efficiency of our approach, highlighting its capability to reach high accuracy levels without necessitating extensive training data.

Figure S1.2 showcases a systematic evaluation of a circular pattern classification model across a spectrum of architectural depths, ranging from 1 to 5 layers. The graphical analysis reveals that models with a solitary layer lag in performance compared to those with increased layer counts, marking a clear trend: as the number of layers escalates, so does the model's classification accuracy. Specifically, a single-layer setup achieves a peak accuracy of 61.9%, whereas a more complex five-layer configuration significantly elevates this metric to 88.8%, even when limited to only 35 training samples. This observation underscores a critical insight—enhancing the model's depth systematically improves its predictive capabilities, a phenomenon consistent with the advantages afforded by the data reuploading strategy integral to our approach. Given this marked improvement in model efficacy with layer augmentation, the paper prioritizes an in-depth investigation and discourse on the five-layer model's architecture, focusing on its ability to optimize classification accuracy with efficient utilization of training data.

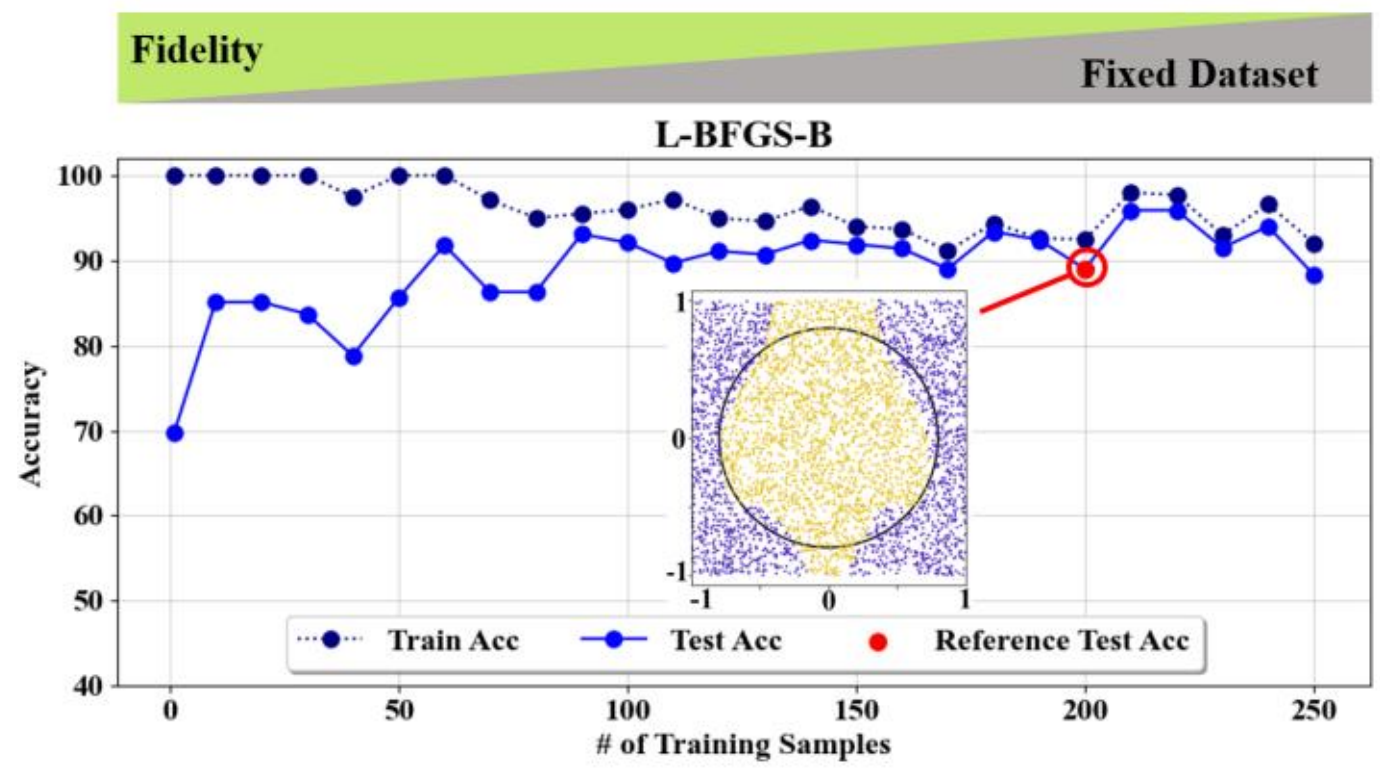

Figure S1.1 Train and test accuracy of fidelity for the 5-layer model of circle classification and fixed dataset for L-BFGS-B minimization method. The inset graph shows the visualization of a nonlinear classification reported on[1].

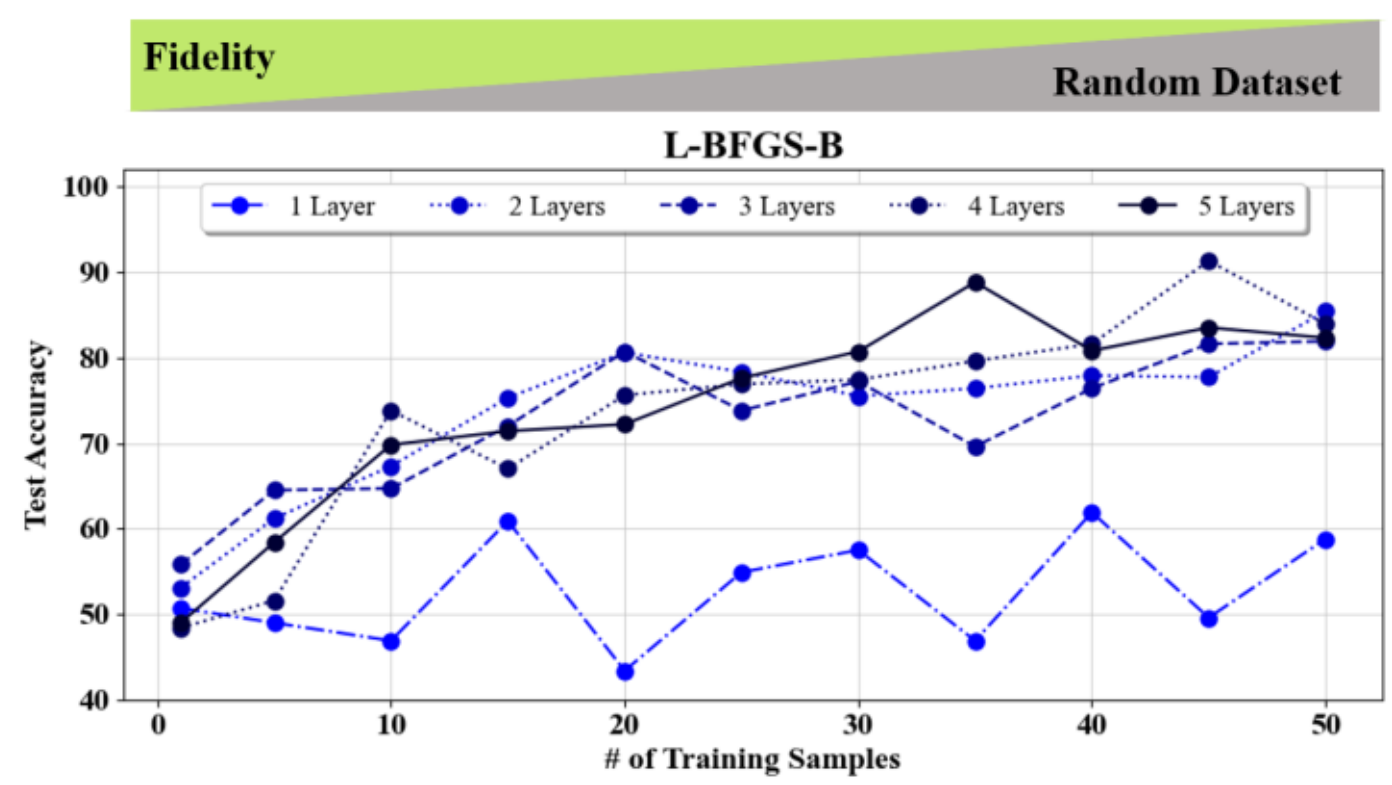

Figure S1.2. Evaluate the test accuracy of fidelity for circle classification and random dataset for L-BFGS-B minimization method, ranging from 1 to 5 layers.

## Supplementary Note 2: Evaluating non-linear and linear classification approaches for fidelity cost function in fixed and random datasets for 1-qubit classifier for four different minimization methods

Figure 2 illustrates a comparison of four distinct optimization techniques, namely L-BFGS-B, COBYLA, Nelder-Mead, and SLSQP, applied to the task of classifying the circle pattern. The comparison evaluates both training and test accuracies using a fixed dataset of 4000 test samples and 5 layers. Initially, all algorithms demonstrate a perfect training accuracy of 100% with just a single sample, a result that aligns with expectations. However, as we increase the sample size, a divergence in performance becomes evident for these four minimization methods. The L- BFGS-B method maintains a training accuracy close to 90%, showcasing its robustness against overfitting. In contrast, COBYLA, Nelder-Mead, and SLSQP show significant variability and a decline in training accuracy, indicating a susceptibility to overfitting. Interestingly, the peak accuracy for COBYLA, Nelder-Mead, and SLSQP is achieved with merely 50 samples, beyond which overfitting becomes a significant issue. This observation suggests that, unlike L-BFGS-B, which requires a minimum of 100 samples to achieve an accuracy of 92%, the other three methods can attain over 95% accuracy with only 50 samples. L-BFGS-B does not reach this high accuracy level at 100 samples, and its performance slightly declines with an increase in training samples after 150 training samples. This analysis highlights the critical importance of carefully selecting the number of training samples based on the minimization method used. The right choice can effectively prevent overfitting, thereby enhancing classification accuracy. This insight is crucial for optimizing machine learning models and ensuring their generalizability and efficiency in practical applications.

Figure 3 delves into the accuracy of these four distinct minimization methods —L-BFGS-B, COBYLA, Nelder-Mead, and SLSQP— when applied to a fidelity cost function and a random dataset for circle classification. This analysis underscores a consistent trend across all methods: an initial increase in test accuracy corresponding to the rise in the number of training samples, yet fails to surpass a peak accuracy of 90%. This trend highlights the inherent challenges faced by these minimization methods when dealing with random datasets. In the L-BFGS-B method as depicted in figure 3(a), showcases a notable performance, achieving its highest test accuracy of 88.8% with 35 training samples. This point also marks the narrowest gap of 5% between training and test accuracy, indicating a relatively high level of model efficiency and generalization at this sample size. However, as the analysis progresses, it becomes apparent that increasing the number of training samples beyond this optimal point does not translate to improved performance. The gap between the train and test accuracy remains notably constant at around 10% even as the sample size is increased to 70 training samples. Transitioning to the COBYLA method, as depicted in figure 3(b), a different performance pattern emerges. Contrary to L-BFGS-B, COBYLA achieves its best test accuracy at 84.8% with a higher training sample equal to 70. This method experiences fluctuations, yet it is noteworthy that the gap between training and test accuracies exhibits a decreasing trend, suggesting a gradual improvement in model generalization compared to the initial stability seen with L-BFGS-B. Figure 3(c) focuses on the Nelder-Mead method, highlighting a decrease in the gap between training and test accuracies as the number of training samples increases, culminating in a maximum accuracy of 86.9% with 60 training samples. Figure 3(d) examines the SLSQP method, which shows an increase in test accuracy up to 50 training samples before demonstrating a decline in both training and test accuracies. This shows the SLSQP method is more prone to overfitting. The SLSQP method reaches a maximum accuracy of 86.7% when applied to a dataset of 50 samples. These results, as detailed in figure 6, provide vital insights into the performance of various minimization methods when working with a fidelity cost function and a random dataset. The diverse outcomes emphasize the importance of choosing an optimal number of training samples to prevent overfitting and enhance accuracy. This underlines the delicate balance needed to fully leverage these computational methods in practical scenarios.

Figure 4 illustrates a comparison of four different optimization techniques applied to the task of classifying line patterns, using fidelity-based cost function and the fixed dataset. The subplot (a) focuses on the performance of the L-BFGS-B method. Here, the training accuracy starts at a perfect 100% and impressively remains above 97% even as the number of training samples increases. Conversely, the test accuracy initiates at a relatively lower rate of 62.2% with just a single sample yet it progressively improves, reaching approximately 95% accuracy with 75 training samples and slightly declines for larger training samples. An initial notable gap between the training and test accuracy is evident, but this gap diminishes significantly as the dataset expands with more training data, indicating an improvement in the model's ability to generalize from the training to the unseen test data. The subplot (b) depicts the results obtained using the COBYLA algorithm, which exhibits a performance pattern similar to that of the L-BFGS-B method, consistently achieving 100% accuracy on the training data. The accuracy on the test set starts at 66.9% and steadily improves as more training samples are added, ultimately reaching 95% when 125 samples are used for training. The disparity between training and test set accuracies mirrors the pattern observed with the L-BFGS-B method, consistently manifesting across all training dataset sizes. The Nelder-Mead approach, shown in figure 4(c), achieves a notable test accuracy of 97.7% with 125 training samples. The inset provides a graphical visualization of line classification using this minimization method at this specific point, illustrating that the line classification performance is exceptionally well. The visualization clearly demonstrates the method's effectiveness in accurately separating the data points into distinct classes, highlighting the Nelder-Mead method's precision and robustness in handling line classification tasks with a substantial number of training samples. Furthermore, the training and test accuracy curves show a notably smaller gap, converging to the same value with training sets of 100 and 125 samples. The final subplot (d) evaluates the performance of the SLSQP method, which closely aligns with the results from the COBYLA method. The test set accuracy exhibits a progressive increase, rising from 62.7% to 96.6%. The disparity between the training and test accuracies is similar to that observed with the COBYLA method. In summary, all four optimization techniques demonstrate a reduction in overfitting as the training dataset size increases, ultimately achieving a test accuracy of at least 95% when training with 125 samples for this line classification task.

Figure 5 showcases an analysis of the classification accuracy obtained using the same minimization methods across random datasets. Consistently, a rise in the number of training samples correlates with an increase in test accuracy across all methods evaluated. Notably, with just 50 training samples, all methods surpass the 90% accuracy threshold. Specifically, in figure 5(a), the L-BFGS-B method reaches the peak accuracy of 92.8% with 50 training samples. It was observed that as the number of samples increased, the disparity between train and test accuracies for the L-BFGS-B method began to narrow, although this gap persisted in being slightly wider than that observed in the other methods. Figure 5(b) demonstrates that the COBYLA method, with the same number of samples, attains a superior accuracy of 93.5%. This suggests that COBYLA not only reaches high classification accuracy with a minimal dataset but also demonstrates better generalization compared to L-BFGS-B, as reflected by the narrower gap between its training and test accuracies. Figure 5(c) examines the Nelder-Mead method, showing its peak accuracy of 93% with 40 training samples, after which its accuracy slightly declines. Interestingly, the smallest disparity between training and test accuracies—only 1.8%—occurs in 50 training samples. Despite slightly lower accuracy at this point, this smallest gap signifies that the Nelder-Mead method achieves a remarkable balance between learning from the training data and generalizing to unseen data, highlighting its efficiency and potential for precise model tuning. Figure 5(d) illustrates that the SLSQP method achieves an impressive peak test accuracy of 96.4% for line classification using a random dataset, attained with 45 training samples. At this juncture, the discrepancy between training and test accuracies is notably small, indicating a high level of model precision and generalization. Like the Nelder-Mead method, the SLSQP method exhibits a nonmonotonic increment in test accuracy as a function of training samples, as indicated by the irregular slope of test accuracy. This fluctuation suggests that for these methods, adding more training samples does not straightforwardly translate to higher test accuracies, highlighting the complexity of optimizing model performance across different minimization techniques.

A comparison of figures 2 and 4 reveals that the accuracy curves for line classification are more stable and consistent across all optimization techniques when compared to those for circle classification. The accuracy values for classifying circle patterns display greater variability and fluctuations than those observed in the line classification task. The observed differences in performance between circle and line classification could stem from several technical factors: (1) Line classification likely represents a more straightforward pattern that aligns better with the linear decision boundaries most classifiers are adept at identifying. In contrast, circle classification involves recognizing more complex, non-LCP, which can challenge the classifiers' ability to generalize from the training data without overfitting or underfitting. (2) The algorithms applied for circle classification might be more prone to getting trapped in local minima due to the more intricate decision boundaries required to accurately classify circular patterns. This can hinder the optimization process, leading to increased fluctuations in classification accuracy as the model struggles to find the global optimum. (3) The differences in performance may also reflect the inherent adaptability of the algorithms to the specific types of classification tasks with the geometric properties. A comparative analysis of Figures 6 and 8 indicates that the specific characteristics of the classification problem significantly affect the potential to attain higher accuracy with fewer samples. The fluctuations in the line classification pattern are less pronounced than those in the circle classification pattern. This observation underscores the importance of selecting appropriate optimization methods based on the complexity of the classification problem.

## Supplementary Note 3: Evaluating non-linear and linear classification approaches for trace distance cost function in fixed and random datasets for 1-qubit classifier

Figure 6 showcases the effectiveness of the trace distance cost function in classifying circular patterns within a fixed dataset. In subplot (a), the L-BFGS-B minimization method achieves its highest test accuracy at 79.2% with a dataset comprising 100 training samples. Subplot (b) examines the performance of the COBYLA method, which displays greater variability in training accuracy than L-BFGS-B but ultimately achieves a higher peak test accuracy of 84.6%, also with 100 training samples. Notably, COBYLA demonstrates enhanced generalization capabilities relative to other methods, as indicated by the narrower margin between its training and testing accuracies. This performance suggests that, when applied alongside the trace distance cost function, the COBYLA method is particularly adept at optimizing parameters for improved generalization to unseen testing data. An accompanying visualization within the inset illustrates the classification of circular patterns at this accuracy peak. In subplot (c), the analysis shifts to the performance of the Nelder-Mead method, which records its optimal test accuracy at 72.6% utilizing 60 training samples. This method exhibits signs of overfitting, a condition where the model learns the training data too closely and fails to generalize well to new, unseen data. Despite a narrowing gap between training and testing accuracies as the number of training samples grows, a concurrent decline in training accuracy is observed, which adversely affects the overall test accuracy. This pattern suggests a limitation in the Nelder-Mead method's capacity to effectively handle the trace distance cost function, likely due to its inherent characteristics such as reliance on simplex-based optimization, which might struggle with the complexity of the trace distance landscape. Consequently, this method appears less suited for tasks requiring robust generalization from the trace distance cost function, particularly in scenarios demanding accurate classification of complex patterns with a limited dataset. In subplot (d), the focus turns to the SLSQP method which attains its peak test accuracy at 83.6% with a dataset of 100 training samples. The disparity between training and testing accuracy contracts by increasing the training samples, indicating an improvement in the model's ability to generalize from the training to the testing dataset. However, even at the point of 100 training samples, the gap between training and testing accuracies, while reduced, remains significant. This persistent gap suggests that while the SLSQP method is effective at learning and generalizing from the given data, there is still a margin for optimization to further bridge the difference in accuracies. Each optimization technique successfully minimizes the cost function and attains perfect accuracy on the training set using a comparatively small number of samples. However, their performance varies considerably when it comes to generalizing to the test set. This highlights the crucial role played by the choice of optimization algorithm in determining the overall effectiveness of the model. In conclusion, when considering the fixed dataset and the trace distance cost function, the COBYLA method demonstrates superior performance in optimizing the parameters to generalize effectively to unseen test data. Compared to the other techniques evaluated, it necessitates fewer training samples to achieve satisfactory accuracy on the test set.

Figure 7 illustrates how the accuracy on both the training and test sets evolves as the number of training samples grows, specifically for the task of classifying circular patterns using the trace distance cost function, evaluated on a randomly generated dataset. Similar to all scenarios analyzed so far, a common pattern emerges where test accuracy begins at a relatively low level for all minimization methods but demonstrates a consistent increase as more training data is provided. This trend highlights the methods' capacity to effectively learn distinguishing features, thereby enhancing their ability to generalize to unseen data. Specifically, in subplot (a), the L-BFGS-B method illustrates impressive learning efficiency, with test accuracy exceeding 70% after incorporating just 40 training samples and achieving its highest test accuracy of 77.8% with 45 training samples. In subplot (b), the COBYLA method's performance is slightly lower compared to L-BFGS-B, plateauing at a test accuracy of 72.8% with 45 training samples. This performance indicates that while COBYLA may be susceptible to some degree of overfitting, it nonetheless achieves a reasonable level of generalization. Subplot (c) explores the Nelder-Mead method, which reaches its peak test accuracy of 75.1% with 50 training samples. Subplot (d) utilizes the SLSQP method, which shows fluctuations in its training accuracy remaining above 80%. The test accuracy for SLSQP was enhanced significantly, reaching 74.6% with 50 samples. This fluctuation and eventual rise in test accuracy underscores the method's potential for optimizing classification tasks, despite the initial variability. In sum, the L-BFGS-B method stands out for achieving the highest test accuracy among the methods evaluated, requiring only 45 training samples to reach this optimum on a random dataset. Summarily, employing the trace distance cost function across these various minimization strategies yields test accuracy ranging from 65% to 78% on the random dataset, illustrating the function's effectiveness and the distinct performance capabilities of each minimization method.

Figure 8 offers a comparative analysis of the accuracy achieved by four different optimization methods when applied to a trace distance cost function for line pattern classification using a fixed dataset. Subplot (a) highlights the L-BFGS-B method, showcasing its high level of stability in training accuracy. The test accuracy shows a steady increase, reaching 91.8% with 100 training samples. While there is a substantial gap between the accuracies of the training and test sets at the outset, this difference gradually narrows as more training samples are introduced. This highlights the L-BFGS-B method's capacity to adapt and learn more complex patterns effectively, demonstrating robustness and in leveraging larger datasets for improved generalization. The subplot (b) illustrates the results obtained using the COBYLA method. In contrast to the L-BFGS-B approach, the accuracy of the training set shows greater fluctuations, even experiencing a drop to 56.9% at one instance before rebounding. The test accuracy follows a similar pattern to that seen in L-BFGS-B, beginning at 49.8% and increasing to 87.4%. Once the training set size reaches 80 samples, both the training and test accuracies seem to reach a plateau, slightly below the 90% mark. In subplot (c), the Nelder-Mead method starts with a modest test accuracy of 55.3%, which significantly improves to 87% with the addition of 60 training samples demonstrating a similar trend as the L-BFGS-B method. Initially, a pronounced gap exists between training and test accuracies, which persists until the dataset is expanded to include 80 training samples. Beyond this point, the sign of overfitting emerges, as demonstrated by a decline in training accuracy while test accuracy plateaus. For 100 training samples, the test accuracy interestingly becomes 2% higher than the training accuracy, indicating a unique inversion where the model performs slightly better on unseen data than on the training set itself, a rare occurrence that may suggest the model has reached a point of optimization where it generalizes exceptionally well to new data. The subplot (d) of figure 11 presents the results of the SLSQP method. Notably, this technique achieves the highest accuracy on the test set, reaching 93.3% using just 40 training examples. The SLSQP method appears to be the most appropriate choice for trace distance classification tasks, as it exhibits a smaller discrepancy between its performance on the training and test datasets. The inset provides a visual representation of the SLSQP's performance at this specific point. To summarize, all optimization methods demonstrate an upward trajectory in test accuracy as the size of the training dataset increases, suggesting enhanced generalization capabilities of the model. Among the four techniques evaluated, the SLSQP method seems to strike the most favorable balance between its performance on the training and test sets.

Figure 9 presents a comparison of different optimization techniques when applied to the task of classifying line pattern using a randomly generated dataset and a cost function based on trace distance. In subplot (a), we examine the performance of the L-BFGS-B method, which attains its peak test accuracy of 86.3% with 55 training samples. Before reaching this point, the method's test accuracy demonstrated considerable variability, oscillating between 70% and 80% as the number of training samples ranged from 20 to 50. However, a notable improvement occurs when the dataset is expanded to 55 training samples, at which the test accuracy leaps to 86.3%, effectively surpassing the earlier fluctuation band. This pivotal moment also marks the occurrence of the smallest gap between training and test accuracies, showcasing a significant enhancement in the model's ability to generalize from the training dataset to unseen data, thereby achieving an optimal balance at this specific training sample size. Subplot (b) delves into the efficacy of the COBYLA optimization method, which achieves its highest test accuracy of 86.8% with a relatively smaller dataset of 35 training samples. Beyond this optimal threshold, signs of overfitting become apparent, as both training and test accuracies start to decline. This pattern suggests that while the COBYLA method is highly effective up to a certain point, adding more training samples beyond this number paradoxically hampers the model's performance. The decline in accuracy indicates that the model begins to memorize the training data rather than learning to generalize, leading to a decrease in its ability to accurately predict outcomes on unseen data. This observation underscores the importance of identifying the ideal number of training samples to maximize the effectiveness of the COBYLA method without crossing into the territory of overfitting. In subplot (c), the focus is on the Nelder-Mead optimization method, which shows some fluctuations in performance before reaching its maximum test accuracy. It successfully achieves a test accuracy of 88.1% with 40 training samples. However, akin to the pattern observed with the COBYLA method, the Nelder-Mead method also sees a decline in both training and test accuracies when additional training samples are added beyond this optimal number. This decline serves as a clear indication of the onset of overfitting, suggesting that while the Nelder-Mead method can efficiently utilize a certain number of training samples to improve its predictive accuracy, exceeding this number leads to a reduction in model performance. In subplot (d), a more continuous and stable increase in test accuracy is observed with each increase in the number of training samples. This trend results in the highest test accuracy being recorded at 88.3% with 55 training samples. Unlike the previous methods discussed, this subplot suggests a method that maintains its efficiency and ability to generalize well without showing immediate signs of overfitting up to this point. The gradual and consistent improvement in test accuracy highlights the method's effective learning curve and suggests an optimal balance between learning from the training data and applying this knowledge to unseen data.

## Supplementary Note 4: performance comparison of 5-Layer single-qubit quantum classifiers using fidelity and trace distance cost functions across various classification tasks and dataset types

Figure S4.1 offers a comparative analysis of the highest accuracies achieved for two distinct classification patterns – linear (line) and non-linear (circle) – across the four distinct minimization methods when applied to both random and fixed datasets within the context of a fidelity cost function. The analysis reveals a notable trend: in circle classification tasks, the fixed dataset consistently yields higher accuracies than their random counterparts for all tested minimization methods. This suggests that the inherent geometric complexities of non-LCP may align more closely with the simpler structure of fixed datasets, thereby facilitating more accurate classification. Similarly, for line classification, the fixed dataset leads to enhanced accuracies with the L-BFGS-B and SLSQP methods, indicating these methods' effectiveness in leveraging structured data to accurately discern linear relationships. However, the random dataset achieves better accuracy when classified using the Nelder-Mead method. This could suggest that the Nelder-Mead method, known for its simplicity and direct search approach, might be particularly adept at navigating the stochastic nature of random datasets to identify linear patterns. Across all algorithms, the task of classifying non-LCP, especially within random datasets, emerges as inherently challenging. This complexity likely stems from the algorithms' varying abilities to parse and learn from the unpredictable variance found in random datasets, as well as the added difficulty of accurately modeling non-linear relationships. The findings underscore the critical importance of selecting the appropriate minimization method based on the dataset's nature and the classification task's geometric complexity to optimize classification accuracy.

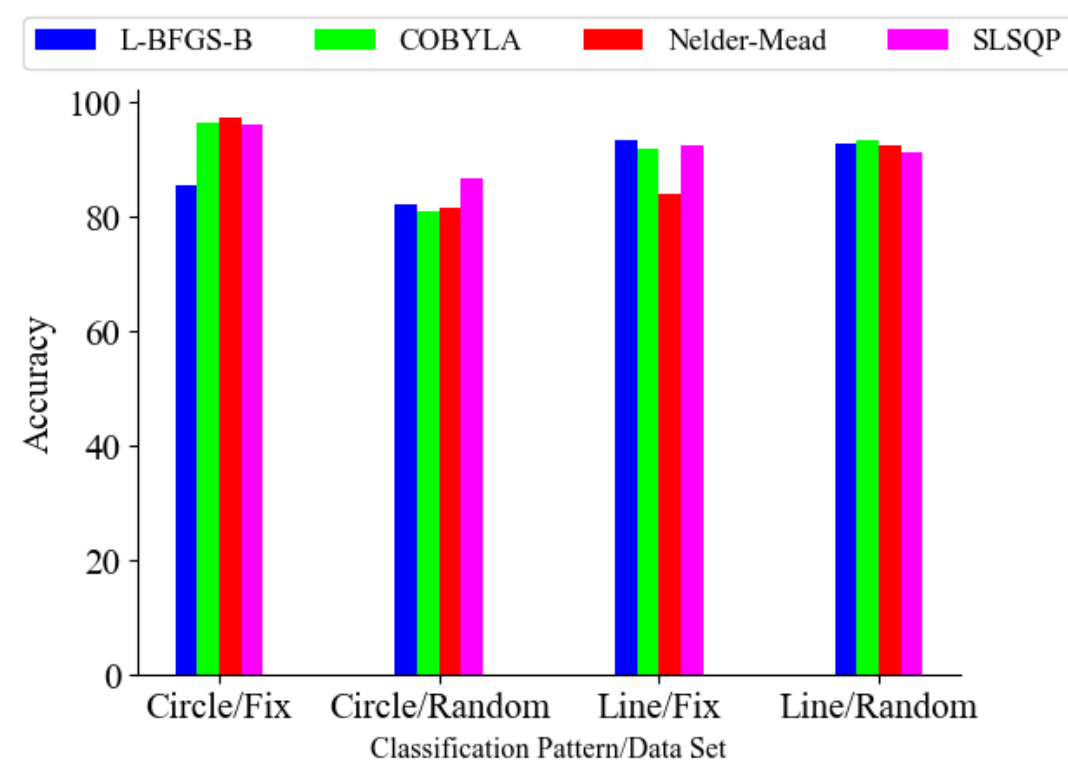

Figure S4.1. Evaluating of Fidelity cost function test accuracy of 5-layer model across 50 samples for LCP and non-LCP problems for random and fixed datasets in four minimization methods.

Figure S4.2 provides the performance comparison of two distinct classification patterns—line and circle—across four different minimization methods when applied to both random and fixed datasets, this time employing the trace distance cost function. A pivotal observation emerges when comparing the performance of circle classification with a fixed dataset (circle/fixed) against the fidelity cost function results presented in figure S4.1. It is evident that the accuracies achieved using the trace distance cost function are notably lower across all minimization methods compared to those obtained with the fidelity cost function. This discrepancy highlights the inherent challenges and differences in how each cost function interacts with the underlying data and the classification task at hand. The trace distance cost function, known for quantifying the distinguishability between quantum states, may present a more complex landscape for optimization, particularly when applied to classical data patterns such as lines and circles. This complexity could lead to lower classification accuracy as the minimization methods struggle to navigate the nuances of the trace distance landscape effectively. Such an observation underscores the importance of cost function selection in machine learning tasks, emphasizing that the choice of cost function can significantly impact the model's ability to learn and generalize from the data. The comparative analysis in figure S4.2 serves as a testament to the nuanced interplay between cost functions, dataset types (fixed vs. random), and the geometric nature of the classification patterns, offering valuable insights into optimizing classification accuracy through strategic method and cost function selection.

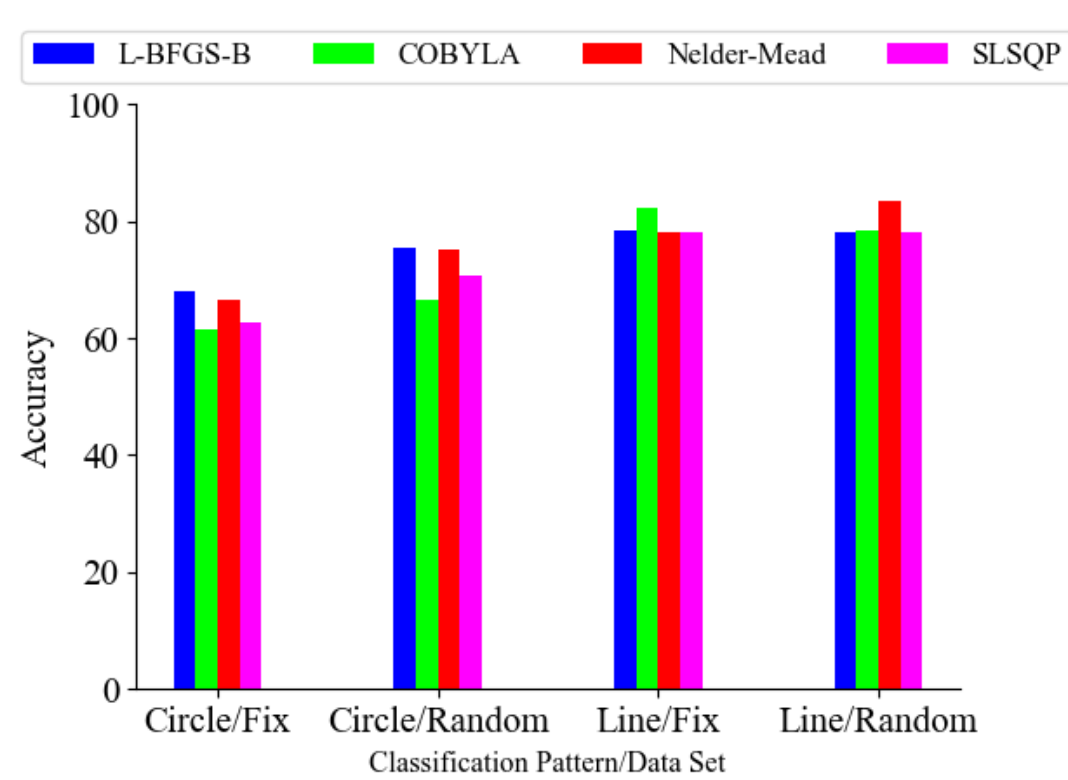

Figure S4.2. Evaluating of trace distance test accuracy of 5-layer model across 50 samples for LCP and non-LCP problems for random and fixed datasets in four minimization methods.

In addition, the fixed dataset achieves superior accuracy specifically when employing the COBYLA minimization method, indicating a unique synergy between COBYLA's optimization strategy and the structured nature of fixed datasets for LCP. Conversely, for the random dataset, there's a notable trend where it consistently outperforms the fixed dataset across all other minimization methods, suggesting that the stochastic characteristics of random datasets may be better suited to the optimization landscapes these methods navigate, particularly for LCP. In circle classification tasks, the random dataset not only demonstrates improved accuracy over the fixed dataset for all minimization methods but also reinforces the observation that random datasets generally offer a more favorable context for the trace distance cost function across both classification patterns. This enhancement in accuracy with random datasets could be attributed to the trace distance cost function's sensitivity to the variances within the dataset, allowing for more effective differentiation and classification of non-LCP like circles when the data is less predictable.

## Supplementary Note 5: Evaluating non-linear and linear classification approaches for fidelity in fixed and random datasets for 2-qubit and 2-qubit entangled classifiers

Focusing on figure 10(a), we observe the performance of a single-qubit system applied to a LCP pattern. The system demonstrates a steep initial learning curve, with accuracy rapidly increasing from 51.6% to 92% after just 75 training samples. This sharp rise highlights the single-qubit system's ability to efficiently learn and generalize from a relatively small dataset. The notable jump in accuracy suggests that a properly trained single-qubit classifier can capture the essential features of the LCP task with high precision. After reaching 92% accuracy at 75 training samples, the system stabilizes, maintaining a test accuracy consistently in the range of 92% to 97.7% as the training sample size increases to 125. The minimal fluctuation in accuracy indicates a robust performance, with the single-qubit system effectively avoiding overfitting even as the training data expands. The stable test accuracy underscores the system's reliability and suitability for LCP tasks where computational simplicity and consistent performance are crucial. In terms of computational cost, as shown in figure 1(d), the single-qubit system exhibits a gentle increase in computational time, reaching 62.15 seconds for 250 training samples. This computational efficiency, coupled with the system's stable accuracy, makes the single-qubit classifier an appealing option for linear problems, particularly in scenarios where computational resources are limited but high accuracy is still required.

In figure 10(b), the performance of the 2-qubit classifier in a LCP task shows a more gradual improvement in accuracy compared to the single-qubit system. The initial accuracy is relatively high, starting at 73.2% with just one training sample, which suggests that the additional qubit provides a more robust representation of the problem space even with minimal training. As the number of training samples increases to 75, the accuracy rises steadily, reaching 94.1%. This gradual improvement, as opposed to the sharp jump seen in the single-qubit system, highlights the ability of the 2-qubit classifier to build on its already strong initial performance with increasing training data. Beyond 50 training samples, the 2-qubit classifier continues to demonstrate incremental gains, eventually peaking at around 95.7% test accuracy with 175 training samples. Notably, the test accuracy fluctuates between 92% and 96% throughout this range, suggesting that while the system performs consistently well, there are slight variations in how the test data is classified as more training samples are introduced. These fluctuations could indicate that the system is sensitive to the nature of the training data or potentially approaching the limits of its capacity for linear classification. From a computational perspective, shown in Figure 1(e), the 2-qubit classifier exhibits a significant increase in computational time as the number of training samples grows. By the time the training sample size reaches 250, the computational time extends to around 260 seconds. This is a sharp contrast to the single-qubit system, illustrating the tradeoff between the enhanced accuracy and robustness offered by the 2-qubit classifier and the increased computational demands. For LCP tasks, this suggests that while the 2-qubit classifier provides higher initial accuracy and steady performance improvements, it comes at the cost of a much higher computational burden, making it potentially less suitable for scenarios where time or resources are constrained.

Examining figure 10(c), the performance of the 2-qubit entangled classifier in a LCP task reveals a distinctive pattern when compared to non-entangled systems. The initial accuracy is relatively low, starting at 51.3% with just one training sample. This suggests that the entanglement introduces complexities that make the system less effective in identifying patterns from very limited data. However, as the number of training samples increases to 75, the system exhibits a steep improvement in accuracy, reaching 93.3%. This rapid climb indicates that while the entangled system may struggle with very small datasets, it quickly capitalizes on additional training samples to enhance its classification performance. As the training samples continue to increase beyond 75, the 2-qubit entangled classifier shows notable fluctuations in accuracy, ranging between 88% and 97.5%. These fluctuations, which are more pronounced than those seen in the single-qubit or non-entangled 2-qubit classifier, suggest that entanglement introduces both benefits and challenges. On one hand, the system achieves the highest peak accuracy (97.5%) among all three systems, demonstrating its potential for superior performance. On the other hand, the variability in test accuracy highlights the sensitivity of the entangled system to the training data, possibly indicating overfitting or instability when processing larger datasets. In terms of computational cost, as shown in figure 10(f), the 2-qubit entangled classifier mirrors the trend seen in the non-entangled 2-qubit classifier, with computational time increasing significantly as the number of training samples rises. At 250 training samples, the computational time reaches 260 seconds, similar to the non-entangled classifier. Despite this computational burden, the 2-qubit entangled classifier offers a potential advantage in terms of peak accuracy, making it a compelling choice for applications where achieving the highest possible accuracy is paramount, even if it comes with the tradeoff of greater computational complexity and variability in performance.

In comparing the classifier, we observe clear tradeoffs between simplicity, stability, and computational complexity. The single-qubit classifier is the most stable and computationally efficient but may not reach the same peak accuracies as the more complex systems. The 2-qubit classifier offers higher initial accuracy and consistent improvement but requires significantly more computational resources. Finally, the 2-qubit entangled system, while achieving the highest peak accuracy, also introduces greater instability and computational demands, making it best suited for scenarios where peak performance is the priority, and computational cost is less of a concern. Ultimately, the choice of system depends on the specific requirements of the classification task, such as whether stability, computational efficiency, or peak accuracy is the primary objective.

Figure 11 presents a comprehensive analysis of two quantum classifiers - a 2-qubit classifier and a 2-qubit entangled classifier for non-LCP. The results are displayed across six subplots, labeled (a) through (f), which provide insights into the performance and characteristics of these classifiers under various conditions. Subplots (a) and (b) show the train and test accuracies as a function of the number of training samples for the 2-qubit and the 2-qubit entangled classifiers, respectively. In both cases, we observe that the accuracies generally improve as the number of training samples increases. However, the 2-qubit classifier (a) shows higher initial test accuracy, 73.5%, and a more stable performance across different sample sizes. The 2-qubit entangled classifier (b) starts with lower test accuracy, 47.6% but shows significant improvement as the sample size increases. Both classifiers seem to converge in terms of train and test accuracy around 175 training samples, which explains why this number was chosen for subsequent analyses. Subplots (c) and (d) illustrate how the number of layers in the quantum circuit affects the accuracies of the classifiers for a specific number of training samples. For the 2-qubit classifier (c), we see a general upward trend in both train and test accuracies as the number of layers increases, with some fluctuations. The 2-qubit entangled classifier (d) shows a more pronounced improvement with increasing layers, especially in the early stages. Both classifiers appear to reach a plateau in performance after about 12-15 layers, suggesting that further increases in circuit depth may not yield significant improvements. Subplots (e) and (f) depict the computational time required as the number of layers increases for the 2-qubit and the 2-qubit entangled classifiers, respectively. Both show a clear exponential growth in computational time as the number of layers increases. This trend is consistent across both classifiers, indicating that the computational cost scales similarly regardless of whether entanglement is used. Comparing the classifiers overall, we can see that the 2-qubit classifier generally achieves higher accuracies with fewer training samples and maintains more consistent performance across different numbers of layers. The 2-qubit entangled classifier, while starting with lower accuracy, shows more dramatic improvements as both the number of training samples and layers increase. This suggests that entanglement might provide additional expressive power to the classifier, allowing it to capture more complex patterns in the data as the circuit depth increases. However, this potential advantage comes at the cost of increased sensitivity to the number of training samples and layers, as evidenced by the more volatile accuracy curves in subplots (b) and (d). The computational time plots (e) and (f) remind us that increasing the number of layers quickly becomes computationally expensive for both classifiers, which is an important consideration in practical applications. In conclusion, these results provide valuable insights into the trade-offs between accuracy, circuit complexity, and computational cost for quantum classifiers, highlighting the potential benefits and challenges of using entanglement in quantum machine learning tasks.

Figure 12 presents a comparative analysis of four optimization algorithms (COBYLA, L-BFGS-B, NELDER MEAD, and SLSQP) applied to a LCP using a quantum circuit with 2 qubits. The experiment uses a random dataset with 250 training samples and employs a fidelity cost function to measure the performance. The figure includes subplots depicting accuracy and computational time for both 2-qubit and 2-qubit entangled classifiers. In terms of accuracy, both training and test accuracies are generally high across all algorithms. However, there are subtle differences between the algorithms. As shown in figure 12(a), for the 2-qubit entangled classifier, the average test accuracy is approximately 2% higher than the 2-qubit non-entangled classifier. In terms of individual performance, the L-BFGS-B minimization method consistently achieves the highest test accuracy, reaching 96.3% for non-entangled and 97% for entangled classifiers. The overall variation in test accuracy between the highest and lowest performing algorithms is 2.3%. For 2-qubit non-entangled classifier, COBYLA exhibits the lowest test accuracy at 94%, while for 2-qubit entangled classifier, NELDER MEAD achieves the lowest test accuracy of 95.3%. Computational time analysis reveals interesting patterns across both classifiers. In figure 12(c) the 2-qubit classifier, computational time varies widely from 9 to 90 minutes. COBYLA stands out as the fastest method, completing the task in just 9 minutes, while L-BFGS-B and NELDER_MEAD are the most time-consuming at 90 and 89 minutes respectively. SLSQP occupies a middle ground, requiring 45 minutes. In figure 12(d) the 2-qubit entangled classifier generally shows improved computational efficiency. While COBYLA maintains its swift performance at 9 minutes, other methods see reduced execution times. Most notably, L-BFGS-B improves from 90 to 71 minutes, a significant reduction, while NELDER_MEAD and SLSQP methods remain at 87 and 44 minutes respectively. In conclusion, this analysis reveals that the 2-qubit entangled classifier generally outperforms the 2-qubit non-entangled classifier in both accuracy and computational efficiency. The L-BFGS-B method consistently provides the highest accuracy, albeit at a higher computational cost. COBYLA emerges as a well-balanced option, offering good accuracy with minimal computational time, particularly in the 2-qubit entangled classifier. These findings underscore the significant impact of minimization method selection on both accuracy and computational time in quantum machine learning tasks. Furthermore, the 2-qubit entangled classifier's closer alignment of train and test accuracies suggests enhanced generalization capabilities, a crucial factor in practical machine learning applications.

Figure 13 shows a comprehensive comparison of different optimization methods for non-LCP using both 2-qubit and 2-qubit entangled classifiers for a specific random dataset. This analysis encompasses four optimization techniques: COBYLA, L-BFGS-B, NELDER_MEAD, and SLSQP, evaluating their performance based on accuracy and computational time for 250 number of training samples. In the accuracy graphs (a) and (b), we observe distinct performance patterns between the 2-qubit and 2-qubit entangled classifiers. For the 2-qubit classifier, L-BFGS-B demonstrates the highest accuracy, with both train and test accuracies exceeding 90%. COBYLA shows the lowest performance, with a test accuracy of 76.7% and train accuracy 81.4%. NELDER_MEAD and SLSQP exhibit intermediate performance, with test accuracies in the 82-87% range. The 2-qubit entangled classifier, depicted in graph (b), shows overall improved accuracy across all methods. L-BFGS-B maintains its superior performance, while COBYLA shows significant improvement, reaching accuracies to 85.4%. Notably, the gap between train and test accuracies is generally smaller in the 2-qubit entangled classifier, suggesting better generalization. The computational time graphs (c) and (d) reveal interesting efficiency patterns. In the 2-qubit classifier, COBYLA is the fastest method, requiring only 9 minutes. L-BFGS-B, despite its high accuracy, is the most time-consuming at 130 minutes. NELDER_MEAD takes 89 minutes, while SLSQP requires 45 minutes. The 2-qubit entangled classifier (graph d) shows generally reduced computational times. COBYLA remains the fastest, maintaining its 9-minute runtime. L-BFGS-B shows the most dramatic improvement, reducing its time to 81 minutes. Interestingly, NELDER_MEAD in the 2-qubit entangled classifier takes slightly longer than L-BFGS-B, at 88 minutes. SLSQP maintains a consistent performance of about 42 minutes in both systems. These results highlight the trade-offs between accuracy and computational efficiency in quantum machine learning tasks. The 2-qubit entangled classifier demonstrates superior performance in both accuracy and computational time across all methods. L-BFGS-B consistently provides the highest accuracy but at a higher computational cost, especially in the 2-qubit classifier. COBYLA emerges as a balanced option, offering good accuracy with minimal computational time, particularly in the entangled system. This analysis underscores the importance of choosing appropriate optimization methods and leveraging entanglement to enhance the performance of quantum classification tasks.

## Supplementary Note 6: Method

Quantum computing manipulates quantum systems to enhance information processing, leveraging superposition to simultaneously operate on multiple states for faster and more complex computation. At its core is the qubit, represented in a two-dimensional Hilbert space, with operations governed by quantum gates. These gates, essential for altering quantum states, must be unitary to ensure the conservation of probability, a fundamental principle of quantum dynamics[2].

The framework of a quantum circuit unfolds in three key phases: encoding classical data into quantum format, manipulating the quantum state using quantum gates, and measuring the quantum state post-transformation. This process transitions from preparing an initial quantum state, through strategic alterations via quantum gates affecting computation outcomes, to a final probabilistic measurement—distinguishing quantum computing's potential and challenges from deterministic classical computing.

Achieving optimal performance in quantum computing requires a nuanced understanding of these phases, including the initial state preparation, the strategic selection and application of quantum gates, and the final measurement process. Each component must be meticulously optimized to perform specific tasks, such as classification, highlighting the intricate interplay between quantum mechanics and computational logic in the design and execution of quantum algorithms.

### *A. RE-UPLOADING CLASSICAL INFORMATION AND PROCESSING*

The integration of classical information into quantum computing represents a groundbreaking approach to data processing and analysis. This process begins with the strategic encoding of data into the initial wave function's coefficients within a quantum circuit[3]. In simpler terms, data is initially uploaded through the manipulation of qubits via rotational operations on a computational basis. This foundational step sets the stage for executing sophisticated quantum algorithms, including those designed for classification tasks.

The most successful programming paradigm in machine learning is predicated on artificial neural networks, which represent a highly abstracted and simplified model inspired by the human brain [4]. An artificial neural network comprises interconnected units or nodes known as artificial neurons, often arranged in layers [5]. These networks are characterized by their diverse architectures and the ability to learn from data through the adjustment of a vast network of parameters during the training phase. Among the various types of neural networks, feed-forward neural networks exemplify the process of sequential data processing, where input data is transformed layer by layer, simulating a form of data re-uploading at each neuron. This mechanism of data re-uploading and processing in ANNs provides a parallel to the innovative approach of constructing a universal quantum classifier using a single qubit. The essence of this quantum computing strategy lies in the repeated introduction of classical data at different stages of computation, analogous to the data processing in a single hidden layer neural network. This process can be visualized diagrammatically, as shown in figure 14 in the main paper. the neural network architecture is depicted, where data points are fed into individual processing units, analogous to neurons within the hidden layer. These neurons collectively process these input data, culminating in the activation of a final neuron responsible for constructing the output for subsequent analysis. Similarly, in the quantum domain, the single-qubit classifier incorporates data points into each stage of the computation through unitary rotations. These rotations are not isolated; rather, each one builds upon the transformations applied by its predecessors, effectively integrating the input data multiple times throughout the computation. The culmination of this process is a quantum state that encapsulates the computational outcome.

To construct a universal quantum classifier with only a single qubit, a complex integration of data input and computational processing within a single quantum circuit is crucial. We achieve this objective through the deployment of parametrized quantum circuits (PQCs). In these circuits, certain rotational angles are meticulously adjusted based on classical parameters, which are refined through an optimization process aimed at minimizing a specifically defined cost function.

The cost function plays a pivotal role in the operational efficacy of the quantum classifier. It quantitatively assesses the circuit's performance in segregating data points into distinct categories, which are represented as separate regions on the Bloch sphere. Each of these regions corresponds to a different class, and the classifier's goal is to assign data points to the correct class based on their features.

### B. Applying Cost Functions

In the realm of quantum computing, a quantum circuit is distinguished by its processing angles $\theta_i$ and associated weights $w_i$, leading to the generation of a final state $|\psi>$. The measurement outcomes from this state are used to compute a classification error metric, defined as $\chi^2$. The goal is to minimize this error metric by adjusting the circuit's classical parameters, a process that can be effectively managed through various supervised machine learning techniques.

At the heart of using quantum measurement for classification tasks lies the approach of optimally aligning observed outputs with specific target classes. This alignment is primarily facilitated by the principle of maximizing orthogonality between the output states, ensuring clear distinction[6]. In the context of binary (dichotomous) classification, this means categorizing each observation into one of two predefined classes—referred to here as class A and class B. The decision criterion involves comparing the probabilities of observing the quantum state $P(0)$ for outcome 0 and $P(1)$ for outcome 1. If $P(0) > P(1)$, the observation is assigned to class A; otherwise, it falls under class B. To enhance this binary classification scheme, one can introduce a bias $(\lambda)$, adjusting the threshold for classification such that observation is deemed part of class A if $P(0)$ is greater than $\lambda$, and class B if it falls below. The value of $\lambda$ is chosen to maximize classification accuracy on a training dataset. The effectiveness of this approach is then confirmed through evaluation on a separate validation dataset.

Viewed through a geometric lens, the single-qubit classifier operates within a 2-dimensional Hilbert space —the Bloch sphere—where data encoding and classification decisions are delineated through specific rotational parameters. Any operation $L(i)$ is a rotation on the Bloch sphere surface. From this viewpoint, any point can be classified using just one unitary operation. Consequently, we can transfer any point to another point on the Bloch sphere by precisely selecting the rotation angles. However, when dealing with multiple data points, a single rotation may not suffice due to differing optimal rotation requirements. The solution lies in introducing additional layers into the quantum circuit, enabling distinct rotation and fostering a richer feature map. Within this enhanced feature space, data points can be effectively segregated into their respective classes based on their positioning within the Bloch sphere's regions, thereby enabling a sophisticated and adaptable approach to quantum classification.

#### 1) FIDELITY COST FUNCTION

The goal is to align the quantum states ($|\psi(\vec{\theta}, \vec{w}, \vec{x})>$) as closely as possible to a designated target state on the Bloch sphere, as outlined in [1]. This alignment can be quantitatively assessed by measuring the angular distance between the labeled state and the data state, using the metric of relative fidelity [7]. The primary objective focuses on maximizing the average fidelity between the quantum states produced by the circuit and the label states corresponding to their respective classes. To facilitate this, a cost function is introduced and mathematically formulated as Equation 1:

$$\chi_f^2(\vec{\theta}, \vec{\omega}) = \sum_{\mu-1}^{M} \left( 1 - \left| \left\langle \tilde{\psi}_s | \psi\left( \vec{\theta}, \vec{\omega}, \vec{x}_\mu \right) \right\rangle \right|^2 \right) \qquad (1)$$

where $|\tilde{\psi}_s\rangle$ is the correct label state of the $\mu$ data point, which will correspond to one of the classes.

#### 2) TRACE DISTANCE COST FUNCTION

In quantum information theory, quantifying the dissimilarity between two quantum states is a fundamental problem. Various distance measures have been proposed, each with its unique properties and applications. One such measure is the trace distance, which captures the distinguishability between two quantum states [7]. Perez-Salinas et al. have analyzed the fidelity cost function with data re-uploading [1]. However, the authors do not consider the case of the trace distance cost function, which is what we focus on in this section. We will explore the definition and properties of the trace distance, particularly in the context of single-qubit systems, and discuss its potential as a cost function for quantum classifiers.

Despite the different mathematical formulations of trace distance and fidelity, these two measures share many similar properties and are widely used in the quantum computing and quantum information community. However, depending on the specific application, one measure may be more convenient or easier to work with than the other. This versatility and widespread adoption of both trace distance and fidelity in the field motivates our decision to discuss and compare these two important distance measures in the context of quantum classifiers. The trace distance between quantum states $\rho$ and $\sigma$ can be defined as,

$$D(\rho,\sigma) \equiv \frac{1}{2} tr\left|\rho - \sigma\right|^2 \qquad (2)$$

The trace distance between two single-qubit states, represented by their respective Bloch vectors $\vec{r}$ and $\vec{s}$, is equal to one-half of the Euclidean distance between these vectors on the Bloch sphere. [7]

$$D(\rho,\sigma) = \frac{\left|\vec{r} - \vec{s}\right|}{2}. \qquad (3)$$

This relation provides a geometric interpretation of the trace distance for single-qubit systems, linking it to the intuitive notion of distance in three-dimensional space.

### *C. From Universality of the Single-Qubit Classifier to the Expansion into Multi-Qubit Quantum Classification*

A key challenge in Quantum Machine Learning (QML) involves creating quantum circuits that efficiently handle complex tasks like classification without excessive use of quantum resources. The Universal Approximation Theorem (UAT) [8] is crucial for tackling this issue, demonstrating that a single-layer neural network with an appropriate activation function can approximate any continuous function to a desired accuracy, assuming enough hidden neurons are available. This UAT finds a compelling parallel in the quantum computing domain, particularly when considering the dynamics of quantum circuits. Here, the classical activation function is analogously performed by a unitary rotation acting upon a qubit. Specifically, a single-qubit quantum classifier, enhanced by the technique of data re-uploading, emerges as a universal approximator for any conceivable classification function. This universality hinges on the frequency of data re-uploading throughout the circuit's span [1], underscoring that even a solitary qubit is capable of encoding and processing multifaceted high-dimensional data. This is achieved through the execution of multiple rotations, each characterized by distinct angles and weights. The culmination of these processes is a final quantum state, which is then analyzed against a predefined target state correlating to each class. Optimization of the circuit's parameters is pursued through the minimization of a cost function, which is indicative of the fidelity or trace distance between the comparative states.

By establishing the UAT within the context of quantum classifiers, a robust theoretical foundation is laid, alongside practical guidelines for the design and implementation of quantum circuits adept at sophisticated and non-LCP tasks with minimal quantum resource expenditure. This breakthrough not only forges a theoretical link between quantum circuits and neural networks but also paves the way for innovative approaches in QML. Through this lens, quantum circuits are envisioned not merely as computational tools but as entities with the potential to parallel, and possibly surpass, the capabilities of their classical neural network counterparts, inspiring a new wave of methodologies in the realm of QML.

To enhance the performance of the single-qubit classifier, it is proposed to extend it to a multi-qubit system. Adding more qubits, especially with entanglement, can improve the classifier's effectiveness, similar to how adding layers enhances neural networks. Entanglement may provide a quantum advantage in classification, though the analogy between multi-qubit classifiers and neural networks with entanglement is not fully understood and requires further exploration. Perez et al. propose a measurement strategy for multi-qubit classifiers, which extends the single-qubit approach. These strategies utilize a fidelity-based cost function.

## Supplementary Note 7: Optimization Methods

In practice, deploying a parameterized quantum classifier involves a process of minimizing within the parameter space that delineates the circuit's configuration. The process is often termed a hybrid algorithm, denoting the symbiotic relationship and advantages derived from combining quantum logic and classical logic. In particular, the ensemble of angles ($\theta_i$) and weights ($w_i$) defines a parameter space that requires systematic exploration to achieve the minimization of $\chi^2$.

The occurrence of local minima is unavoidable e. The arrangement of rotation gates results in an intricate multiplication of independent trigonometric functions, suggesting that our problem is characterized by a widespread distribution of minima.

The primary challenge boils down to minimizing a function that is defined by a vast array of parameters. In the case of a single-qubit classifier, the total number of parameters can be expressed as, where represents the problem's dimension (that is, the dimension of), and signifies the number of layers. Among these parameters, three are rotational angles, while the rest pertain to the weight [1]. To identify the most effective solution, we evaluate the performance of four distinct minimization techniques: the L-BFGS-B method, the COBYLA method, the Nelder-Mead method, and the Sequential Least Squares Programming (SLSQP) method.

The key challenge in optimizing a single-qubit classifier involves minimizing a function across a complex parameter space, calculated as $(3+d)N$, where "d" is the problem's dimension and "N" is the number of layers. Also, in addition, we need to consider rotational angles and the weight ($\vec{w}_i$) corresponding to the dimension [1]. To discover the optimal solution, we delve into the efficiency of four diverse minimization strategies: the L-BFGS-B, COBYLA, Nelder-Mead, and Sequential Least Squares Programming (SLSQP) methods.

### A. L-BFGS-B METHOD

The L-BFGS-B technique, part of the quasi-Newton optimization methods, refines the Broyden–Fletcher–Goldfarb–Shanno (BFGS) approach by efficiently using limited computer memory [10]. Its design excels in handling optimization tasks involving numerous variables, offering a linear memory usage advantage, making it highly effective for large-scale problems [11].

The L-BFGS-B method is widely recognized as a cornerstone technique across various advanced applications in the field of graphics [12,13]. It specializes in minimizing a scalar function of one or several variables by initiating with a preliminary estimate of the optimum value. Through iterative refinement, it progressively improves upon this initial estimate to approach an optimal solution. The method employs function derivatives to determine the steepest descent's direction and to approximate the function's Hessian matrix (its second derivative), showcasing exceptional efficiency in matrix-vector multiplication operations [14].

### B. CONSTRAINED OPTIMIZATION BY LINEAR APPROXIMATION METHOD

COBYLA (Constrained Optimization BY Linear Approximation) is an optimization algorithm designed to minimize a scalar objective function that depends on one or more variables, subject to constraints [15,16]. One of the key features of COBYLA is that it does not require the calculation of derivatives, such as gradients or Hessians, of the objective function or constraints. This makes COBYLA particularly useful in situations where the derivatives are unknown, unreliable, or computationally expensive to obtain [15]. By relying on linear approximations of the objective function and constraints, COBYLA can effectively navigate the optimization landscape and find solutions even in the absence of explicit derivative information.

COBYLA has been effectively utilized in quantum computing, especially as a classical optimization routine within Variational Hybrid Quantum-Classical Algorithms (VHQCAs) [17]. These algorithms employ a parameterized quantum circuit, or ansatz, which is refined through a dynamic interchange between a classical computer and a quantum device. The classical computer adjusts the ansatz's parameters to minimize a cost function, which the quantum device efficiently evaluates. Through iterative updates based on the cost function outcomes, the VHQCA aims to discover the most effective ansatz configuration for specific problems. The derivative-free characteristic of COBYLA makes it particularly advantageous for this setting, where the cost functions often lack easily computable or analytically defined derivatives.

### C. NELDER-MEAD METHOD

The Nelder-Mead algorithm, introduced by John Nelder and Roger Mead in 1965, is a widely used direct search method for unconstrained optimization problems [18]. The algorithm operates by maintaining a simplex of n+1 points in an n-dimensional space, iteratively moving the simplex toward the optimal solution through a series of transformations, including reflection, expansion, contraction, and shrinkage [18].

Recent studies have focused on enhancing the Nelder-Mead algorithm to improve its efficiency and adaptability. Gao and Han [19]proposed an implementation of the Nelder-Mead algorithm with adaptive parameters, which can automatically adjust the parameter values based on the optimization progress. This adaptive approach has been shown to improve the algorithm's convergence speed and solution quality [19].

Its capacity to address problems in which derivative information is not readily accessible renders it a favorable option for numerous applications in QML. However, it is essential to conduct comprehensive evaluations to scrutinize the method's accuracy, efficiency, and sensitivity to the initial guess for each unique application [20,21].

### D. SEQUANTIAL LEAST SQUARES PROGRAMMING METHOD

The Sequential Least Squares Programming (SLSQP) method is an optimization technique that minimizes functions while adhering to specific constraints [22]. It is based on Sequential Quadratic Programming (SQP), which simplifies the optimization problem into a series of smaller, more manageable quadratic subproblems. In each subproblem, a quadratic approximation of the objective function and constraints is constructed using a second-order parabolic curve to model the function's behavior near a specific point. SLSQP updates this approximation using the quasi-Newton method.

Additionally, SLSQP applies a least-squares method to solve these quadratic subproblems, striving to minimize the total squared deviations between the approximation and actual function values. This method can handle both equality and inequality constraints, including variable bounds, by integrating a penalty function that imposes additional costs for any constraint or bound violations. SLSQP ensures efficient convergence by terminating the optimization process upon meeting a predefined convergence criterion, typically related to changes in the objective function value or the gradient vector's norm. This safeguard prevents indefinite computations, ensuring timely solutions.

Local minima are common challenges in both neural networks and quantum classifiers due to their complex mathematical structures—neural networks with compounded nonlinear functions and quantum circuits with prevalent trigonometric functions. This complexity increases the likelihood of encountering local minima during optimization. Moreover, with smaller training sets, the choice of optimization method is crucial. For instance, the Nelder-Mead method is noted for its robustness, particularly its lower susceptibility to local minima.

It is also critical to recognize that minimization methods are sensitive to noise, which can significantly impact their effectiveness, especially in practical quantum computing applications [17].

## Data Availability

All data generated or analyzed during this study are included in this published article and its supplementary information files.

**Supplementary Note 8: Comparing the developed code for this research with original reference.**

```python
# coding=utf-8
#############################################################
#Quantum classifier
#Sara Aminpour, Mike Banad, Sarah Sharif
#September 25th 2024

#School of Electrical and Computer Engineering/ Center for Quantum and Technology, University of Oklahoma, Norman, OK 73019 USA,

#############################################################
#IMPORTANT_NOTE:
#The code on the left was developed by Sara Aminpour, while the code on the right serves as the reference implementation by Adrián Pérez-Salinas.
#The code on the left has been restructured to handle random data. So some certain sections has been deleted from the reference code.
#Additionally, our code on the left developed to analyze trace distance cost function and linear classification problem
#as well as necessary modification to apply COBYLA, L-BFGS-B, NELDER-MEAD, and SLSQP minimization methods.
#############################################################
# so that the usage is automatized
import datetime
from data_gen import data_generator
from problem_gen import problem_generator, representatives, representatives_tr
from fidelity_minimization import fidelity_minimization
from trace_minimization import trace_minimization
from weighted_fidelity_minimization import weighted_fidelity_minimization
from test_data import Accuracy_test, tester
from save_data import write_summary, read_summary, name_folder, samples_paint, samples_paint_worldmap, laea_x, laea_y

from save_data import write_epochs_file, write_epoch, close_epochs_file, create_folder, write_epochs_error_rate

import numpy as np
import matplotlib.pyplot as plt
from circuitery import code_coords, circuit
from matplotlib.cm import get_cmap
from matplotlib.colors import Normalize

def minimizer(chi, problem, qubits, entanglement, layers, method, name,
              epochs=3000, batch_size=20,  eta=0.1):

    """
    This function creates data and minimizes whichever problem (from the selected ones)
    INPUT:
        -chi: cost function, to choose between 'fidelity_chi' or 'weighted_fidelity_chi'
        -problem: name of the problem, to choose among
            ['circle', '3 circles', 'hypersphere', 'tricrown', 'non convex', 'crown', 'sphere', 'squares', 'wavy lines']
        -qubits: number of qubits, must be an integer
        -entanglement: whether there is entanglement or not in the Ansätze, just 'y'/'n'
        -layers: number of layers, must be an integer. If layers == 1, entanglement is not taken in account

        -method: minimization method, to choose among ['SGD', another valid for function scipy.optimize.minimize]

        -name: a name we want for our our files to be save with
        -seed: seed of numpy.random, needed for replicating results
        -epochs: number of epochs for a 'SGD' method. If there is another method, this input has got no importance

        -batch_size: size of the batches for stochastic gradient descent, only for 'SGD' method
        -eta: learning rate, only for 'SGD' method
    OUTPUT:
        This function has got no outputs, but several files are saved in an appropiate folder. The files are

        -summary.txt: Saves useful information for the problem
        -theta.txt: saves the theta parameters as a flat array
        -alpha.txt: saves the alpha parameters as a flat array
        -weight.txt: saves the weights as a flat array if they exist
    """

    data, drawing = data_generator(problem)
    if problem == 'sphere':
        train_data = data[:500]
        test_data = data[500:]
    elif problem == 'hypersphere':
        train_data = data[:1000]
        test_data = data[1000:]
    else:
        train_data = data[:250]
        test_data = data[250:]

    if chi == 'fidelity_chi':
        Accuracy_tr=0
        Accuracy_te=0
        i=1
        while i<21:
            qubits_lab = qubits
            theta, alpha, reprs = problem_generator(problem,qubits, layers, chi,
                                                    qubits_lab=qubits_lab)

            theta, alpha, f = fidelity_minimization(theta, alpha, train_data, reprs,
                                                    entanglement, method,
                                                    batch_size, eta, epochs)

            acc_train = tester(theta, alpha, train_data, reprs, entanglement, chi)
            Accuracy_tr+=acc_train

            acc_test = tester(theta, alpha, test_data, reprs, entanglement, chi)
            Accuracy_te+=acc_test

            text_file_nn = open('acc.txt', mode='a+')
            text_file_nn.write(str(i) + problem +'_'+ chi +'_'+ str(qubits) +'Qubits_' + str(layers) +'Layers_' + entanglement +'_'+ method +'_'+'acc_train'+' = '+ str(acc_train))
            text_file_nn.write('\n')
            text_file_nn.write(str(i) + problem +'_'+ chi +'_'+ str(qubits) +'Qubits_' + str(layers) +'Layers_' + entanglement +'_'+ method +'_'+'acc_test'+' = '+ str(acc_test))
            text_file_nn.write('\n')
            text_file_nn.write('======================================================')
            text_file_nn.write('\n')
            text_file_nn.close()

            i+=1
        print(i-1)
        atr=Accuracy_tr/(i-1)
        ate=Accuracy_te/(i-1)

        text_file_nn = open('AverageAcc.txt', mode='a+')
        text_file_nn.write(problem +'_'+ chi +'_'+ str(qubits) +'Qubits_' + str(layers) +'Layers_' + entanglement +'_'+ method +'_'+'Ave_acc_train'+' = '+ str(atr))
        text_file_nn.write('\n')
        text_file_nn.write(problem +'_'+ chi +'_'+ str(qubits) +'Qubits_' + str(layers) +'Layers_' + entanglement +'_'+ method +'_'+'Ave_acc_test'+' = '+ str(ate))
        text_file_nn.write('\n')
        text_file_nn.write('======================================================')
        text_file_nn.write('\n')
        text_file_nn.close()

        write_summary(chi, problem, qubits, entanglement, layers, method, name,
                theta, alpha, 0, f, atr, ate, epochs=epochs)
    elif chi == 'trace_chi':

        Accuracy_tr=0
        Accuracy_te=0
        i=1
        while i<21:
            qubits_lab = qubits
            theta, alpha, reprs = problem_generator(problem,qubits, layers, chi,
                                                    qubits_lab=qubits_lab)
            theta, alpha, f = trace_minimization(theta, alpha, train_data, reprs,
                                                    entanglement, method,
                                                    batch_size, eta, epochs)


            acc_train = tester(theta, alpha, train_data, reprs, entanglement, chi)
            Accuracy_tr+=acc_train

            acc_test = tester(theta, alpha, test_data, reprs, entanglement, chi)
            Accuracy_te+=acc_test

            text_file_nn = open('acc.txt', mode='a+')
            text_file_nn.write(str(i) + problem +'_'+ chi +'_'+ str(qubits) +'Qubits_' + str(layers) +'Layers_' + entanglement +'_'+ method +'_'+'acc_train'+' = '+ str(acc_train))
            text_file_nn.write('\n')
            text_file_nn.write(str(i) + problem +'_'+ chi +'_'+ str(qubits) +'Qubits_' + str(layers) +'Layers_' + entanglement +'_'+ method +'_'+'acc_test'+' = '+ str(acc_test))
            text_file_nn.write('\n')
            text_file_nn.write('======================================================')
            text_file_nn.write('\n')
            text_file_nn.close()


            i+=1
        print(i-1)
        atr=Accuracy_tr/(i-1)
        ate=Accuracy_te/(i-1)


        text_file_nn = open('AverageAcc.txt', mode='a+')
        text_file_nn.write(problem +'_'+ chi +'_'+ str(qubits) +'Qubits_' + str(layers) +'Layers_' + entanglement +'_'+ method +'_'+'Ave_acc_train'+' = '+ str(atr))
        text_file_nn.write('\n')
        text_file_nn.write(problem +'_'+ chi +'_'+ str(qubits) +'Qubits_' + str(layers) +'Layers_' + entanglement +'_'+ method +'_'+'Ave_acc_test'+' = '+ str(ate))
        text_file_nn.write('\n')
        text_file_nn.write('======================================================')
        text_file_nn.write('\n')
        text_file_nn.close()

        write_summary(chi, problem, qubits, entanglement, layers, method, name,
                theta, alpha, 0, f, atr, ate, epochs=epochs)
    elif chi == 'weighted_fidelity_chi':


        Accuracy_tr=0
        Accuracy_te=0
        i=1
        while i<21:
            qubits_lab = 1
            theta, alpha, weight, reprs = problem_generator(problem,qubits, layers, chi,
                                                            qubits_lab=qubits_lab)
            theta, alpha, weight, f = weighted_fidelity_minimization(theta, alpha, weight, train_data, reprs,

                                                    entanglement, method)



            acc_train = tester(theta, alpha, train_data, reprs, entanglement, chi, weights=weight)
            Accuracy_tr+=acc_train

            acc_test = tester(theta, alpha, test_data, reprs, entanglement, chi, weights=weight)
            Accuracy_te+=acc_test

            text_file_nn = open('acc.txt', mode='a+')
            text_file_nn.write(str(i) + problem +'_'+ chi +'_'+ str(qubits) +'Qubits_' + str(layers) +'Layers_' + entanglement +'_'+ method +'_'+'acc_train'+' = '+ str(acc_train))
            text_file_nn.write('\n')
            text_file_nn.write(str(i) + problem +'_'+ chi +'_'+ str(qubits) +'Qubits_' + str(layers) +'Layers_' + entanglement +'_'+ method +'_'+'acc_test'+' = '+ str(acc_test))
            text_file_nn.write('\n')
            text_file_nn.write('======================================================')
            text_file_nn.write('\n')
            text_file_nn.close()


            i+=1
        print(i-1)
        atr=Accuracy_tr/(i-1)
        ate=Accuracy_te/(i-1)

        text_file_nn = open('AverageAcc.txt', mode='a+')
        text_file_nn.write(problem +'_'+ chi +'_'+ str(qubits) +'Qubits_' + str(layers) +'Layers_' + entanglement +'_'+ method +'_'+'Ave_acc_train'+' = '+ str(atr))
        text_file_nn.write('\n')
        text_file_nn.write(problem +'_'+ chi +'_'+ str(qubits) +'Qubits_' + str(layers) +'Layers_' + entanglement +'_'+ method +'_'+'Ave_acc_test'+' = '+ str(ate))
        text_file_nn.write('\n')
        text_file_nn.write('======================================================')
        text_file_nn.write('\n')
        text_file_nn.close()

        write_summary(chi, problem, qubits, entanglement, layers, method, name,
                theta, alpha, weight, f, acc_train, acc_test, epochs=epochs)




def painter(chi, problem, qubits, entanglement, layers, method, name,
        standard_test = True, samples = 4000, bw = False, err = False):
    a=datetime.datetime.now()
    """
    This function takes written text files and paint the results of the problem
    INPUT:
        -chi: cost function, to choose between 'fidelity_chi' or 'weighted_fidelity_chi'
        -problem: name of the problem, to choose among
            ['circle', '3 circles', 'hypersphere', 'tricrown', 'non convex', 'crown', 'sphere', 'squares', 'wavy lines']
        -qubits: number of qubits, must be an integer
        -entanglement: whether there is entanglement or not in the Ansätze, just 'y'/'n'
        -layers: number of layers, must be an integer. If layers == 1, entanglement is not taken in account

        -method: minimization method, to choose among ['SGD', another valid for function scipy.optimize.minimize]

        -name: a name we want for our our files to be save with
        -seed: seed of numpy.random, needed for replicating results
        -standard_test: Whether we want to paint the set test used for checking when minimizing. If True, seed and samples are not taken in account
        -samples: number of samples of the test set
        -bw: painting in black and white
    OUTPUT:
        This function has got no outputs, but a file containing the representation of the test set is created

    """


    if chi == 'fidelity_chi':
        qubits_lab = qubits

    elif chi == 'trace_chi':
        qubits_lab = qubits

    elif chi == 'weighted_fidelity_chi':
        qubits_lab = 1

    if standard_test == True:
        data, drawing = data_generator(problem)
        if problem == 'sphere':
            test_data = data[500:]
        elif problem == 'hypersphere':
            test_data = data[1000:]
        else:
            test_data = data[250:]

    elif standard_test == False:
        test_data, drawing = data_generator(problem, samples = samples)

    if problem in ['circle','line', '2 lines', 'wavy circle','sphere', 'non convex', 'crown', 'hypersphere']:
        classes = 2
    if problem in ['tricrown']:
        classes = 3
    if problem in ['6squares']:
        classes = 6
    elif problem in ['3 circles','wavy lines','squares']:
        classes = 4

    #reprs = representatives(classes, qubits_lab)

    params = read_summary(chi, problem, qubits, entanglement, layers, method, name)

    if chi == 'fidelity_chi':
        reprs = representatives(classes, qubits_lab)
        theta, alpha = params
        sol_test, acc_test = Accuracy_test(theta, alpha, test_data, reprs, entanglement, chi)


    if chi == 'trace_chi':
        reprs = representatives_tr(classes, qubits_lab)
        theta, alpha = params
        sol_test, acc_test = Accuracy_test(theta, alpha, test_data, reprs, entanglement, chi)

    if chi == 'weighted_fidelity_chi':
        reprs = representatives(classes, qubits_lab)
        theta, alpha, weight = params
        sol_test, acc_test = Accuracy_test(theta, alpha, test_data, reprs,
                                           entanglement, chi, weights = weight)

    foldname = name_folder(chi, problem, qubits, entanglement, layers, method)
    samples_paint(problem, drawing, sol_test, foldname, name, bw)



    b=datetime.datetime.now()
    c=b-a
    text_file_nn = open('time.txt', mode='a+')
    text_file_nn.write(problem +'_'+ chi +'_'+ str(layers) +'Layers' +'_'+ 'painter' +' = '+ str(c))
    text_file_nn.write('\n')
    text_file_nn.close()
'''
def paint_world(chi, problem, qubits, entanglement, layers, method, name,
        seed = 30, standard_test = True, samples = 4000, bw = False, err = False):
    np.random.seed(seed)

    if chi == 'fidelity_chi':
        qubits_lab = qubits
    if chi == 'trace_chi':
        qubits_lab = qubits
    elif chi == 'weighted_fidelity_chi':
        qubits_lab = 1

    if standard_test == True:
        data, drawing = data_generator(problem)
        if problem == 'sphere':
            test_data = data[500:]
        elif problem == 'hypersphere':
            test_data = data[1000:]
        else:
            test_data = data[:250]

    elif standard_test == False:
        test_data, drawing = data_generator(problem, samples=samples)

    if problem in ['circle', 'line', '2 lines', 'wavy circle', 'sphere', 'non convex', 'crown', 'hypersphere']:
        classes = 2
    if problem in ['tricrown']:
        classes = 3
    if problem in ['6squares']:
        classes = 6
    elif problem in ['3 circles', 'wavy lines', 'squares']:
        classes = 4

    #reprs = representatives(classes, qubits_lab)

    params = read_summary(chi, problem, qubits, entanglement, layers, method, name)

    if chi == 'fidelity_chi':
        reprs = representatives(classes, qubits_lab)
        theta, alpha = params
        sol_test, acc_test = Accuracy_test(theta, alpha, test_data, reprs, entanglement, chi)

    if chi == 'trace_chi':
        reprs = representatives_tr(classes, qubits_lab)
        theta, alpha = params
        sol_test, acc_test = Accuracy_test(theta, alpha, test_data, reprs, entanglement, chi)

    if chi == 'weighted_fidelity_chi':
        reprs = representatives(classes, qubits_lab)
        theta, alpha, weight = params
        sol_test, acc_test = Accuracy_test(theta, alpha, test_data, reprs,
                                           entanglement, chi, weights=weight)

    foldname = name_folder(chi, problem, qubits, entanglement, layers, method)
    angles = np.zeros((len(sol_test), 2))
    for i, x in enumerate(sol_test[:, :2]):
        theta_aux = code_coords(theta, alpha, x)
        C = circuit(theta_aux, entanglement)
        angles[i, 0] = np.arccos(np.abs(C.psi[0])**2 - np.abs(C.psi[1])**2) - np.pi/2
        angles[i, 1] = np.angle(C.psi[1] / C.psi[0])
        print(angles[i])

    if bw == False:
        colors_classes = get_cmap('plasma')
        norm_class = Normalize(vmin=-.5, vmax=np.max(sol_test[:, -3]) + .5)

        colors_rightwrong = get_cmap('RdYlGn')
        norm_rightwrong = Normalize(vmin=-.1, vmax=1.1)

    if bw == True:
        colors_classes = get_cmap('Greys')
        norm_class = Normalize(vmin=-.1, vmax=np.max(sol[:, -3]) + .1)

        colors_rightwrong = get_cmap('Greys')
        norm_rightwrong = Normalize(vmin=-.1, vmax=1.1)

    fig, ax = plt.subplots(nrows=2)
    ax[0].plot(laea_x(np.pi, np.linspace(0, np.pi)), laea_y(np.pi, np.linspace(0, np.pi)), color='k')
    ax[0].plot(laea_x(-np.pi, np.linspace(0, -np.pi)), laea_y(-np.pi, np.linspace(0, -np.pi)), color='k')

    ax[1].plot(laea_x(np.pi, np.linspace(0, np.pi)), laea_y(np.pi, np.linspace(0, np.pi)), color='k')
    ax[1].plot(laea_x(-np.pi, np.linspace(0, -np.pi)), laea_y(-np.pi, np.linspace(0, -np.pi)), color='k')

    ax[0].scatter(laea_x(angles[:, 1], angles[:, 0]), laea_y(angles[:, 1], angles[:, 0]), c=sol_test[:, -2],

                  cmap=colors_classes, s=2, norm=norm_class)
    ax[1].scatter(laea_x(angles[:, 1], angles[:, 0]), laea_y(angles[:, 1], angles[:, 0]), c=sol_test[:,-1], cmap = colors_rightwrong, s=2, norm=norm_rightwrong)
    plt.show()
'''

def SGD_step_by_step_minimization(problem, qubits, entanglement, layers, name,
                                  epochs = 3000, batch_size = 20, eta = .1, err=False):
    """
    This function creates data and minimizes whichever problem using a step by step SGD and saving all results from accuracies for training and test sets
    INPUT:
        -problem: name of the problem, to choose among
            ['circle', '3 circles', 'hypersphere', 'tricrown', 'non convex', 'crown', 'sphere', 'squares', 'wavy lines']

        -qubits: number of qubits, must be an integer
        -entanglement: whether there is entanglement or not in the Ansätze, just 'y'/'n'
        -layers: number of layers, must be an integer. If layers == 1, entanglement is not taken in account

        -method: minimization method, to choose among ['SGD', another valid for function scipy.optimize.minimize]

        -name: a name we want for our our files to be save with
        -seed: seed of numpy.random, needed for replicating results
        -epochs: number of epochs for a 'SGD' method. If there is another method, this input has got no importance

        -batch_size: size of the batches for stochastic gradient descent, only for 'SGD' method
        -eta: learning rate, only for 'SGD' method
    OUTPUT:
        This function has got no outputs, but several files are saved in an appropiate folder. The files are

        -summary.txt: Saves useful information for the problem
        -theta.txt: saves the theta parameters as a flat array
        -alpha.txt: saves the alpha parameters as a flat array
        -error_rates: accuracies for training and test sets as flat arrays
    """
    chi = 'fidelity_chi'
    method = 'SGD'


    data, drawing = data_generator(problem, err=err)
    if problem == 'sphere':
        train_data = data[:500]
        test_data = data[500:]
    elif problem == 'hypersphere':
        train_data = data[:1000]
        test_data = data[1000:]
    else:
        train_data = data[:250]
        test_data = data[250:]

    if chi == 'fidelity_chi':
        qubits_lab = qubits
    if chi == 'trace_chi':
        qubits_lab = qubits
    elif chi == 'weighted_fidelity_chi':
        qubits_lab = 1

    theta, alpha, reprs = problem_generator(problem, qubits, layers, chi,
                                            qubits_lab=qubits_lab)
    accs_test=[]
    accs_train=[]
    chis=[]
    acc_test_sol = 0
    acc_train_sol = 0
    fid_sol = 0
    best_epoch = 0
    theta_sol = theta.copy()
    alpha_sol = alpha.copy()

    file_text = write_epochs_file(chi, problem, qubits, entanglement, layers, method, name)
    for e in range(epochs):
        theta, alpha, fid = fidelity_minimization(theta, alpha, train_data, reprs,
                                                  entanglement, method, batch_size, eta, 1)

        acc_train = tester(theta, alpha, train_data, reprs, entanglement, chi)
        acc_test = tester(theta, alpha, test_data, reprs, entanglement, chi)
        accs_test.append(acc_test)
        accs_train.append(acc_train)
        chis.append(fid)

        write_epoch(file_text, e, theta, alpha, fid, acc_train, acc_test)

        if acc_test > acc_test_sol:

            acc_test_sol = acc_test
            acc_train_sol = acc_train
            fid_sol = fid
            theta_sol = theta
            alpha_sol = alpha
            best_epoch = e

    close_epochs_file(file_text, best_epoch)
    write_summary(chi, problem, qubits, entanglement, layers, method, name,
          theta_sol, alpha_sol, None, fid_sol, acc_train_sol, acc_test_sol, epochs)
    write_epochs_error_rate(chi, problem, qubits, entanglement, layers, method, name,
                            accs_train, accs_test)

def overlearning_paint(chi, problem, qubits, entanglement, layers, method, name):
    """
    This function takes overlearning functions and paints them
    INPUT:
        -chi: cost function, just 'fidelity_chi'
        -problem: name of the problem, to choose among
            ['circle', '3 circles', 'hypersphere', 'tricrown', 'non convex', 'crown', 'sphere', 'squares', 'wavy lines']

        -qubits: number of qubits, must be an integer
        -entanglement: whether there is entanglement or not in the Ansätze, just 'y'/'n'
        -layers: number of layers, must be an integer. If layers == 1, entanglement is not taken in account

        -method: minimization method, 'SGD'
        -name: a name we want for our our files to be save with
    OUTPUT:
        This function has got no outputs, but saves a picture with the information of the overlearning rates

    """
    foldname = name_folder(chi, problem, qubits, entanglement, layers, method)
    create_folder(foldname)
    filename_train = foldname + '/' + name + '_train.txt'
    filename_test = foldname + '/' + name + '_test.txt'

    train_err_rate = np.loadtxt(filename_train)
    test_err_rate = np.loadtxt(filename_test)
    fig, ax = plt.subplots()
    ax.plot(range(len(train_err_rate)), train_err_rate, label = 'Training set')
    ax.plot(range(len(test_err_rate)), test_err_rate, label = 'Test set')
    ax.set_xlabel('Epochs', fontsize=16)
    ax.set_ylabel('Error rate', fontsize=16)
    ax.legend()
    filename = foldname + '/' + name + '_overlearning'
    fig.savefig(filename)
    plt.close('all')
```

```python
#############################################################
#Quantum classifier
#Adrián Pérez-Salinas, Alba Cervera-Lierta, Elies Gil, J. Ignacio Latorre
#Code by APS
#Code-checks by ACL
#June 3rd 2019


#Universitat de Barcelona / Barcelona Supercomputing Center/Institut de Ciències del Cosmos

#############################################################
#This file is a file taking many different functions from other files and mixing them all together
# so that the usage is automatized
from data_gen import data_generator
from problem_gen import problem_generator, representatives
from fidelity_minimization import fidelity_minimization
from weighted_fidelity_minimization import weighted_fidelity_minimization
from test_data import Accuracy_test, tester
from save_data import write_summary, read_summary, name_folder, samples_paint, samples_paint_worldmap, laea_x, laea_y
from save_data import write_epochs_file, write_epoch, close_epochs_file, create_folder, write_epochs_error_rate
import numpy as np
import matplotlib.pyplot as plt
from circuitery import code_coords, circuit
from matplotlib.cm import get_cmap
from matplotlib.colors import Normalize

def minimizer(chi, problem, qubits, entanglement, layers, method, name,
              seed = 30, epochs=3000, batch_size=20,  eta=0.1):

    """
    This function creates data and minimizes whichever problem (from the selected ones)
    INPUT:
        -chi: cost function, to choose between 'fidelity_chi' or 'weighted_fidelity_chi'
        -problem: name of the problem, to choose among
            ['circle', '3 circles', 'hypersphere', 'tricrown', 'non convex', 'crown', 'sphere', 'squares', 'wavy lines']
        -qubits: number of qubits, must be an integer
        -entanglement: whether there is entanglement or not in the Ansätze, just 'y'/'n'
        -layers: number of layers, must be an integer. If layers == 1, entanglement is not taken in account
        -method: minimization method, to choose among ['SGD', another valid for function scipy.optimize.minimize]
        -name: a name we want for our our files to be save with
        -seed: seed of numpy.random, needed for replicating results
        -epochs: number of epochs for a 'SGD' method. If there is another method, this input has got no importance
        -batch_size: size of the batches for stochastic gradient descent, only for 'SGD' method
        -eta: learning rate, only for 'SGD' method
    OUTPUT:
        This function has got no outputs, but several files are saved in an appropiate folder. The files are
        -summary.txt: Saves useful information for the problem
        -theta.txt: saves the theta parameters as a flat array
        -alpha.txt: saves the alpha parameters as a flat array
        -weight.txt: saves the weights as a flat array if they exist
    """
    np.random.seed(seed)
    data, drawing = data_generator(problem)
    if problem == 'sphere':
        train_data = data[:500]
        test_data = data[500:]
    elif problem == 'hypersphere':
        train_data = data[:1000]
        test_data = data[1000:]
    else:
        train_data = data[:200]
        test_data = data[200:]

    if chi == 'fidelity_chi':
        qubits_lab = qubits
        theta, alpha, reprs = problem_generator(problem,qubits, layers, chi,
                                                qubits_lab=qubits_lab)
        theta, alpha, f = fidelity_minimization(theta, alpha, train_data, reprs,
                                                entanglement, method,
                                                batch_size, eta, epochs)
        acc_train = tester(theta, alpha, train_data, reprs, entanglement, chi)
        acc_test = tester(theta, alpha, test_data, reprs, entanglement, chi)
        write_summary(chi, problem, qubits, entanglement, layers, method, name,
                theta, alpha, 0, f, acc_train, acc_test, seed, epochs=epochs)
    elif chi == 'weighted_fidelity_chi':
        qubits_lab = 1
        theta, alpha, weight, reprs = problem_generator(problem,qubits, layers, chi,
                                                        qubits_lab=qubits_lab)
        theta, alpha, weight, f = weighted_fidelity_minimization(theta, alpha, weight, train_data, reprs,
                                                    entanglement, method)
        acc_train = tester(theta, alpha, train_data, reprs, entanglement, chi, weights=weight)
        acc_test = tester(theta, alpha, test_data, reprs, entanglement, chi, weights=weight)
        write_summary(chi, problem, qubits, entanglement, layers, method, name,
                theta, alpha, weight, f, acc_train, acc_test, seed, epochs=epochs)


def painter(chi, problem, qubits, entanglement, layers, method, name,
        seed = 30, standard_test = True, samples = 4000, bw = False, err = False):
    """
    This function takes written text files and paint the results of the problem
    INPUT:
        -chi: cost function, to choose between 'fidelity_chi' or 'weighted_fidelity_chi'
        -problem: name of the problem, to choose among
            ['circle', '3 circles', 'hypersphere', 'tricrown', 'non convex', 'crown', 'sphere', 'squares', 'wavy lines']
        -qubits: number of qubits, must be an integer
        -entanglement: whether there is entanglement or not in the Ansätze, just 'y'/'n'
        -layers: number of layers, must be an integer. If layers == 1, entanglement is not taken in account
        -method: minimization method, to choose among ['SGD', another valid for function scipy.optimize.minimize]
        -name: a name we want for our our files to be save with
        -seed: seed of numpy.random, needed for replicating results
        -standard_test: Whether we want to paint the set test used for checking when minimizing. If True, seed and samples are not taken in account
        -samples: number of samples of the test set
        -bw: painting in black and white
    OUTPUT:
        This function has got no outputs, but a file containing the representation of the test set is created
    """
    np.random.seed(seed)

    if chi == 'fidelity_chi':
        qubits_lab = qubits
    elif chi == 'weighted_fidelity_chi':
        qubits_lab = 1


    if standard_test == True:
        data, drawing = data_generator(problem)
        if problem == 'sphere':
            test_data = data[500:]
        elif problem == 'hypersphere':
            test_data = data[1000:]
        else:
            test_data = data[200:]

    elif standard_test == False:
        test_data, drawing = data_generator(problem, samples = samples)

    if problem in ['circle','wavy circle','sphere', 'non convex', 'crown', 'hypersphere']:
        classes = 2
    if problem in ['tricrown']:
        classes = 3
    elif problem in ['3 circles','wavy lines','squares']:
        classes = 4

    reprs = representatives(classes, qubits_lab)

    params = read_summary(chi, problem, qubits, entanglement, layers, method, name)

    if chi == 'fidelity_chi':
        theta, alpha = params
        sol_test, acc_test = Accuracy_test(theta, alpha, test_data, reprs, entanglement, chi)

    if chi == 'weighted_fidelity_chi':
        theta, alpha, weight = params
        sol_test, acc_test = Accuracy_test(theta, alpha, test_data, reprs,
                                           entanglement, chi, weights = weight)

    foldname = name_folder(chi, problem, qubits, entanglement, layers, method)
    samples_paint(problem, drawing, sol_test, foldname, name, bw)

def paint_world(chi, problem, qubits, entanglement, layers, method, name,
        seed = 30, standard_test = True, samples = 4000, bw = False, err = False):
    np.random.seed(seed)

    if chi == 'fidelity_chi':
        qubits_lab = qubits
    elif chi == 'weighted_fidelity_chi':
        qubits_lab = 1

    if standard_test == True:
        data, drawing = data_generator(problem)
        if problem == 'sphere':
            test_data = data[500:]
        elif problem == 'hypersphere':
            test_data = data[1000:]
        else:
            test_data = data[200:]

    elif standard_test == False:
        test_data, drawing = data_generator(problem, samples=samples)

    if problem in ['circle', 'wavy circle', 'sphere', 'non convex', 'crown', 'hypersphere']:
        classes = 2
    if problem in ['tricrown']:
        classes = 3
    elif problem in ['3 circles', 'wavy lines', 'squares']:
        classes = 4

    reprs = representatives(classes, qubits_lab)

    params = read_summary(chi, problem, qubits, entanglement, layers, method, name)

    if chi == 'fidelity_chi':
        theta, alpha = params
        sol_test, acc_test = Accuracy_test(theta, alpha, test_data, reprs, entanglement, chi)

    if chi == 'weighted_fidelity_chi':
        theta, alpha, weight = params
        sol_test, acc_test = Accuracy_test(theta, alpha, test_data, reprs,
                                           entanglement, chi, weights=weight)

    foldname = name_folder(chi, problem, qubits, entanglement, layers, method)
    angles = np.zeros((len(sol_test), 2))
    for i, x in enumerate(sol_test[:, :2]):
        theta_aux = code_coords(theta, alpha, x)
        C = circuit(theta_aux, entanglement)
        angles[i, 0] = np.arccos(np.abs(C.psi[0])**2 - np.abs(C.psi[1])**2) - np.pi/2
        angles[i, 1] = np.angle(C.psi[1] / C.psi[0])
        print(angles[i])

    if bw == False:
        colors_classes = get_cmap('plasma')
        norm_class = Normalize(vmin=-.5, vmax=np.max(sol_test[:, -3]) + .5)

        colors_rightwrong = get_cmap('RdYlGn')
        norm_rightwrong = Normalize(vmin=-.1, vmax=1.1)

    if bw == True:
        colors_classes = get_cmap('Greys')
        norm_class = Normalize(vmin=-.1, vmax=np.max(sol[:, -3]) + .1)

        colors_rightwrong = get_cmap('Greys')
        norm_rightwrong = Normalize(vmin=-.1, vmax=1.1)

    fig, ax = plt.subplots(nrows=2)
    ax[0].plot(laea_x(np.pi, np.linspace(0, np.pi)), laea_y(np.pi, np.linspace(0, np.pi)), color='k')
    ax[0].plot(laea_x(-np.pi, np.linspace(0, -np.pi)), laea_y(-np.pi, np.linspace(0, -np.pi)), color='k')
    ax[1].plot(laea_x(np.pi, np.linspace(0, np.pi)), laea_y(np.pi, np.linspace(0, np.pi)), color='k')
    ax[1].plot(laea_x(-np.pi, np.linspace(0, -np.pi)), laea_y(-np.pi, np.linspace(0, -np.pi)), color='k')
    ax[0].scatter(laea_x(angles[:, 1], angles[:, 0]), laea_y(angles[:, 1], angles[:, 0]), c=sol_test[:, -2],
                  cmap=colors_classes, s=2, norm=norm_class)
    ax[1].scatter(laea_x(angles[:, 1], angles[:, 0]), laea_y(angles[:, 1], angles[:, 0]), c=sol_test[:,-1], cmap = colors_rightwrong, s=2, norm=norm_rightwrong)
    plt.show()

def SGD_step_by_step_minimization(problem, qubits, entanglement, layers, name,
                                  seed = 30, epochs = 3000, batch_size = 20, eta = .1, err=False):
    """
    This function creates data and minimizes whichever problem using a step by step SGD and saving all results from accuracies for training and test sets
    INPUT:
        -problem: name of the problem, to choose among
            ['circle', '3 circles', 'hypersphere', 'tricrown', 'non convex', 'crown', 'sphere', 'squares', 'wavy lines']
        -qubits: number of qubits, must be an integer
        -entanglement: whether there is entanglement or not in the Ansätze, just 'y'/'n'
        -layers: number of layers, must be an integer. If layers == 1, entanglement is not taken in account
        -method: minimization method, to choose among ['SGD', another valid for function scipy.optimize.minimize]
        -name: a name we want for our our files to be save with
        -seed: seed of numpy.random, needed for replicating results
        -epochs: number of epochs for a 'SGD' method. If there is another method, this input has got no importance
        -batch_size: size of the batches for stochastic gradient descent, only for 'SGD' method
        -eta: learning rate, only for 'SGD' method
    OUTPUT:
        This function has got no outputs, but several files are saved in an appropiate folder. The files are
        -summary.txt: Saves useful information for the problem
        -theta.txt: saves the theta parameters as a flat array
        -alpha.txt: saves the alpha parameters as a flat array
        -error_rates: accuracies for training and test sets as flat arrays
    """
    chi = 'fidelity_chi'
    method = 'SGD'

    np.random.seed(seed)
    data, drawing = data_generator(problem, err=err)
    if problem == 'sphere':
        train_data = data[:500]
        test_data = data[500:]
    elif problem == 'hypersphere':
        train_data = data[:1000]
        test_data = data[1000:]
    else:
        train_data = data[:200]
        test_data = data[200:]

    if chi == 'fidelity_chi':
        qubits_lab = qubits
    elif chi == 'weighted_fidelity_chi':
        qubits_lab = 1

    theta, alpha, reprs = problem_generator(problem, qubits, layers, chi,
                                            qubits_lab=qubits_lab)
    accs_test=[]
    accs_train=[]
    chis=[]
    acc_test_sol = 0
    acc_train_sol = 0
    fid_sol = 0
    best_epoch = 0
    theta_sol = theta.copy()
    alpha_sol = alpha.copy()

    file_text = write_epochs_file(chi, problem, qubits, entanglement, layers, method, name)
    for e in range(epochs):
        theta, alpha, fid = fidelity_minimization(theta, alpha, train_data, reprs,
                                                  entanglement, method, batch_size, eta, 1)

        acc_train = tester(theta, alpha, train_data, reprs, entanglement, chi)
        acc_test = tester(theta, alpha, test_data, reprs, entanglement, chi)
        accs_test.append(acc_test)
        accs_train.append(acc_train)
        chis.append(fid)

        write_epoch(file_text, e, theta, alpha, fid, acc_train, acc_test)

        if acc_test > acc_test_sol:

            acc_test_sol = acc_test
            acc_train_sol = acc_train
            fid_sol = fid
            theta_sol = theta
            alpha_sol = alpha
            best_epoch = e

    close_epochs_file(file_text, best_epoch)
    write_summary(chi, problem, qubits, entanglement, layers, method, name,
          theta_sol, alpha_sol, None, fid_sol, acc_train_sol, acc_test_sol, seed, epochs)
    write_epochs_error_rate(chi, problem, qubits, entanglement, layers, method, name,
                            accs_train, accs_test)

def overlearning_paint(chi, problem, qubits, entanglement, layers, method, name):
    """
    This function takes overlearning functions and paints them
    INPUT:
        -chi: cost function, just 'fidelity_chi'
        -problem: name of the problem, to choose among
            ['circle', '3 circles', 'hypersphere', 'tricrown', 'non convex', 'crown', 'sphere', 'squares', 'wavy lines']
        -qubits: number of qubits, must be an integer
        -entanglement: whether there is entanglement or not in the Ansätze, just 'y'/'n'
        -layers: number of layers, must be an integer. If layers == 1, entanglement is not taken in account
        -method: minimization method, 'SGD'
        -name: a name we want for our our files to be save with
    OUTPUT:
        This function has got no outputs, but saves a picture with the information of the overlearning rates
    """
    foldname = name_folder(chi, problem, qubits, entanglement, layers, method)
    create_folder(foldname)
    filename_train = foldname + '/' + name + '_train.txt'
    filename_test = foldname + '/' + name + '_test.txt'

    train_err_rate = np.loadtxt(filename_train)
    test_err_rate = np.loadtxt(filename_test)
    fig, ax = plt.subplots()
    ax.plot(range(len(train_err_rate)), train_err_rate, label = 'Training set')
    ax.plot(range(len(test_err_rate)), test_err_rate, label = 'Test set')
    ax.set_xlabel('Epochs', fontsize=16)
    ax.set_ylabel('Error rate', fontsize=16)
    ax.legend()
    filename = foldname + '/' + name + '_overlearning'
    fig.savefig(filename)
    plt.close('all')
```

Text Compare

```
# coding=utf-8
###########################################################################
#Quantum classifier
#Sara Aminpour, Mike Banad, Sarah Sharif
#September 25th 2024

#School of Electrical and Computer Engineering/ Center for Quantum and Technology, University of Oklahoma, Norman, OK 73019 USA,
###########################################################################
#IMPORTANT_NOTE:
#The code on the left was developed by Sara Aminpour, while the code on the right serves as the reference implementation by Adrián Pérez-Salinas.
#The code on the left has been restructured to handle random data. So some certain sections has been deleted from the reference code.
#Additionally, our code on the left developed to analyze trace distance cost function and linear classification problem
#as well as necessary modification to apply COBYLA, L-BFGS-B, NELDER-MEAD, and SLSQP minimization methods.
###########################################################################

## This file creates the data points for the different problems to be tackled by the quantum classifier



import numpy as np

problems = ['circle', 'line', '3 circles', 'wavy circle', 'hypersphere', 'tricrown', 'non convex', 'crown', 'sphere', 'squares', 'wavy lines']

def data_generator(problem, samples=None):
    """
    This function generates the data for a problem
    INPUT:
        -problem: Name of the problem, one of: 'circle', '3 circles', 'hypersphere', 'tricrown', 'non convex', 'crown', 'sphere', 'squares', 'wavy lines'
        -samples Number of samples for the data
    OUTPUT:
        -data: set of training and test data
        -settings: things needed for drawing
    """
    problem = problem.lower()
    if problem not in problems:
        raise ValueError('problem must be one of {}'.format(problems))
    if samples == None:
        if problem == 'sphere':
            samples = 4500
        elif problem == 'hypersphere':
            samples = 5000
        else:
            samples = 4250

    if problem == 'circle':
        data, settings = _circle(samples)

    if problem == '3 circles':
        data, settings = _3_circles(samples)

    if problem == 'wavy lines':
        data, settings = _wavy_lines(samples)

    if problem == 'squares':
        data, settings = _squares(samples)

    if problem == 'sphere':
        data, settings = _sphere(samples)

    if problem == 'non convex':
        data, settings = _non_convex(samples)

    if problem == 'crown':
        data, settings = _crown(samples)

    if problem == 'tricrown':
        data, settings = _tricrown(samples)

    if problem == 'hypersphere':
        data, settings = _hypersphere(samples)
#==========================================================================
    if problem == 'line':
        data, settings = _line(samples)
#==========================================================================

    return data, settings

def _circle(samples):
    centers = np.array([[0, 0]])
    radii = np.array([np.sqrt(2/np.pi)])
    data=[]
    dim = 2
    for i in range(samples):
        x = 2 * (np.random.rand(dim)) - 1
        y = 0
        for c, r in zip(centers, radii):
            if np.linalg.norm(x - c) < r:
                y = 1

        data.append([x, y])

    return data, (centers, radii)

def _3_circles(samples):
    centers = np.array([[-1, 1], [1, 0], [-.5, -.5]])
    radii = np.array([1, np.sqrt(6/np.pi - 1), 1/2])
    data=[]
    dim = 2
    for i in range(samples):
        x = 2 * (np.random.rand(dim)) - 1
        y = 0
        for j, (c, r) in enumerate(zip(centers, radii)):
            if np.linalg.norm(x - c) < r:
                y = j + 1

        data.append([x, y])


    return data, (centers, radii)


def _wavy_lines(samples, freq = 1):
    def fun1(s):
        return s + np.sin(freq * np.pi * s)

    def fun2(s):
        return -s + np.sin(freq * np.pi * s)
    data=[]
    dim=2
    for i in range(samples):
        x = 2 * (np.random.rand(dim)) - 1
        if x[1] < fun1(x[0]) and x[1] < fun2(x[0]): y = 0
        if x[1] < fun1(x[0]) and x[1] > fun2(x[0]): y = 1
        if x[1] > fun1(x[0]) and x[1] < fun2(x[0]): y = 2
        if x[1] > fun1(x[0]) and x[1] > fun2(x[0]): y = 3
        data.append([x, y])

    return data, freq

def _squares(samples):
    data=[]
    dim=2
    for i in range(samples):
        x = 2 * (np.random.rand(dim)) - 1
        if x[0] < 0 and x[1] < 0: y = 0
        if x[0] < 0 and x[1] > 0: y = 1
        if x[0] > 0 and x[1] < 0: y = 2
        if x[0] > 0 and x[1] > 0: y = 3
        data.append([x, y])

    return data, None

#==========================================================================
def _line(samples):
    data=[]
    dim=2
    for i in range(samples):
        x = 2 * np.random.rand(dim) -1
        #x = np.random.rand(dim)
        if x[0] < x[1] : y = 0
        if x[0] > x[1] : y = 1

        data.append([x, y])

    return data, None
#==========================================================================

def _non_convex(samples, freq = 1, x_val = 2, sin_val = 1.5):
    def fun(s):
        return -x_val * s + sin_val * np.sin(freq * np.pi * s)

    data = []
    dim = 2
    for i in range(samples):
        x = 2 * (np.random.rand(dim)) - 1
        if x[1] < fun(x[0]): y = 0
        if x[1] > fun(x[0]): y = 1
        data.append([x, y])

    return data, (freq, x_val, sin_val)

def _crown(samples):
    c = [[0,0],[0,0]]
    r = [np.sqrt(.8), np.sqrt(.8 - 2/np.pi)]
    data = []
    dim = 2
    for i in range(samples):
        x = 2 * (np.random.rand(dim)) - 1
        if np.linalg.norm(x - c[0]) < r[0] and np.linalg.norm(x - c[1]) > r[1]:
            y = 1
        else:
            y=0
        data.append([x, y])

    return data, (c, r)


def _tricrown(samples):
    centers = [[0,0],[0,0]]
    radii = [np.sqrt(.8 - 2/np.pi), np.sqrt(.8)]
    data = []
    dim = 2
    for i in range(samples):
        x = 2 * (np.random.rand(dim)) - 1
        y=0
        for j,(r,c) in enumerate(zip(radii, centers)):
            if np.linalg.norm(x - c) > r:
                y = j + 1
        data.append([x, y])

    return data, (centers, radii)

def _sphere(samples):
    centers = np.array([[0, 0, 0]])
    radii = np.array([(3/np.pi)**(1/3)])
    data=[]
    dim = 3
    for i in range(samples):
        x = 2 * (np.random.rand(dim)) - 1
        y = 0
        for c, r in zip(centers, radii):
            if np.linalg.norm(x - c) < r:
                y = 1

        data.append([x, y])

    return data, (centers, radii)

def _hypersphere(samples):
    centers = np.array([[0, 0, 0, 0]])
    radii = np.array([(2/np.pi)**(1/2)])
    data=[]
    dim = 4
    for i in range(samples):
        x = 2 * (np.random.rand(dim)) - 1
        y = 0
        for c, r in zip(centers, radii):
            if np.linalg.norm(x - c) < r:
                y = 1

        data.append([x, y])

    return data, (centers, radii)
```

```
###########################################################################
#Quantum classifier
#Adrián Pérez-Salinas, Alba Cervera-Lierta, Elies Gil, J. Ignacio Latorre
#Code by APS
#Code-checks by ACL
#June 3rd 2019


#Universitat de Barcelona / Barcelona Supercomputing Center/Institut de Ciències del Cosmos

###########################################################################

## This file creates the data points for the different problems to be tackled by the quantum classifier



import numpy as np

problems = ['circle', '3 circles', 'wavy circle', 'hypersphere', 'tricrown', 'non convex', 'crown', 'sphere', 'squares', 'wavy lines']

def data_generator(problem, samples=None):
    """
    This function generates the data for a problem
    INPUT:
        -problem: Name of the problem, one of: 'circle', '3 circles', 'hypersphere', 'tricrown', 'non convex', 'crown', 'sphere', 'squares', 'wavy lines'
        -samples Number of samples for the data
    OUTPUT:
        -data: set of training and test data
        -settings: things needed for drawing
    """
    problem = problem.lower()
    if problem not in problems:
        raise ValueError('problem must be one of {}'.format(problems))
    if samples == None:
        if problem == 'sphere':
            samples = 4500
        elif problem == 'hypersphere':
            samples = 5000
        else:
            samples = 4200

    if problem == 'circle':
        data, settings = _circle(samples)

    if problem == '3 circles':
        data, settings = _3_circles(samples)

    if problem == 'wavy lines':
        data, settings = _wavy_lines(samples)

    if problem == 'squares':
        data, settings = _squares(samples)

    if problem == 'sphere':
        data, settings = _sphere(samples)

    if problem == 'non convex':
        data, settings = _non_convex(samples)

    if problem == 'crown':
        data, settings = _crown(samples)

    if problem == 'tricrown':
        data, settings = _tricrown(samples)

    if problem == 'hypersphere':
        data, settings = _hypersphere(samples)


    return data, settings

def _circle(samples):
    centers = np.array([[0, 0]])
    radii = np.array([np.sqrt(2/np.pi)])
    data=[]
    dim = 2
    for i in range(samples):
        x = 2 * (np.random.rand(dim)) - 1
        y = 0
        for c, r in zip(centers, radii):
            if np.linalg.norm(x - c) < r:
                y = 1

        data.append([x, y])

    return data, (centers, radii)

def _3_circles(samples):
    centers = np.array([[-1, 1], [1, 0], [-.5, -.5]])
    radii = np.array([1, np.sqrt(6/np.pi - 1), 1/2])
    data=[]
    dim = 2
    for i in range(samples):
        x = 2 * (np.random.rand(dim)) - 1
        y = 0
        for j, (c, r) in enumerate(zip(centers, radii)):
            if np.linalg.norm(x - c) < r:
                y = j + 1

        data.append([x, y])


    return data, (centers, radii)


def _wavy_lines(samples, freq = 1):
    def fun1(s):
        return s + np.sin(freq * np.pi * s)

    def fun2(s):
        return -s + np.sin(freq * np.pi * s)
    data=[]
    dim=2
    for i in range(samples):
        x = 2 * (np.random.rand(dim)) - 1
        if x[1] < fun1(x[0]) and x[1] < fun2(x[0]): y = 0
        if x[1] < fun1(x[0]) and x[1] > fun2(x[0]): y = 1
        if x[1] > fun1(x[0]) and x[1] < fun2(x[0]): y = 2
        if x[1] > fun1(x[0]) and x[1] > fun2(x[0]): y = 3
        data.append([x, y])

    return data, freq

def _squares(samples):
    data=[]
    dim=2
    for i in range(samples):
        x = 2 * (np.random.rand(dim)) - 1
        if x[0] < 0 and x[1] < 0: y = 0
        if x[0] < 0 and x[1] > 0: y = 1
        if x[0] > 0 and x[1] < 0: y = 2
        if x[0] > 0 and x[1] > 0: y = 3
        data.append([x, y])

    return data, None


def _non_convex(samples, freq = 1, x_val = 2, sin_val = 1.5):
    def fun(s):
        return -x_val * s + sin_val * np.sin(freq * np.pi * s)

    data = []
    dim = 2
    for i in range(samples):
        x = 2 * (np.random.rand(dim)) - 1
        if x[1] < fun(x[0]): y = 0
        if x[1] > fun(x[0]): y = 1
        data.append([x, y])

    return data, (freq, x_val, sin_val)

def _crown(samples):
    c = [[0,0],[0,0]]
    r = [np.sqrt(.8), np.sqrt(.8 - 2/np.pi)]
    data = []
    dim = 2
    for i in range(samples):
        x = 2 * (np.random.rand(dim)) - 1
        if np.linalg.norm(x - c[0]) < r[0] and np.linalg.norm(x - c[1]) > r[1]:
            y = 1
        else:
            y=0
        data.append([x, y])

    return data, (c, r)


def _tricrown(samples):
    centers = [[0,0],[0,0]]
    radii = [np.sqrt(.8 - 2/np.pi), np.sqrt(.8)]
    data = []
    dim = 2
    for i in range(samples):
        x = 2 * (np.random.rand(dim)) - 1
        y=0
        for j,(r,c) in enumerate(zip(radii, centers)):
            if np.linalg.norm(x - c) > r:
                y = j + 1
        data.append([x, y])

    return data, (centers, radii)

def _sphere(samples):
    centers = np.array([[0, 0, 0]])
    radii = np.array([(3/np.pi)**(1/3)])
    data=[]
    dim = 3
    for i in range(samples):
        x = 2 * (np.random.rand(dim)) - 1
        y = 0
        for c, r in zip(centers, radii):
            if np.linalg.norm(x - c) < r:
                y = 1

        data.append([x, y])

    return data, (centers, radii)

def _hypersphere(samples):
    centers = np.array([[0, 0, 0, 0]])
    radii = np.array([(2/np.pi)**(1/2)])
    data=[]
    dim = 4
    for i in range(samples):
        x = 2 * (np.random.rand(dim)) - 1
        y = 0
        for c, r in zip(centers, radii):
            if np.linalg.norm(x - c) < r:
                y = 1

        data.append([x, y])

    return data, (centers, radii)
```

Text Compare

```python
#Quantum classifier
#Sara Aminpour, Mike Banad, Sarah Sharif
#September 25th 2024

#School of Electrical and Computer Engineering/ Center for Quantum and Technology, University of Oklahoma, Norman, OK 73019 USA,
###########################################################################
#IMPORTANT_NOTE:
#The code on the left was developed by Sara Aminpour, while the code on the right serves as the reference implementation by Adrián Pérez-Salinas.
#The code on the left has been restructured to handle random data. So some certain sections has been deleted from the reference code.
#Additionally, our code on the left developed to analyze trace distance cost function and linear classification problem
#as well as necessary modification to apply COBYLA, L-BFGS-B, NELDER-MEAD, and SLSQP minimization methods.
from big_functions import minimizer, painter, SGD_step_by_step_minimization, overlearning_paint

import datetime
qubits = 2  #integer, number of qubits
layers = 5 #integer, number of layers (time we reupload data)
chi = 'fidelity_chi' #Cost function; choose between ['fidelity_chi', 'trace_chi']

entanglement = 'y' #entanglement y/n
name = 'run' #However you want to name your files
seed = 30 #random seed
#epochs=3000 #number of epochs, only for SGD methods


problem=['circle', 'line'] #name of the problem, choose among ['circle', 'wavy circle', '3 circles', 'wavy lines', 'sphere', 'non convex', 'crown']
for problem in problem:

            method = ['l-bfgs-b', 'cobyla', 'nelder-mead', 'slsqp'] #minimization methods between ['l-bfgs-b', 'cobyla', 'nelder-mead', 'slsqp']
            for method in method:
                a=datetime.datetime.now()
                #SGD_step_by_step_minimization(problem, qubits, entanglement, layers, name)
                minimizer(chi, problem, qubits, entanglement, layers, method, name)
                painter(chi, problem, qubits, entanglement, layers, method, name, standard_test=True)

                #paint_world(chi, problem, qubits, entanglement, layers, method, name, standard_test=True)

                b=datetime.datetime.now()
                c=b-a

                text_file_nn = open('time.txt', mode='a+')
                text_file_nn.write(problem +'_'+ chi +'_'+ method +'_'+ str(qubits) +'Qubits_' + entanglement +'_'+ str(layers) +'Layers_' + method + "__" + 'total_time'+' = '+ str(c))
                text_file_nn.write('\n')
                text_file_nn.write('==================================================================================')
                text_file_nn.write('\n')
                text_file_nn.close()
```

```python
from big_functions import minimizer, painter, SGD_step_by_step_minimization, overlearning_paint, paint_world

qubits = 1  #integer, number of qubits
layers = 5 #integer, number of layers (time we reupload data)
chi = 'fidelity_chi' #Cost function; choose between ['fidelity_chi', 'weighted_fidelity_chi']
problem='wavy lines' #name of the problem, choose among ['circle', 'wavy circle', '3 circles', 'wavy lines', 'sphere', 'non convex', 'crown']
entanglement = 'y' #entanglement y/n
method = 'L-BFGS-B' #minimization methods, scipy methods or 'SGD'
name = 'run' #However you want to name your files
seed = 30 #random seed
#epochs=3000 #number of epochs, only for SGD methods

#SGD_step_by_step_minimization(problem, qubits, entanglement, layers, name)
minimizer(chi, problem, qubits, entanglement, layers, method, name, seed = seed)
painter(chi, problem, qubits, entanglement, layers, method, name, standard_test=True, seed=seed)
paint_world(chi, problem, qubits, entanglement, layers, method, name, standard_test=True, seed=seed)
```

Text Compare

```python
# coding=utf-8
###########################################################################
#Quantum classifier
#Sara Aminpour, Mike Banad, Sarah Sharif
#September 25th 2024

#School of Electrical and Computer Engineering/ Center for Quantum and Technology, University of Oklahoma, Norman, OK 73019 USA,
###########################################################################
#IMPORTANT_NOTE:
#The code on the left was developed by Sara Aminpour, while the code on the right serves as the reference implementation by Adrián Pérez-Salinas.
#The code on the left has been restructured to handle random data. So some certain sections has been deleted from the reference code.
#Additionally, our code on the left developed to analyze trace distance cost function and linear classification problem
#as well as necessary modification to apply COBYLA, L-BFGS-B, NELDER-MEAD, and SLSQP minimization methods.
###########################################################################

## This file creates the problems and their settings
import numpy as np

def problem_generator(problem, qubits, layers, chi, qubits_lab=1):
    """
    This function generates everything needed for solving the problem
    INPUT:
        -chi: cost function, to choose between 'fidelity_chi' or 'weighted_fidelity_chi'
        -problem: name of the problem, to choose among
            ['circle', '3 circles', 'hypersphere', 'tricrown', 'non convex', 'crown', 'sphere', 'squares', 'wavy lines']
        -qubits: number of qubits, must be an integer
        -layers: number of layers, must be an integer. If layers == 1, entanglement is not taken in account


    OUTPUT:
        -theta: set of parameters needed for the circuit. It is an array with shape (qubits, layers, 3)
        -alpha: set of parameters needed for the circuit. It is an array with shape (qubits, layers, dimension of data)
        -weight: set of parameters needed fot the circuit only if chi == 'weighted_fidelity_chi'. It is an array with shape (classes, qubits)
        -reprs: variable encoding the label states of the different classes
    """
    chi = chi.lower()
    if chi in ['fidelity', 'weighted_fidelity','trace']: chi += '_chi'
    if chi not in ['fidelity_chi', 'weighted_fidelity_chi','trace_chi']:
        raise ValueError('Figure of merit is not valid')

    if chi == 'weighted_fidelity_chi' and qubits_lab != 1:
        qubits_lab = 1
        print('WARNING: number of qubits for the label states has been changed to 1')

    problem = problem.lower()
    if problem == 'circle':
        theta, alpha, reprs = _circle(qubits, layers, qubits_lab, chi)
    elif problem == '3 circles':
        theta, alpha, reprs = _3_circles(qubits, layers, qubits_lab, chi)
    elif problem == 'wavy lines':
        theta, alpha, reprs = _wavy_lines(qubits, layers, qubits_lab, chi)
    elif problem == 'squares':
        theta, alpha, reprs = _squares(qubits, layers, qubits_lab, chi)
    elif problem == 'sphere':
        theta, alpha, reprs = _sphere(qubits, layers, qubits_lab, chi)
    elif problem == 'non convex':
        theta, alpha, reprs = _non_convex(qubits, layers, qubits_lab, chi)
    elif problem == 'crown':
        theta, alpha, reprs = _crown(qubits, layers, qubits_lab, chi)
    elif problem == 'tricrown':
        theta, alpha, reprs = _tricrown(qubits, layers, qubits_lab, chi)
    elif problem == 'hypersphere':
        theta, alpha, reprs = _hypersphere(qubits, layers, qubits_lab, chi)
#========================================================================
    elif problem == 'line':
        theta, alpha, reprs = _line(qubits, layers, qubits_lab, chi)
#========================================================================
    else:
        raise ValueError('Problem is not valid')

    if chi == 'fidelity_chi':
        return theta, alpha, reprs
    elif chi == 'trace_chi':
        return theta, alpha, reprs
    elif chi == 'weighted_fidelity_chi':
        weights = np.ones((len(reprs), qubits))
        return theta, alpha, weights, reprs

#All these are auxiliary functions for problem_generator
def _circle(qubits, layers, qubits_lab, chi):
    classes = 2
    if chi == 'trace_chi':
        reprs = representatives_tr(classes, qubits_lab)
    else:
        reprs = representatives(classes, qubits_lab)

    theta = np.random.rand(qubits, layers, 3)
    alpha = np.random.rand(qubits, layers, 2)
    return theta, alpha, reprs

def _3_circles(qubits, layers, qubits_lab, chi):
    classes = 4
    reprs = representatives(classes, qubits_lab)
    theta = np.random.rand(qubits, layers, 3)
    alpha = np.random.rand(qubits, layers, 2)
    return theta, alpha, reprs

def _wavy_lines(qubits, layers, qubits_lab, chi):
    classes = 4
    reprs = representatives(classes, qubits_lab)
    theta = np.random.rand(qubits, layers, 3)
    alpha = np.random.rand(qubits, layers, 2)
    return theta, alpha, reprs

def _squares(qubits, layers, qubits_lab, chi):
    classes = 4
    reprs = representatives(classes, qubits_lab)
    theta = np.random.rand(qubits, layers, 3)
    alpha = np.random.rand(qubits, layers, 2)
    return theta, alpha, reprs
#========================================================================
def _line(qubits, layers, qubits_lab, chi):
    classes = 2
    if chi == 'trace_chi':
        reprs = representatives_tr(classes, qubits_lab)
    else:
        reprs = representatives(classes, qubits_lab)

    theta = np.random.rand(qubits, layers, 3)
    alpha = np.random.rand(qubits, layers, 2)
    return theta, alpha, reprs
#========================================================================
def _non_convex(qubits, layers, qubits_lab, chi):
    classes = 2
    if chi == 'trace_chi':
        reprs = representatives_tr(classes, qubits_lab)
    else:
        reprs = representatives(classes, qubits_lab)

    theta = np.random.rand(qubits, layers, 3)
    alpha = np.random.rand(qubits, layers, 2)
    return theta, alpha, reprs

def _crown(qubits, layers, qubits_lab, chi):
    classes = 2
    if chi == 'trace_chi':
        reprs = representatives_tr(classes, qubits_lab)
    else:
        reprs = representatives(classes, qubits_lab)

    theta = np.random.rand(qubits, layers, 3)
    alpha = np.random.rand(qubits, layers, 2)
    return theta, alpha, reprs

def _tricrown(qubits, layers, qubits_lab, chi):
    classes = 3
    reprs = representatives(classes, qubits_lab)
    theta = np.random.rand(qubits, layers, 3)
    alpha = np.random.rand(qubits, layers, 2)
    return theta, alpha, reprs

def _sphere(qubits, layers, qubits_lab, chi):
    classes = 2
    reprs = representatives(classes, qubits_lab)
    theta = np.random.rand(qubits, layers, 3)
    alpha = np.random.rand(qubits, layers, 3)
    return theta, alpha, reprs

def _hypersphere(qubits, layers, qubits_lab, chi):
    classes = 2
    reprs = representatives(classes, qubits_lab)
    theta = np.random.rand(qubits, layers, 6)
    alpha = np.random.rand(qubits, layers, 4)
    return theta, alpha, reprs


def representatives_tr(classes, qubits_lab):
    """
    This function creates the label states for the classification task
    INPUT:
        -classes: number of classes of our problem
        -qubits_lab: how many qubits will store the labels
    OUTPUT:
        -reprs: the label states
    """
    #reprs = np.zeros((classes, 2**qubits_lab), dtype = 'complex')
    reprs = np.zeros((classes, 3), dtype = 'complex')
    if qubits_lab == 1:
        if classes == 0:
            raise ValueError('Nonsense classifier')
        if classes == 1:
            raise ValueError('Nonsense classifier')
        if classes == 2:
            #reprs[0] = np.array([1, 0])
            reprs[0] = np.array([0.2938926261462367, -0.5090369604551273, 0.8090169943749473])
            reprs[1] = np.array([-0.2938926261462367, 0.5090369604551273, -0.8090169943749473])
        if classes == 3:
            reprs[0] = np.array([1, 0])
            reprs[1] = np.array([1 / 2, np.sqrt(3) / 2])
            reprs[2] = np.array([1 / 2, -np.sqrt(3) / 2])
        if classes == 4:
            reprs[0] = np.array([1, 0])
            reprs[1] = np.array([1 / np.sqrt(3), np.sqrt(2 / 3)])
            reprs[2] = np.array([1 / np.sqrt(3), np.exp(1j * 2 * np.pi / 3) * np.sqrt(2 / 3)])
            reprs[3] = np.array([1 / np.sqrt(3), np.exp(-1j * 2 * np.pi / 3) * np.sqrt(2 / 3)])
        if classes == 6:
            reprs[0] = np.array([0.2938926261462367, -0.5090369604551273, 0.8090169943749473])
            reprs[1] = np.array([-0.2938926261462367, 0.5090369604551273, -0.8090169943749473])
            reprs[2] = np.array([-0.7006292692220369, -0.4045084971874737, 0.5877852522924729])
            reprs[3] = np.array([0.7006292692220369, 0.4045084971874737, -0.5877852522924729])
            reprs[4] = np.array([0.4045084971874736, -0.7006292692220369, 0.5877852522924729])
            reprs[5] = np.array([0.7006292692220369, 0.4045084971874737, -0.5877852522924729])

    if qubits_lab == 2:
        if classes == 0:
            raise ValueError('Nonsense classifier')
        if classes == 1:
            raise ValueError('Nonsense classifier')
        if classes == 2:
            reprs[0] = np.array([0.29, -0.5, 0.8])
            reprs[1] = np.array([-0.29, 0.5, -0.8])
        if classes == 3:
            reprs[0] = np.array([1, 0, 0, 0])
            reprs[1] = np.array([0, 1, 0, 0])
            reprs[2] = np.array([0, 0, 1, 0])
        if classes == 4:
            reprs[0] = np.array([1, 0, 0, 0])
            reprs[1] = np.array([0, 1, 0, 0])
            reprs[2] = np.array([0, 0, 1, 0])
            reprs[3] = np.array([0, 0, 0, 1])

    return reprs

def representatives(classes, qubits_lab):
    """
    This function creates the label states for the classification task
    INPUT:
        -classes: number of classes of our problem
        -qubits_lab: how many qubits will store the labels
    OUTPUT:
        -reprs: the label states
    """
    reprs = np.zeros((classes, 2**qubits_lab), dtype = 'complex')
    if qubits_lab == 1:
        if classes == 0:
            raise ValueError('Nonsense classifier')
        if classes == 1:
            raise ValueError('Nonsense classifier')
        if classes == 2:
            reprs[0] = np.array([1, 0])
            reprs[1] = np.array([0, 1])
        if classes == 3:
            reprs[0] = np.array([1, 0])
            reprs[1] = np.array([1 / 2, np.sqrt(3) / 2])
            reprs[2] = np.array([1 / 2, -np.sqrt(3) / 2])
        if classes == 4:
            reprs[0] = np.array([1, 0])
            reprs[1] = np.array([1 / np.sqrt(3), np.sqrt(2 / 3)])
            reprs[2] = np.array([1 / np.sqrt(3), np.exp(1j * 2 * np.pi / 3) * np.sqrt(2 / 3)])
            reprs[3] = np.array([1 / np.sqrt(3), np.exp(-1j * 2 * np.pi / 3) * np.sqrt(2 / 3)])
        if classes == 6:
            reprs[0] = np.array([1, 0])
            reprs[1] = np.array([0, 1])
            reprs[2] = 1 / np.sqrt(2) * np.array([1, 1])
            reprs[3] = 1 / np.sqrt(2) * np.array([1, -1])
            reprs[4] = 1 / np.sqrt(2) * np.array([1, 1j])
            reprs[5] = 1 / np.sqrt(2) * np.array([1, -1j])

    if qubits_lab == 2:
        if classes == 0:
            raise ValueError('Nonsense classifier')
        if classes == 1:
            raise ValueError('Nonsense classifier')
        if classes == 2:
            reprs[0] = np.array([1, 0, 0, 0])
            reprs[1] = np.array([0, 0, 0, 1])
        if classes == 3:
            reprs[0] = np.array([1, 0, 0, 0])
            reprs[1] = np.array([0, 1, 0, 0])
            reprs[2] = np.array([0, 0, 1, 0])
        if classes == 4:
            reprs[0] = np.array([1, 0, 0, 0])
            reprs[1] = np.array([0, 1, 0, 0])
            reprs[2] = np.array([0, 0, 1, 0])
            reprs[3] = np.array([0, 0, 0, 1])

    return reprs

```

```python
###########################################################################
#Quantum classifier
#Adrián Pérez-Salinas, Alba Cervera-Lierta, Elies Gil, J. Ignacio Latorre
#Code by APS
#Code-checks by ACL
#June 3rd 2019


#Universitat de Barcelona / Barcelona Supercomputing Center/Institut de Ciències del Cosmos

###########################################################################

## This file creates the problems and their settings
import numpy as np

def problem_generator(problem, qubits, layers, chi, qubits_lab=1):
    """
    This function generates everything needed for solving the problem
    INPUT:
        -chi: cost function, to choose between 'fidelity_chi' or 'weighted_fidelity_chi'
        -problem: name of the problem, to choose among
            ['circle', '3 circles', 'hypersphere', 'tricrown', 'non convex', 'crown', 'sphere', 'squares', 'wavy lines']
        -qubits: number of qubits, must be an integer
        -layers: number of layers, must be an integer. If layers == 1, entanglement is not taken in account


    OUTPUT:
        -theta: set of parameters needed for the circuit. It is an array with shape (qubits, layers, 3)
        -alpha: set of parameters needed for the circuit. It is an array with shape (qubits, layers, dimension of data)
        -weight: set of parameters needed fot the circuit only if chi == 'weighted_fidelity_chi'. It is an array with shape (classes, qubits)
        -reprs: variable encoding the label states of the different classes
    """
    chi = chi.lower()
    if chi in ['fidelity', 'weighted_fidelity']: chi += '_chi'
    if chi not in ['fidelity_chi', 'weighted_fidelity_chi']:
        raise ValueError('Figure of merit is not valid')

    if chi == 'weighted_fidelity_chi' and qubits_lab != 1:
        qubits_lab = 1
        print('WARNING: number of qubits for the label states has been changed to 1')

    problem = problem.lower()
    if problem == 'circle':
        theta, alpha, reprs = _circle(qubits, layers, qubits_lab, chi)
    elif problem == '3 circles':
        theta, alpha, reprs = _3_circles(qubits, layers, qubits_lab, chi)
    elif problem == 'wavy lines':
        theta, alpha, reprs = _wavy_lines(qubits, layers, qubits_lab, chi)
    elif problem == 'squares':
        theta, alpha, reprs = _squares(qubits, layers, qubits_lab, chi)
    elif problem == 'sphere':
        theta, alpha, reprs = _sphere(qubits, layers, qubits_lab, chi)
    elif problem == 'non convex':
        theta, alpha, reprs = _non_convex(qubits, layers, qubits_lab, chi)
    elif problem == 'crown':
        theta, alpha, reprs = _crown(qubits, layers, qubits_lab, chi)
    elif problem == 'tricrown':
        theta, alpha, reprs = _tricrown(qubits, layers, qubits_lab, chi)
    elif problem == 'hypersphere':
        theta, alpha, reprs = _hypersphere(qubits, layers, qubits_lab, chi)

    else:
        raise ValueError('Problem is not valid')

    if chi == 'fidelity_chi':
        return theta, alpha, reprs
    elif chi == 'weighted_fidelity_chi':
        weights = np.ones((len(reprs), qubits))
        return theta, alpha, weights, reprs

#All these are auxiliary functions for problem_generator
def _circle(qubits, layers, qubits_lab, chi):
    classes = 2
    reprs = representatives(classes, qubits_lab)
    theta = np.random.rand(qubits, layers, 3)
    alpha = np.random.rand(qubits, layers, 2)
    return theta, alpha, reprs

def _3_circles(qubits, layers, qubits_lab, chi):
    classes = 4
    reprs = representatives(classes, qubits_lab)
    theta = np.random.rand(qubits, layers, 3)
    alpha = np.random.rand(qubits, layers, 2)
    return theta, alpha, reprs

def _wavy_lines(qubits, layers, qubits_lab, chi):
    classes = 4
    reprs = representatives(classes, qubits_lab)
    theta = np.random.rand(qubits, layers, 3)
    alpha = np.random.rand(qubits, layers, 2)
    return theta, alpha, reprs

def _squares(qubits, layers, qubits_lab, chi):
    classes = 4
    reprs = representatives(classes, qubits_lab)
    theta = np.random.rand(qubits, layers, 3)
    alpha = np.random.rand(qubits, layers, 2)
    return theta, alpha, reprs

def _non_convex(qubits, layers, qubits_lab, chi):
    classes = 2
    reprs = representatives(classes, qubits_lab)
    theta = np.random.rand(qubits, layers, 3)
    alpha = np.random.rand(qubits, layers, 2)
    return theta, alpha, reprs

def _crown(qubits, layers, qubits_lab, chi):
    classes = 2
    reprs = representatives(classes, qubits_lab)
    theta = np.random.rand(qubits, layers, 3)
    alpha = np.random.rand(qubits, layers, 2)
    return theta, alpha, reprs

def _tricrown(qubits, layers, qubits_lab, chi):
    classes = 3
    reprs = representatives(classes, qubits_lab)
    theta = np.random.rand(qubits, layers, 3)
    alpha = np.random.rand(qubits, layers, 2)
    return theta, alpha, reprs

def _sphere(qubits, layers, qubits_lab, chi):
    classes = 2
    reprs = representatives(classes, qubits_lab)
    theta = np.random.rand(qubits, layers, 3)
    alpha = np.random.rand(qubits, layers, 3)
    return theta, alpha, reprs

def _hypersphere(qubits, layers, qubits_lab, chi):
    classes = 2
    reprs = representatives(classes, qubits_lab)
    theta = np.random.rand(qubits, layers, 6)
    alpha = np.random.rand(qubits, layers, 4)
    return theta, alpha, reprs

def representatives(classes, qubits_lab):
    """
    This function creates the label states for the classification task
    INPUT:
        -classes: number of classes of our problem
        -qubits_lab: how many qubits will store the labels
    OUTPUT:
        -reprs: the label states
    """
    reprs = np.zeros((classes, 2**qubits_lab), dtype = 'complex')
    if qubits_lab == 1:
        if classes == 0:
            raise ValueError('Nonsense classifier')
        if classes == 1:
            raise ValueError('Nonsense classifier')
        if classes == 2:
            reprs[0] = np.array([1, 0])
            reprs[1] = np.array([0, 1])
        if classes == 3:
            reprs[0] = np.array([1, 0])
            reprs[1] = np.array([1 / 2, np.sqrt(3) / 2])
            reprs[2] = np.array([1 / 2, -np.sqrt(3) / 2])
        if classes == 4:
            reprs[0] = np.array([1, 0])
            reprs[1] = np.array([1 / np.sqrt(3), np.sqrt(2 / 3)])
            reprs[2] = np.array([1 / np.sqrt(3), np.exp(1j * 2 * np.pi / 3) * np.sqrt(2 / 3)])
            reprs[3] = np.array([1 / np.sqrt(3), np.exp(-1j * 2 * np.pi / 3) * np.sqrt(2 / 3)])
        if classes == 6:
            reprs[0] = np.array([1, 0])
            reprs[1] = np.array([0, 1])
            reprs[2] = 1 / np.sqrt(2) * np.array([1, 1])
            reprs[3] = 1 / np.sqrt(2) * np.array([1, -1])
            reprs[4] = 1 / np.sqrt(2) * np.array([1, 1j])
            reprs[5] = 1 / np.sqrt(2) * np.array([1, -1j])

    if qubits_lab == 2:
        if classes == 0:
            raise ValueError('Nonsense classifier')
        if classes == 1:
            raise ValueError('Nonsense classifier')
        if classes == 2:
            reprs[0] = np.array([1, 0, 0, 0])
            reprs[1] = np.array([0, 0, 0, 1])
        if classes == 3:
            reprs[0] = np.array([1, 0, 0, 0])
            reprs[1] = np.array([0, 1, 0, 0])
            reprs[2] = np.array([0, 0, 1, 0])
        if classes == 4:
            reprs[0] = np.array([1, 0, 0, 0])
            reprs[1] = np.array([0, 1, 0, 0])
            reprs[2] = np.array([0, 0, 1, 0])
            reprs[3] = np.array([0, 0, 0, 1])

    return reprs
```

Text Compare

```python
# coding=utf-8
###########################################################################
#Quantum classifier
#Sara Aminpour, Mike Banad, Sarah Sharif
#September 25th 2024

#School of Electrical and Computer Engineering/ Center for Quantum and Technology, University of Oklahoma, Norman, OK 73019 USA,
###########################################################################
#IMPORTANT_NOTE:
#The code on the left was developed by Sara Aminpour, while the code on the right serves as the reference implementation by Adrián Pérez-Salinas.
#The code on the left has been restructured to handle random data. So some certain sections has been deleted from the reference code.
#Additionally, our code on the left developed to analyze trace distance cost function and linear classification problem

#as well as necessary modification to apply COBYLA, L-BFGS-B, NELDER-MEAD, and SLSQP minimization methods.
###########################################################################


## This is an auxiliary file. It provides the tools needed for simulating quantum
# circuits.

import numpy as np
class QCircuit(object):
    def __init__(self,qubits):
        self.num_qubits = qubits
        self.psi = [0]*2**self.num_qubits
        self.psi[0] = 1
        self.E_x=0
        self.E_y=0
        self.E_z=0
        self.r=np.array([0,0,0])

    def Ry(self,i,theta):
        if i>=self.num_qubits: raise ValueError('There are not enough qubits')
        c = np.cos(theta/2)
        s = np.sin(theta/2)
        for k in range(2**(self.num_qubits-1)):
            S = k%(2**i) + 2*(k - k%(2**i))
            S_=S + 2**i
            a=c*self.psi[S] - s*self.psi[S_];
            b=s*self.psi[S] + c*self.psi[S_];
            self.psi[S]=a; self.psi[S_]=b;

    def Rx(self,i,theta):
        if i>=self.num_qubits: raise ValueError('There are not enough qubits')
        c = np.cos(theta/2)
        s = np.sin(theta/2)
        for k in range(2**(self.num_qubits-1)):
            S = k%(2**i) + 2*(k - k%(2**i))
            S_=S + 2**i
            a=c*self.psi[S] - 1j*s*self.psi[S_];
            b=-1j*s*self.psi[S] + c*self.psi[S_];
            self.psi[S]=a; self.psi[S_]=b;

    def U2(self,i,phi,lamb):
        if i >= self.num_qubits: raise ValueError('There are not enough qubits')
        f = np.exp(1j*phi)
        l = np.exp(-1j*lamb)
        for k in range(2**(self.num_qubits-1)):
            S = k%(2**i) + 2*(k - k%(2**i))
            S_=S + 2**i
            a=1/np.sqrt(2)*(self.psi[S] - l*self.psi[S_]);
            b=1/np.sqrt(2)*(f*self.psi[S] + f*l*self.psi[S_]);
            self.psi[S]=a; self.psi[S_]=b;

    def U3(self, i, theta3):
        if i >= self.num_qubits: raise ValueError('There are not enough qubits')
        c = np.cos(theta3[0] / 2)
        s = np.sin(theta3[0] / 2)
        e_phi = np.exp(1j * theta3[1] / 2)
        e_phi_s = np.conj(e_phi)
        e_lambda = np.exp(1j * theta3[2] / 2)
        e_lambda_s = np.conj(e_lambda)

        for k in range(2 ** (self.num_qubits - 1)):
            S = k % (2 ** i) + 2 * (k - k % (2 ** i))
            S_ = S + 2 ** i
            a = c * e_phi * e_lambda * self.psi[S] - s * e_phi * e_lambda_s * self.psi[S_];

            b = s * e_phi_s * e_lambda * self.psi[S] + c * e_phi_s * e_lambda_s * self.psi[S_];

            self.psi[S] = a;
            self.psi[S_] = b;

        theta_f=np.arccos(np.abs(self.psi[S])**2 - np.abs(self.psi[S_])**2) - np.pi/2
        phi_f=np.angle(self.psi[S_] / self.psi[S])
        self.r=np.array([np.sin(theta_f)*np.cos(phi_f),np.sin(phi_f)*np.sin(theta_f),np.cos(theta_f)])

    def Rz(self,i,theta):
        if i>=self.num_qubits: raise ValueError('There are not enough qubits')
        ex = np.exp(1j*theta)
        for k in range(2**(self.num_qubits-1)):
            S = k%(2**i) + 2*(k - k%(2**i)) + 2**i
            self.psi[S]=ex*self.psi[S];

    def Hx(self,i):
        if i>=self.num_qubits: raise ValueError('There are not enough qubits')
        for k in range(2**(self.num_qubits-1)):
            S = k%(2**i) + 2*(k - k%(2**i))
            S_=S + 2**i
            a=1/np.sqrt(2)*self.psi[S] + 1/np.sqrt(2)*self.psi[S_];
            b=1/np.sqrt(2)*self.psi[S] - 1/np.sqrt(2)*self.psi[S_];
            self.psi[S] = a
            self.psi[S_] = b

    def Hy(self,i):
        if i>=self.num_qubits: raise ValueError('There are not enough qubits')
        for k in range(2**(self.num_qubits-1)):
            S = k%(2**i) + 2*(k - k%(2**i))
            S_=S + 2**i
            a =1/np.sqrt(2)*self.psi[S] -1j/np.sqrt(2)*self.psi[S_];
            b =-1j/np.sqrt(2)*self.psi[S] + 1/np.sqrt(2)*self.psi[S_];
            self.psi[S] = a
            self.psi[S_] = b

    def HyT(self,i):
        if i>=self.num_qubits: raise ValueError('There are not enough qubits')
        for k in range(2**(self.num_qubits-1)):
            S = k%(2**i) + 2*(k - k%(2**i))
            S_=S + 2**i
            a=1/np.sqrt(2)*self.psi[S] +1j/np.sqrt(2)*self.psi[S_];
            b=1j/np.sqrt(2)*self.psi[S] + 1/np.sqrt(2)*self.psi[S_];
            self.psi[S]=a; self.psi[S_]=b;

    def Cz(self,i,j):
        if i>=self.num_qubits: raise ValueError('There are not enough qubits')
        if j>=self.num_qubits: raise ValueError('There are not enough qubits')
        if i==j: raise ValueError('Control and target qubits are the same')
        if j<i: a=i; i=j; j=a;
        for k in range(2**(self.num_qubits-2)):
            S = k%2**i + (
                ( k - k%2**i)*2)%2**j + 2*(
                    (k-k%2**i)*2-((2*(k-k%2**i))%2**j)) + 2**i + 2**j;
            self.psi[S]=-self.psi[S]

    def SWAP(self,i,j):
        if i>=self.num_qubits: raise ValueError('There are not enough qubits')
        if j>=self.num_qubits: raise ValueError('There are not enough qubits')
        if i==j: raise ValueError('Control and target qubits are the same')
        for k in range(2**(self.num_qubits-2)):
            S = k%2**i + (
                ( k - k%2**i)*2)%2**j + 2*(
                    (k-k%2**i)*2-((2*(k-k%2**i))%2**j)) + 2**j;
            S_ = S + 2**i - 2**j
            a=self.psi[S_]
            self.psi[S_] = self.psi[S]
            self.psi[S] = a


    def Cx(self,i,j):
        #i = control
        #j = target
        if i>=self.num_qubits: raise ValueError('There are not enough qubits')
        if j>=self.num_qubits: raise ValueError('There are not enough qubits')
        if i==j: raise ValueError('Control and target qubits are the same')
        for k in range(2**(self.num_qubits-2)):
            S = k%2**i + (
                ( k - k%2**i)*2)%2**j + 2*(
                    (k-k%2**i)*2-((2*(k-k%2**i))%2**j)) + 2**i;
            S_ = S + 2**j
            '''
            a=self.psi[S_]
            self.psi[S_] = self.psi[S]
            self.psi[S] = a
            '''
            self.psi[S],self.psi[S_] = self.psi[S_],self.psi[S]
    def Cy(self,i,j):
        if i>=self.num_qubits: raise ValueError('There are not enough qubits')
        if j>=self.num_qubits: raise ValueError('There are not enough qubits')
        if i==j: raise ValueError('Control and target qubits are the same')
        for k in range(2**(self.num_qubits-2)):
            S = k%2**i + (
                ( k - k%2**i)*2)%2**j + 2*(
                    (k-k%2**i)*2-((2*(k-k%2**i))%2**j)) + 2**i;
            S_ = S + 2**j
            self.psi[S],self.psi[S_] = 1j*self.psi[S_],-1j*self.psi[S]

    def MeasureZ(self):
        self.E_z = 0;
        for h in range(2 ** self.num_qubits):
            s = np.binary_repr(h, width=self.num_qubits)
            self.E_z += np.abs(self.psi[h])**2*(s.count('1')-s.count('0'))

    def MeasureX(self):
        self.E_x = 0;
        for i in range(self.num_qubits):
            self.Hx(i);
        for h in range(2 ** self.num_qubits):
            s = np.binary_repr(h, width=self.num_qubits)
            self.E_x += np.abs(self.psi[h])**2*(s.count('1')-s.count('0'))
        for i in range(self.num_qubits):
            self.Hx(i);

    def MeasureY(self):
        self.E_y = 0;
        for i in range(self.num_qubits):
            self.Hy(i);
        for h in range(2 ** self.num_qubits):
            s = np.binary_repr(h, width=self.num_qubits)
            self.E_y += np.abs(self.psi[h])**2*(s.count('1')-s.count('0'))
        for i in range(self.num_qubits):
            self.HyT(i);

    def reduced_density_matrix(self, q):
        rho = np.zeros((2,2), dtype='complex')
        for i in range(2):
            for j in range(i + 1):
                for k in range(2**(self.num_qubits-1)):
                    S = k%(2**q) + 2*(k - k%(2**q))
                    rho[i,j] += self.psi[S + i*2**q] * np.conj(self.psi[S + j*2**q])
                rho[j,i] = np.conj(rho[i,j])
        return rho


```

```python
###########################################################################
#Quantum classifier
#Adrián Pérez-Salinas, Alba Cervera-Lierta, Elies Gil, J. Ignacio Latorre
#Code by APS
#Code-checks by ACL
#June 3rd 2019


#Universitat de Barcelona / Barcelona Supercomputing Center/Institut de Ciències del Cosmos

###########################################################################


## This is an auxiliary file. It provides the tools needed for simulating quantum
# circuits.

import numpy as np
class QCircuit(object):
    def __init__(self,qubits):
        self.num_qubits = qubits
        self.psi = [0]*2**self.num_qubits
        self.psi[0] = 1
        self.E_x=0
        self.E_y=0
        self.E_z=0

    def Ry(self,i,theta):
        if i>=self.num_qubits: raise ValueError('There are not enough qubits')
        c = np.cos(theta/2)
        s = np.sin(theta/2)
        for k in range(2**(self.num_qubits-1)):
            S = k%(2**i) + 2*(k - k%(2**i))
            S_=S + 2**i
            a=c*self.psi[S] - s*self.psi[S_];
            b=s*self.psi[S] + c*self.psi[S_];
            self.psi[S]=a; self.psi[S_]=b;

    def Rx(self,i,theta):
        if i>=self.num_qubits: raise ValueError('There are not enough qubits')
        c = np.cos(theta/2)
        s = np.sin(theta/2)
        for k in range(2**(self.num_qubits-1)):
            S = k%(2**i) + 2*(k - k%(2**i))
            S_=S + 2**i
            a=c*self.psi[S] - 1j*s*self.psi[S_];
            b=-1j*s*self.psi[S] + c*self.psi[S_];
            self.psi[S]=a; self.psi[S_]=b;

    def U2(self,i,phi,lamb):
        if i >= self.num_qubits: raise ValueError('There are not enough qubits')
        f = np.exp(1j*phi)
        l = np.exp(-1j*lamb)
        for k in range(2**(self.num_qubits-1)):
            S = k%(2**i) + 2*(k - k%(2**i))
            S_=S + 2**i
            a=1/np.sqrt(2)*(self.psi[S] - l*self.psi[S_]);
            b=1/np.sqrt(2)*(f*self.psi[S] + f*l*self.psi[S_]);
            self.psi[S]=a; self.psi[S_]=b;

    def U3(self, i, theta3):
        if i >= self.num_qubits: raise ValueError('There are not enough qubits')
        c = np.cos(theta3[0] / 2)
        s = np.sin(theta3[0] / 2)
        e_phi = np.exp(1j * theta3[1] / 2)
        e_phi_s = np.conj(e_phi)
        e_lambda = np.exp(1j * theta3[2] / 2)
        e_lambda_s = np.conj(e_lambda)
        for k in range(2 ** (self.num_qubits - 1)):
            S = k % (2 ** i) + 2 * (k - k % (2 ** i))
            S_ = S + 2 ** i
            a = c * e_phi * e_lambda * self.psi[S] - s * e_phi * e_lambda_s * self.psi[S_];
            b = s * e_phi_s * e_lambda * self.psi[S] + c * e_phi_s * e_lambda_s * self.psi[S_];
            self.psi[S] = a;
            self.psi[S_] = b;

    def Rz(self,i,theta):
        if i>=self.num_qubits: raise ValueError('There are not enough qubits')
        ex = np.exp(1j*theta)
        for k in range(2**(self.num_qubits-1)):
            S = k%(2**i) + 2*(k - k%(2**i)) + 2**i
            self.psi[S]=ex*self.psi[S];

    def Hx(self,i):
        if i>=self.num_qubits: raise ValueError('There are not enough qubits')
        for k in range(2**(self.num_qubits-1)):
            S = k%(2**i) + 2*(k - k%(2**i))
            S_=S + 2**i
            a=1/np.sqrt(2)*self.psi[S] + 1/np.sqrt(2)*self.psi[S_];
            b=1/np.sqrt(2)*self.psi[S] - 1/np.sqrt(2)*self.psi[S_];
            self.psi[S] = a
            self.psi[S_] = b

    def Hy(self,i):
        if i>=self.num_qubits: raise ValueError('There are not enough qubits')
        for k in range(2**(self.num_qubits-1)):
            S = k%(2**i) + 2*(k - k%(2**i))
            S_=S + 2**i
            a =1/np.sqrt(2)*self.psi[S] -1j/np.sqrt(2)*self.psi[S_];
            b =-1j/np.sqrt(2)*self.psi[S] + 1/np.sqrt(2)*self.psi[S_];
            self.psi[S] = a
            self.psi[S_] = b

    def HyT(self,i):
        if i>=self.num_qubits: raise ValueError('There are not enough qubits')
        for k in range(2**(self.num_qubits-1)):
            S = k%(2**i) + 2*(k - k%(2**i))
            S_=S + 2**i
            a=1/np.sqrt(2)*self.psi[S] +1j/np.sqrt(2)*self.psi[S_];
            b=1j/np.sqrt(2)*self.psi[S] + 1/np.sqrt(2)*self.psi[S_];
            self.psi[S]=a; self.psi[S_]=b;

    def Cz(self,i,j):
        if i>=self.num_qubits: raise ValueError('There are not enough qubits')
        if j>=self.num_qubits: raise ValueError('There are not enough qubits')
        if i==j: raise ValueError('Control and target qubits are the same')
        if j<i: a=i; i=j; j=a;
        for k in range(2**(self.num_qubits-2)):
            S = k%2**i + (
                ( k - k%2**i)*2)%2**j + 2*(
                    (k-k%2**i)*2-((2*(k-k%2**i))%2**j)) + 2**i + 2**j;
            self.psi[S]=-self.psi[S]

    def SWAP(self,i,j):
        if i>=self.num_qubits: raise ValueError('There are not enough qubits')
        if j>=self.num_qubits: raise ValueError('There are not enough qubits')
        if i==j: raise ValueError('Control and target qubits are the same')
        for k in range(2**(self.num_qubits-2)):
            S = k%2**i + (
                ( k - k%2**i)*2)%2**j + 2*(
                    (k-k%2**i)*2-((2*(k-k%2**i))%2**j)) + 2**j;
            S_ = S + 2**i - 2**j
            a=self.psi[S_]
            self.psi[S_] = self.psi[S]
            self.psi[S] = a


    def Cx(self,i,j):
        #i = control
        #j = target
        if i>=self.num_qubits: raise ValueError('There are not enough qubits')
        if j>=self.num_qubits: raise ValueError('There are not enough qubits')
        if i==j: raise ValueError('Control and target qubits are the same')
        for k in range(2**(self.num_qubits-2)):
            S = k%2**i + (
                ( k - k%2**i)*2)%2**j + 2*(
                    (k-k%2**i)*2-((2*(k-k%2**i))%2**j)) + 2**i;
            S_ = S + 2**j
            '''
            a=self.psi[S_]
            self.psi[S_] = self.psi[S]
            self.psi[S] = a
            '''
            self.psi[S],self.psi[S_] = self.psi[S_],self.psi[S]
    def Cy(self,i,j):
        if i>=self.num_qubits: raise ValueError('There are not enough qubits')
        if j>=self.num_qubits: raise ValueError('There are not enough qubits')
        if i==j: raise ValueError('Control and target qubits are the same')
        for k in range(2**(self.num_qubits-2)):
            S = k%2**i + (
                ( k - k%2**i)*2)%2**j + 2*(
                    (k-k%2**i)*2-((2*(k-k%2**i))%2**j)) + 2**i;
            S_ = S + 2**j
            self.psi[S],self.psi[S_] = 1j*self.psi[S_],-1j*self.psi[S]

    def MeasureZ(self):
        self.E_z = 0;
        for h in range(2 ** self.num_qubits):
            s = np.binary_repr(h, width=self.num_qubits)
            self.E_z += np.abs(self.psi[h])**2*(s.count('1')-s.count('0'))

    def MeasureX(self):
        self.E_x = 0;
        for i in range(self.num_qubits):
            self.Hx(i);
        for h in range(2 ** self.num_qubits):
            s = np.binary_repr(h, width=self.num_qubits)
            self.E_x += np.abs(self.psi[h])**2*(s.count('1')-s.count('0'))
        for i in range(self.num_qubits):
            self.Hx(i);

    def MeasureY(self):
        self.E_y = 0;
        for i in range(self.num_qubits):
            self.Hy(i);
        for h in range(2 ** self.num_qubits):
            s = np.binary_repr(h, width=self.num_qubits)
            self.E_y += np.abs(self.psi[h])**2*(s.count('1')-s.count('0'))
        for i in range(self.num_qubits):
            self.HyT(i);

    def reduced_density_matrix(self, q):
        rho = np.zeros((2,2), dtype='complex')
        for i in range(2):
            for j in range(i + 1):
                for k in range(2**(self.num_qubits-1)):
                    S = k%(2**q) + 2*(k - k%(2**q))
                    rho[i,j] += self.psi[S + i*2**q] * np.conj(self.psi[S + j*2**q])
                rho[j,i] = np.conj(rho[i,j])
        return rho


```

Text Compare

```
# coding=utf-8
##########################################################################
#Quantum classifier
#Sara Aminpour, Mike Banad, Sarah Sharif
#September 25th 2024

#School of Electrical and Computer Engineering/ Center for Quantum and Technology, University of Oklahoma, Norman, OK 73019 USA,
##########################################################################
#IMPORTANT_NOTE:
#The code on the left was developed by Sara Aminpour, while the code on the right serves as the reference implementation by Adrián Pérez-Salinas.
#The code on the left has been restructured to handle random data. So some certain sections has been deleted from the reference code.
#Additionally, our code on the left developed to analyze trace distance cost function and linear classification problem
#as well as necessary modification to apply COBYLA, L-BFGS-B, NELDER-MEAD, and SLSQP minimization methods.
##########################################################################


#This file provides useful tools for painting and saving data according to the problem,
# the minimization style, the number of qubits and layers.

import os
import numpy as np
import matplotlib.pyplot as plt
from matplotlib.cm import get_cmap
from matplotlib.colors import Normalize

def write_summary(chi, problem, qubits, entanglement, layers, method, name,
            theta, alpha, weights, chi_value, acc_train, acc_test, epochs):
    """
    This function takes some informations of a given problem and saves some text files
    with this information and the parameters which are solution of the problem
    INPUT:
        -chi: cost function, to choose between 'fidelity_chi' or 'weighted_fidelity_chi'
        -problem: name of the problem, to choose between
            ['circle', '3 circles', 'hypersphere', 'tricrown', 'non convex', 'crown', 'sphere', 'squares', 'wavy lines']
        -qubits: number of qubits, must be an integer
        -entanglement: whether there is entanglement or not in the Ansätze, just 'y'/'n'
        -layers: number of layers, must be an integer. If layers == 1, entanglement is not taken in account
        -method: minimization method, to choose among ['SGD', another valid for function scipy.optimize.minimize]

        -name: a name we want for our our files to be save with
        -theta: set of parameters needed for the circuit. Must be an array with shape (qubits, layers, 3)
        -alpha: set of parameters needed for the circuit. Must be an array with shape (qubits, layers, dimension of data)
        -weight: set of parameters needed fot the circuit only if chi == 'weighted_fidelity_chi'. Must be an array with shape (classes, qubits)
        -chi_value: Value of the cost function after minimization
        -acc_train: accuracy for the training set
        -acc_test: accuracy for the test set
        -seed: seed of numpy.random, needed for replicating results
        -epochs: number of epochs for a 'SGD' method. If there is another method, this input has got no importance

    OUTPUT:
        This function has got no outputs, but several files are saved in an appropiate folder. The files are
        -summary.txt: Saves useful information for the problem
        -theta.txt: saves the theta parameters as a flat array
        -alpha.txt: saves the alpha parameters as a flat array
        -weight.txt: saves the weights as a flat array if they exist
    """
    foldname = name_folder(chi, problem, qubits, entanglement, layers, method)
    create_folder(foldname)
    file_text = open(foldname + '/' + name + '_summary.txt','w')
    file_text.write('\nFigur of merit = '+chi)
    file_text.write('\nProblem = ' + problem)
    file_text.write('\nNumber of qubits = ' + str(qubits))


    if qubits != 1:
        file_text.write('\nEntanglement = ' + entanglement)
    file_text.write('\nNumber of layers = ' + str(layers))
    file_text.write('\nMinimization method = '+ method)

    if method == 'SGD':
        file_text.write('\nNumber of epochs = '+ str(epochs))
    file_text.write('\n\nBEST RESULT\n\n')
    file_text.write('\nTHETA = \n')
    file_text.write(str(theta))
    file_text.write('\nALPHA = \n')
    file_text.write(str(alpha))



    #==========================================================================

    #==========================================================================



    if chi == 'weighted_fidelity_chi':
        file_text.write('\nWEIGHTS = \n')
        file_text.write(str(weights))
    if chi == 'weighted_trace_chi':
        file_text.write('\nWEIGHTS = \n')
        file_text.write(str(weights))
    file_text.write('\nchi**2 = ' + str(chi_value))
    file_text.write('\nacc_train = ' + str(acc_train))
    file_text.write('\nacc_test = ' + str(acc_test))
    file_text.close()

    np.savetxt(foldname + '/' + name + '_theta.txt', theta.flatten())
    np.savetxt(foldname + '/' + name + '_alpha.txt', alpha.flatten())


    if chi == 'weighted_fidelity_chi':
        np.savetxt(foldname + '/' + name + '_weight.txt', weights.flatten())

    if chi == 'weighted_trace_chi':
        np.savetxt(foldname + '/' + name + '_weight.txt', weights.flatten())



def write_summary_acc(chi, problem,layers, method, name, acc_test):

    foldname_acc = name_folder_acc(chi, layers)
    create_folder(foldname_acc)
    file_text_acc = open(foldname_acc + '/' + name + '_accuracy.txt','a+')
    file_text_acc.write('\nNEW')
    file_text_acc.write('\nProblem = ' + problem + '\n')

    file_text_acc.write('\nacc_test = ' + str(acc_test))


def read_summary(chi, problem, qubits, entanglement, layers, method, name):

    """
    This function reads the files saved by write_summary and returns theta, alpha and weight parameters
    INPUT:
        -chi: cost function, to choose between 'fidelity_chi' or 'weighted_fidelity_chi'
        -problem: name of the problem, to choose among
            ['circle', '3 circles', 'hypersphere', 'tricrown', 'non convex', 'crown', 'sphere', 'squares', 'wavy lines'
        -qubits: number of qubits, must be an integer
        -entanglement: whether there is entanglement or not in the Ansätze, just 'y'/'n'
        -layers: number of layers, must be an integer. If layers == 1, entanglement is not taken in account
        -method: minimization method, to choose among ['SGD', another valid for function scipy.optimize.minimize]

        -name: a name we want for our our files to be save with

    OUTPUT:
        -theta: set of parameters needed for the circuit. It is an array with shape (qubits, layers, 3)
        -alpha: set of parameters needed for the circuit. It is an array with shape (qubits, layers, dimension of data)
        -weight: set of parameters needed fot the circuit only if chi == 'weighted_fidelity_chi'. It is an array with shape (classes, qubits)
    """
    chi = chi.lower().replace(' ','_')
    if chi in ['fidelity', 'weighted_fidelity', 'trace', 'weighted_trace']: chi += '_chi'
    if chi not in ['fidelity_chi', 'weighted_fidelity_chi','trace_chi', 'weighted_trace_chi']:
        raise ValueError('Figure of merit is not valid')
    if chi == 'fidelity_chi':
        foldname = name_folder(chi, problem, qubits, entanglement, layers, method)
        if problem in ['circle', 'line', '2 lines', '6squares', '3 circles', 'wavy circles', 'wavy lines', 'non convex','crown','tricrown','squares']:
            theta = np.loadtxt(foldname + '/' + name + '_theta.txt').reshape((qubits, layers, 3))
            dim = 2
        elif problem == 'sphere':
            theta = np.loadtxt(foldname + '/' + name + '_theta.txt').reshape((qubits, layers, 3))
            dim = 3
        elif problem in ['hypersphere']:
            theta = np.loadtxt(foldname + '/' + name + '_theta.txt').reshape((qubits, layers, 6))
            dim = 4

        alpha = np.loadtxt(foldname + '/' + name + '_alpha.txt').reshape((qubits, layers, dim))
        return theta, alpha
#==========================================================================
#Sara
    if chi == 'trace_chi':
        foldname = name_folder(chi, problem, qubits, entanglement, layers, method)
        if problem in ['circle', 'line', '2 lines', '6squares', '3 circles', 'wavy circles', 'wavy lines', 'non convex','crown','tricrown','squares']:
            theta = np.loadtxt(foldname + '/' + name + '_theta.txt').reshape((qubits, layers, 3))
            dim = 2
        elif problem == 'sphere':
            theta = np.loadtxt(foldname + '/' + name + '_theta.txt').reshape((qubits, layers, 3))
            dim = 3
        elif problem in ['hypersphere']:
            theta = np.loadtxt(foldname + '/' + name + '_theta.txt').reshape((qubits, layers, 6))
            dim = 4

        alpha = np.loadtxt(foldname + '/' + name + '_alpha.txt').reshape((qubits, layers, dim))
        return theta, alpha
#Sara
#==========================================================================
    if chi == 'weighted_fidelity_chi':
        foldname = name_folder(chi, problem, qubits, entanglement, layers, method)
        if problem in ['circle', 'line', '2 lines', '6squares', '3 circles', 'wavy circles', 'wavy lines', 'non convex','crown','tricrown','squares']:
            theta = np.loadtxt(foldname + '/' + name + '_theta.txt').reshape((qubits, layers, 3))
            dim = 2
        elif problem == 'sphere':
            theta = np.loadtxt(foldname + '/' + name + '_theta.txt').reshape((qubits, layers, 3))
            dim = 3
        elif problem in ['hypersphere']:
            theta = np.loadtxt(foldname + '/' + name + '_theta.txt').reshape((qubits, layers, 6))
            dim = 4

        alpha = np.loadtxt(foldname + '/' + name + '_alpha.txt').reshape((qubits, layers, dim))

        if problem in ['3 circles','wavy lines','squares']:
            weight = np.loadtxt(foldname + '/' + name + '_weight.txt').reshape((4, qubits))
        if problem in ['circle', 'line', '2 lines', 'wavy circle','sphere', 'non convex', 'crown', 'hypersphere']:
            weight = np.loadtxt(foldname + '/' + name + '_weight.txt').reshape((2, qubits))
        if problem in ['tricrown']:
            weight = np.loadtxt(foldname + '/' + name + '_weight.txt').reshape((3, qubits))
        if problem in ['6squares']:
            weight = np.loadtxt(foldname + '/' + name + '_weight.txt').reshape((6, qubits))
        return theta, alpha, weight

#==========================================================================
#Sara
    if chi == 'weighted_trace_chi':
        foldname = name_folder(chi, problem, qubits, entanglement, layers, method)
        if problem in ['circle', 'line', '2 lines', '6squares', '3 circles', 'wavy circles', 'wavy lines', 'non convex','crown','tricrown','squares']:
            theta = np.loadtxt(foldname + '/' + name + '_theta.txt').reshape((qubits, layers, 3))
            dim = 2
        elif problem == 'sphere':
            theta = np.loadtxt(foldname + '/' + name + '_theta.txt').reshape((qubits, layers, 3))
            dim = 3
        elif problem in ['hypersphere']:
            theta = np.loadtxt(foldname + '/' + name + '_theta.txt').reshape((qubits, layers, 6))
            dim = 4

        alpha = np.loadtxt(foldname + '/' + name + '_alpha.txt').reshape((qubits, layers, dim))

        if problem in ['3 circles','wavy lines','squares']:
            weight = np.loadtxt(foldname + '/' + name + '_weight.txt').reshape((4, qubits))
        if problem in ['circle', 'line', '2 lines', 'wavy circle','sphere', 'non convex', 'crown', 'hypersphere']:
            weight = np.loadtxt(foldname + '/' + name + '_weight.txt').reshape((2, qubits))
        if problem in ['tricrown']:
            weight = np.loadtxt(foldname + '/' + name + '_weight.txt').reshape((3, qubits))
        if problem in ['6squares']:
            weight = np.loadtxt(foldname + '/' + name + '_weight.txt').reshape((6, qubits))
        return theta, alpha, weight


def write_epochs_file_acc(chi, layers, name):

    """
    This function creates a text file for saving data only in the SGD_step_by_step function
    INPUT:
        -chi: cost function, to choose between 'fidelity_chi' or 'weighted_fidelity_chi'
        -problem: name of the problem, to choose among
            ['circle', '3 circles', 'hypersphere', 'tricrown', 'non convex', 'crown', 'sphere', 'squares', 'wavy lines']
        -qubits: number of qubits, must be an integer
        -entanglement: whether there is entanglement or not in the Ansätze, just 'y'/'n'
        -layers: number of layers, must be an integer. If layers == 1, entanglement is not taken in account
        -method: minimization method, to choose among ['SGD', another valid for function scipy.optimize.minimize]
        -name: a name we want for our our files to be save with
    OUTPUT:
        -file_text: an object which is an open textfile ready to be used
    """
    foldname_acc = name_folder_acc(chi, layers)
    create_folder(foldname_acc)
    filename_acc = foldname_acc + '/' + name + '_epochs.txt'
    file_text_acc = open(filename_acc,'a+')
    return file_text_acc
#Sara
#==========================================================================
def write_epochs_file(chi, problem, qubits, entanglement, layers, method, name):

    """
    This function creates a text file for saving data only in the SGD_step_by_step function
    INPUT:
        -chi: cost function, to choose between 'fidelity_chi' or 'weighted_fidelity_chi'
        -problem: name of the problem, to choose among
            ['circle', '3 circles', 'hypersphere', 'tricrown', 'non convex', 'crown', 'sphere', 'squares', 'wavy lines']
        -qubits: number of qubits, must be an integer
        -entanglement: whether there is entanglement or not in the Ansätze, just 'y'/'n'
        -layers: number of layers, must be an integer. If layers == 1, entanglement is not taken in account
        -method: minimization method, to choose among ['SGD', another valid for function scipy.optimize.minimize]

        -name: a name we want for our our files to be save with
    OUTPUT:
        -file_text: an object which is an open textfile ready to be used
    """
    foldname = name_folder(chi, problem, qubits, entanglement, layers, method)
    create_folder(foldname)
    filename = foldname + '/' + name + '_epochs.txt'
    file_text = open(filename,'w')
    return file_text

def write_epoch(problem, file_text, epoch, theta, alpha, chi_value, acc_train, acc_test):
    """
    This function takes a text file and write information on it
    INPUT:
        -file_text: an object which is an open textfile ready to be used, output of write_epochs_file
        -epoch: the number of epoch providing this information
        -theta: set of parameters needed for the circuit. Must be an array with shape (qubits, layers, 3)
        -alpha: set of parameters needed for the circuit. Must be an array with shape (qubits, layers, dimension of data)
        -weight: set of parameters needed fot the circuit only if chi == 'weighted_fidelity_chi'. Must be an array with shape (classes, qubits)
        -chi_value: Value of the cost function after minimization
        -acc_train: accuracy for the training set
        -acc_test: accuracy for the test set
    OUTPUT:
        -file_text: with more information on it
    """

    file_text.write('\n Epoch = ' + str(epoch))
    file_text.write('\nTHETA = \n')
    file_text.write(str(theta))
    file_text.write('\nALPHA = \n')
    file_text.write(str(alpha))
    file_text.write('\n chi**2 = \n')
    file_text.write(str(chi_value))
    file_text.write('\nacc_train = \n')
    file_text.write(str(acc_train))
    file_text.write('\nacc_test = \n')
    file_text.write(str(acc_test))


def close_epochs_file(file_text, best_epoch):
    """
    This function takes a text file and closes it
    INPUT:
        -file_text: an object which is an open textfile ready to be used, output of write_epochs_file after write_epoch
        -best_epoch: the epoch with the best possible results
    OUTPUT:
        -file_text: closed
    """
    file_text.write('\n\n\nBest epoch = ' + str(best_epoch))
    file_text.close()

def write_epochs_error_rate(chi, problem, qubits, entanglement, layers, method, name,
                    accs_train, accs_test):
    """
    This function takes information from the SGD_step_by_step function and saves the accuracies for training and test sets. It is required for studying the overlearning
    INPUT:
        -chi: cost function, to choose between 'fidelity_chi' or 'weighted_fidelity_chi'
        -problem: name of the problem, to choose among
            ['circle', '3 circles', 'hypersphere', 'tricrown', 'non convex', 'crown', 'sphere', 'squares', 'wavy lines']
        -qubits: number of qubits, must be an integer
        -entanglement: whether there is entanglement or not in the Ansätze, just 'y'/'n'
        -layers: number of layers, must be an integer. If layers == 1, entanglement is not taken in account
        -method: minimization method, to choose among ['SGD', another valid for function scipy.optimize.minimize]

        -name: a name we want for our our files to be save with
        -accs_train: list or array with the accuracies of the training set for all epochs
        -accs_test: list or array with the accuracies of the test set for all epochs
    OUTPUT:
        Two files with the error rates in them
    """
    foldname = name_folder(chi, problem, qubits, entanglement, layers, method)
    create_folder(foldname)
    filename_train = foldname + '/' + name + '_train.txt'
    filename_test = foldname + '/' + name + '_test.txt'

    np.savetxt(filename_train, 1 - np.array(accs_train))
    np.savetxt(filename_test, 1 - np.array(accs_test))

def samples_paint(problem, settings, sol, foldname, filename, bw):
    """
    This function takes the problem and the points when they are already classified, and saves a picture of them

    INPUT:
        -problem: name of the problem, to choose among
            ['circle', '3 circles', 'hypersphere', 'tricrown', 'non convex', 'crown', 'sphere', 'squares', 'wavy lines']
        -settings: parameters the function needs for drawing. Provided by problem_gen.problem_gen
        -sol: solutions of the points alreafy classified
        -foldname : name of the folder where we store results
        -filename: name of the files we will produce
        -bw: black and white, True/False
    OUTPUT:
        a file with the points and their classes, and whether they are right or wrong
    """
    if bw == False:
        colors_classes = get_cmap('Dark2')
        norm_class = Normalize(vmin=-.5,vmax=np.max(sol[:,-3]) + .5)

        colors_rightwrong = get_cmap('RdYlGn')
        norm_rightwrong = Normalize(vmin=-.1,vmax=1.1)

    if bw == True:
        colors_classes = get_cmap('Greys')
        norm_class = Normalize(vmin=-.1,vmax=np.max(sol[:,-3]) + .1)

        colors_rightwrong = get_cmap('Greys')
        norm_rightwrong = Normalize(vmin=-.1,vmax=1.1)

    fig, axs = plt.subplots(ncols = 2, figsize=(10,5))
    ax = axs[0]
    if problem in ['circle', '3 circles', 'crown', 'tricrown']:
        centers, radii = settings
        for c, r in zip(centers, radii):
            ca = plt.Circle(c, r, color='k', fill=False, linewidth=2)
            ax.add_artist(ca)
    elif problem == 'wavy circle':
        centers, radii, wave, freq = settings
        phi = np.linspace(0, 2*np.pi, 1000)
        for (c,r, w, f) in zip(centers, radii, wave, freq):
            ax.plot(c[0] + r*(1 + w * np.cos(f * phi)) * np.cos(phi),
                    c[1] + r*(1 + w * np.cos(f * phi)) * np.sin(phi),
                    'k-')
    elif problem == 'wavy lines':
        freq = settings
        s = np.linspace(-1,1,100)
        ax.plot(s, np.clip(s + np.sin(freq * np.pi * s), -1, 1), 'k-')
        ax.plot(s, -s + np.sin(freq * np.pi * s), 'k-')
    elif problem == 'squares':
        freq = settings
        s = np.linspace(-1,1,10)
        ax.plot(s, np.zeros(10), 'k-')
        ax.plot(np.zeros(10), s, 'k-')
#==========================================================================
    elif problem == 'line':
        freq = settings
        #s = np.linspace(-1,1,10)
        s=np.linspace(-1,1,10)
        #ax.plot(s, np.zeros(10), 'k-')
        ax.plot(s, s, 'k-')
#==========================================================================
    elif problem == '2 lines':
        freq = settings
        s = np.linspace(-1,1,10)
        #ax.plot(s, np.zeros(10), 'k-')
        ax.plot(s, -s, 'k-')
        ax.plot(s, s, 'k-')
#==========================================================================
    elif problem == '6squares':
        freq = settings
        s = np.linspace(-1,1,10)
        ax.plot(s, np.zeros(10), 'k-')
        a=np.array([-0.33,-0.33,-0.33,-0.33,-0.33,-0.33,-0.33,-0.33,-0.33,-0.33])
        b=np.array([0.33,0.33,0.33,0.33,0.33,0.33,0.33,0.33,0.33,0.33])
        ax.plot(a, s, 'k-')
        ax.plot(b, s, 'k-')
#==========================================================================
    elif problem == 'non convex':
        freq, x_val, sin_val = settings
        s = np.linspace(-1,1,100)
        ax.plot(s, np.clip(-x_val * s + sin_val * np.sin(freq * np.pi * s), -1, 1), 'k-')

    ax.scatter(sol[:,0], sol[:,1], c=sol[:,-2], cmap = colors_classes, s=2, norm=norm_class)

    ax.set_xlabel('x', fontsize=16)
    ax.set_ylabel('y', fontsize=16)
    ax.tick_params(axis='both',labelsize=16)
    ax.set_xlim(-1, 1)
    ax.set_ylim(-1, 1)
    ax.margins(0)
    ax.axis('equal')

    bx = axs[1]
    bx.scatter(sol[:,0], sol[:,1], c=sol[:,-1], cmap = colors_rightwrong, s=2, norm=norm_rightwrong)
    if problem in ['circle', '3 circles', 'crown', 'tricrown']:
        centers, radii = settings
        for c, r in zip(centers, radii):
            ca = plt.Circle(c, r, color='k', fill=False, linewidth=2)
            bx.add_artist(ca)
    elif problem == 'wavy circle':
        centers, radii, wave, freq = settings
        phi = np.linspace(0, 2*np.pi, 1000)
        bx.plot(c[0] + r*(1 + wave * np.cos(freq * phi)) * np.cos(phi),
                c[1] + r*(1 + wave * np.cos(freq * phi)) * np.sin(phi),
                'k-')
    elif problem == 'wavy lines':
        freq = settings
        s = np.linspace(-1,1,100)
        bx.plot(s, np.clip(s + np.sin(freq * np.pi * s), -1, 1), 'k-')
        bx.plot(s, -s + np.sin(freq * np.pi * s), 'k-')

    elif problem == 'squares':
        freq = settings
        s = np.linspace(-1,1,10)
        bx.plot(s, np.zeros(10), 'k-')
        bx.plot(np.zeros(10), s, 'k-')
#==========================================================================
    elif problem == 'line':
        freq = settings
        #s = np.linspace(-1,1,10)
        s=np.linspace(-1,1,10)
        #ax.plot(s, np.zeros(10), 'k-')
        bx.plot(s, s, 'k-')
#==========================================================================
    elif problem == '2 lines':
        freq = settings
        s = np.linspace(-1,1,10)
        #ax.plot(s, np.zeros(10), 'k-')
        bx.plot(s, -s, 'k-')
        bx.plot(s, s, 'k-')
#==========================================================================
    elif problem == '6squares':
        freq = settings
        s = np.linspace(-1,1,10)
        ax.plot(s, np.zeros(10), 'k-')
        a=np.array([-0.33,-0.33,-0.33,-0.33,-0.33,-0.33,-0.33,-0.33,-0.33,-0.33])
        b=np.array([0.33,0.33,0.33,0.33,0.33,0.33,0.33,0.33,0.33,0.33])
        ax.plot(a, s, 'k-')
        ax.plot(b, s, 'k-')
#==========================================================================
    elif problem == 'non convex':
        freq, x_val, sin_val = settings
        s = np.linspace(-1,1,100)
        bx.plot(s, np.clip(-x_val * s + sin_val * np.sin(freq * np.pi * s), -1, 1), 'k-')


    bx.set_xlabel('x', fontsize=16)
    bx.tick_params(axis='x', labelsize = 16)
    bx.tick_params(axis='y', labelsize=0)
    bx.set_xlim([-1, 1])
    bx.set_ylim([-1, 1])
    bx.margins(0)
    bx.axis('equal')

    fig.savefig(foldname + '/' + filename)
    plt.close('all')

def laea_x(lamb, phi):
    return 2*np.sqrt(2) * np.cos(phi)*np.sin(lamb / 2) / np.sqrt(1 + np.cos(phi)*np.cos(lamb/2))


def laea_y(lamb, phi):
    return np.sqrt(2) * np.sin(phi) / np.sqrt(1 + np.cos(phi)*np.cos(lamb/2))


def samples_paint_worldmap(problem, settings, sol, foldname, filename, bw):
    """
    This function takes the problem and the points when they are already classified, and saves a picture of them

    INPUT:
        -problem: name of the problem, to choose among
            ['circle', '3 circles', 'hypersphere', 'tricrown', 'non convex', 'crown', 'sphere', 'squares', 'wavy lines']
        -settings: parameters the function needs for drawing. Provided by problem_gen.problem_gen
        -sol: solutions of the points alreafy classified
        -foldname : name of the folder where we store results
        -filename: name of the files we will produce
        -bw: black and white, True/False
    OUTPUT:
        a file with the points and their classes, and whether they are right or wrong
    """
    if bw == False:
        colors_classes = get_cmap('plasma')
        norm_class = Normalize(vmin=-.5,vmax=np.max(sol[:,-3]) + .5)

        colors_rightwrong = get_cmap('RdYlGn')
        norm_rightwrong = Normalize(vmin=-.1,vmax=1.1)

    if bw == True:
        colors_classes = get_cmap('Greys')
        norm_class = Normalize(vmin=-.5,vmax=np.max(sol[:,-3]) + .5)

        colors_rightwrong = get_cmap('Greys')
        norm_rightwrong = Normalize(vmin=-.1,vmax=1.1)

    fig, axs = plt.subplots(nrows = 3, figsize=(5,15))

    line1 = _winkel_map((np.linspace(-np.pi,np.pi), np.zeros(50)))
    line2 = _winkel_map((np.linspace(-np.pi,np.pi), np.ones(50)))
    line3 = _winkel_map((np.linspace(-np.pi,np.pi), -np.ones(50)))
    line4 = _winkel_map((np.zeros(50), np.linspace(-np.pi/2,.5*np.pi)))
    line5 = _winkel_map((np.pi*np.ones(50), np.linspace(-np.pi/2,.5*np.pi)))
    line6 = _winkel_map((-np.pi*np.ones(50), np.linspace(-np.pi/2,.5*np.pi)))
    ax = axs[0]
    ax.plot(line1[0], line1[1], 'k')
    ax.plot(line2[0], line2[1], 'k')
    ax.plot(line3[0], line3[1], 'k')
    ax.plot(line4[0], line4[1], 'k')
    ax.plot(line5[0], line5[1], 'k')
    ax.plot(line6[0], line6[1], 'k')

    X = np.empty((len(sol), 2))
    for i,s in enumerate(sol):
        mapped = _winkel_map(s[:2])
        X[i] = mapped

    ax.scatter(X[:,0], X[:,1], c=sol[:,-3], cmap = colors_classes, s=2, norm=norm_class)

    #ax.set_xlabel('x', fontsize=16)
    #ax.set_ylabel('y', fontsize=16)
    #ax.tick_params(axis='both',labelsize=16)
    #ax.set_xlim(-1, 1)
    #ax.set_ylim(-1, 1)
    #ax.margins(0)
    #ax.axis('equal')

    bx = axs[1]
    bx.scatter(X[:,0], X[:,1], c=sol[:,-2], cmap = colors_classes, s=2, norm=norm_class)

    cx = axs[2]
    cx.scatter(X[:,0], X[:,1], c=sol[:,-1], cmap = colors_rightwrong, s=2, norm=norm_rightwrong)

    #bx.set_xlabel('x', fontsize=16)
    #bx.tick_params(axis='x', labelsize = 16)
    #bx.tick_params(axis='y', labelsize=0)
    #bx.set_xlim([-1, 1])
    #bx.set_ylim([-1, 1])
    #bx.margins(0)
    #bx.axis('equal')

    fig.savefig(foldname + '/' + filename + '_worldmap')
    plt.close('all')

def _winkel_map(angles):

    alpha = np.arccos(np.cos(angles[1])*np.cos(angles[0] / 2))
    x = .5 * (angles[0] * 180 / np.pi + 2 * np.cos(angles[1] * np.sin(.5 * angles[0])) / np.sinc(alpha / np.pi))

    y = .5 * (angles[1] * 180 / np.pi + np.sin(angles[1])/np.sinc(alpha/np.pi))

    return np.array([x,y])


def create_folder(directory):
    """
    Auxiliar function for creating directories with name directory

    """
    try:
        if not os.path.exists(directory):
            os.makedirs(directory)
    except OSError:
        print ('Error: Creating directory. ' + directory)

def name_folder(chi, problem, qubits, entanglement, layers, method):
    """
    This function takes information from the SGD_step_by_step function and saves the accuracies for training and test sets. It is required for studying the overlearning
    INPUT:
        -chi: cost function, to choose between 'fidelity_chi' or 'weighted_fidelity_chi'
        -problem: name of the problem, to choose among
            ['circle', '3 circles', 'hypersphere', 'tricrown', 'non convex', 'crown', 'sphere', 'squares', 'wavy lines']
        -qubits: number of qubits, must be an integer
        -entanglement: whether there is entanglement or not in the Ansätze, just 'y'/'n'
        -layers: number of layers, must be an integer. If layers == 1, entanglement is not taken in account
        -method: minimization method, to choose among ['SGD', another valid for function scipy.optimize.minimize]

        -name: a name we want for our our files to be save with
        -accs_train: list or array with the accuracies of the training set for all epochs
        -accs_test: list or array with the accuracies of the test set for all epochs
    OUTPUT:
        -foldname: A name for a folder
    """
    chi = chi.lower().replace(' ','_')
    if chi in ['fidelity', 'weighted_fidelity', 'trace', 'weighted_trace']: chi += '_chi'
    if chi not in ['fidelity_chi', 'weighted_fidelity_chi', 'trace_chi', 'weighted_trace_chi']:
        raise ValueError('Figure of merit is not valid')
    foldname = chi + '/'
    problem = problem.replace(' ', '_')
    foldname += problem + '/'
    foldname += str(qubits) + '_qubits/'
    if qubits != 1:
        if entanglement.lower()[0] == 'y':
            foldname += 'entangled/'
        if entanglement.lower()[0] == 'n':
            foldname += 'not_entangled/'

    foldname += str(layers) + '_layers/'
    foldname += method

    return foldname

def name_folder_acc(chi, layers):
    """
    This function takes information from the SGD_step_by_step function and saves the accuracies for training and test sets. It is required for studying the overlearning
    INPUT:
        -chi: cost function, to choose between 'fidelity_chi' or 'weighted_fidelity_chi'
        -problem: name of the problem, to choose among
            ['circle', '3 circles', 'hypersphere', 'tricrown', 'non convex', 'crown', 'sphere', 'squares', 'wavy lines']
        -qubits: number of qubits, must be an integer
        -entanglement: whether there is entanglement or not in the Ansätze, just 'y'/'n'
        -layers: number of layers, must be an integer. If layers == 1, entanglement is not taken in account
        -method: minimization method, to choose among ['SGD', another valid for function scipy.optimize.minimize]
        -name: a name we want for our our files to be save with
        -accs_train: list or array with the accuracies of the training set for all epochs
        -accs_test: list or array with the accuracies of the test set for all epochs
    OUTPUT:
        -foldname: A name for a folder
    """
    chi = chi.lower().replace(' ','_')
    if chi in ['fidelity', 'weighted_fidelity', 'trace', 'weighted_trace']: chi += '_chi'
    if chi not in ['fidelity_chi', 'weighted_fidelity_chi', 'trace_chi', 'weighted_trace_chi']:
        raise ValueError('Figure of merit is not valid')
    foldname = chi + '/'
    foldname += str(layers) + '_layers/'

    return foldname
```

```
##########################################################################
#Quantum classifier
#Adrián Pérez-Salinas, Alba Cervera-Lierta, Elies Gil, J. Ignacio Latorre
#Code by APS
#Code-checks by ACL
#June 3rd 2019

#Universitat de Barcelona / Barcelona Supercomputing Center/Institut de Ciències del Cosmos

##########################################################################


#This file provides useful tools for painting and saving data according to the problem,
# the minimization style, the number of qubits and layers.

import os
import numpy as np
import matplotlib.pyplot as plt
from matplotlib.cm import get_cmap
from matplotlib.colors import Normalize

def write_summary(chi, problem, qubits, entanglement, layers, method, name,
            theta, alpha, weights, chi_value, acc_train, acc_test, seed, epochs):
    """
    This function takes some informations of a given problem and saves some text files
    with this information and the parameters which are solution of the problem
    INPUT:
        -chi: cost function, to choose between 'fidelity_chi' or 'weighted_fidelity_chi'
        -problem: name of the problem, to choose between
            ['circle', '3 circles', 'hypersphere', 'tricrown', 'non convex', 'crown', 'sphere', 'squares', 'wavy lines']
        -qubits: number of qubits, must be an integer
        -entanglement: whether there is entanglement or not in the Ansätze, just 'y'/'n'
        -layers: number of layers, must be an integer. If layers == 1, entanglement is not taken in account
        -method: minimization method, to choose among ['SGD', another valid for function scipy.optimize.minimize]
        -name: a name we want for our our files to be save with
        -theta: set of parameters needed for the circuit. Must be an array with shape (qubits, layers, 3)
        -alpha: set of parameters needed for the circuit. Must be an array with shape (qubits, layers, dimension of data)
        -weight: set of parameters needed fot the circuit only if chi == 'weighted_fidelity_chi'. Must be an array with shape (classes, qubits)
        -chi_value: Value of the cost function after minimization
        -acc_train: accuracy for the training set
        -acc_test: accuracy for the test set
        -seed: seed of numpy.random, needed for replicating results
        -epochs: number of epochs for a 'SGD' method. If there is another method, this input has got no importance

    OUTPUT:
        This function has got no outputs, but several files are saved in an appropiate folder. The files are
        -summary.txt: Saves useful information for the problem
        -theta.txt: saves the theta parameters as a flat array
        -alpha.txt: saves the alpha parameters as a flat array
        -weight.txt: saves the weights as a flat array if they exist
    """
    foldname = name_folder(chi, problem, qubits, entanglement, layers, method)
    create_folder(foldname)
    file_text = open(foldname + '/' + name + '_summary.txt','w')
    file_text.write('\nFigur of merit = '+chi)
    file_text.write('\nProblem = ' + problem)
    file_text.write('\nNumber of qubits = ' + str(qubits))

    if qubits != 1:
        file_text.write('\nEntanglement = ' + entanglement)
    file_text.write('\nNumber of layers = ' + str(layers))
    file_text.write('\nMinimization method = '+ method)
    file_text.write('\nRandom seed = '+ str(seed))
    if method == 'SGD':
        file_text.write('\nNumber of epochs = '+ str(epochs))
    file_text.write('\n\nBEST RESULT\n\n')
    file_text.write('\nTHETA = \n')
    file_text.write(str(theta))
    file_text.write('\nALPHA = \n')
    file_text.write(str(alpha))
    if chi == 'weighted_fidelity_chi':
        file_text.write('\nWEIGHTS = \n')
        file_text.write(str(weights))
    file_text.write('\nchi**2 = ' + str(chi_value))
    file_text.write('\nacc_train = ' + str(acc_train))
    file_text.write('\nacc_test = ' + str(acc_test))
    file_text.close()

    np.savetxt(foldname + '/' + name + '_theta.txt', theta.flatten())
    np.savetxt(foldname + '/' + name + '_alpha.txt', alpha.flatten())
    if chi == 'weighted_fidelity_chi':
        np.savetxt(foldname + '/' + name + '_weight.txt', weights.flatten())


def read_summary(chi, problem, qubits, entanglement, layers, method, name):

    """
    This function reads the files saved by write_summary and returns theta, alpha and weight parameters
    INPUT:
        -chi: cost function, to choose between 'fidelity_chi' or 'weighted_fidelity_chi'
        -problem: name of the problem, to choose among
            ['circle', '3 circles', 'hypersphere', 'tricrown', 'non convex', 'crown', 'sphere', 'squares', 'wavy lines'
        -qubits: number of qubits, must be an integer
        -entanglement: whether there is entanglement or not in the Ansätze, just 'y'/'n'
        -layers: number of layers, must be an integer. If layers == 1, entanglement is not taken in account
        -method: minimization method, to choose among ['SGD', another valid for function scipy.optimize.minimize]
        -name: a name we want for our our files to be save with

    OUTPUT:
        -theta: set of parameters needed for the circuit. It is an array with shape (qubits, layers, 3)
        -alpha: set of parameters needed for the circuit. It is an array with shape (qubits, layers, dimension of data)
        -weight: set of parameters needed fot the circuit only if chi == 'weighted_fidelity_chi'. It is an array with shape (classes, qubits)
    """
    chi = chi.lower().replace(' ','_')
    if chi in ['fidelity', 'weighted_fidelity']: chi += '_chi'
    if chi not in ['fidelity_chi', 'weighted_fidelity_chi']:
        raise ValueError('Figure of merit is not valid')
    if chi == 'fidelity_chi':
        foldname = name_folder(chi, problem, qubits, entanglement, layers, method)
        if problem in ['circle', '3 circles', 'wavy circles', 'wavy lines', 'non convex','crown','tricrown','squares']:
            theta = np.loadtxt(foldname + '/' + name + '_theta.txt').reshape((qubits, layers, 3))
            dim = 2
        elif problem == 'sphere':
            theta = np.loadtxt(foldname + '/' + name + '_theta.txt').reshape((qubits, layers, 3))
            dim = 3
        elif problem in ['hypersphere']:
            theta = np.loadtxt(foldname + '/' + name + '_theta.txt').reshape((qubits, layers, 6))
            dim = 4

        alpha = np.loadtxt(foldname + '/' + name + '_alpha.txt').reshape((qubits, layers, dim))
        return theta, alpha

    if chi == 'weighted_fidelity_chi':
        foldname = name_folder(chi, problem, qubits, entanglement, layers, method)
        if problem in ['circle', '3 circles', 'wavy circles', 'wavy lines', 'non convex','crown','tricrown','squares']:
            theta = np.loadtxt(foldname + '/' + name + '_theta.txt').reshape((qubits, layers, 3))
            dim = 2
        elif problem == 'sphere':
            theta = np.loadtxt(foldname + '/' + name + '_theta.txt').reshape((qubits, layers, 3))
            dim = 3
        elif problem in ['hypersphere']:
            theta = np.loadtxt(foldname + '/' + name + '_theta.txt').reshape((qubits, layers, 6))
            dim = 4

        alpha = np.loadtxt(foldname + '/' + name + '_alpha.txt').reshape((qubits, layers, dim))

        if problem in ['3 circles','wavy lines','squares']:
            weight = np.loadtxt(foldname + '/' + name + '_weight.txt').reshape((4, qubits))
        if problem in ['circle','wavy circle','sphere', 'non convex', 'crown', 'hypersphere']:
            weight = np.loadtxt(foldname + '/' + name + '_weight.txt').reshape((2, qubits))
        if problem in ['tricrown']:
            weight = np.loadtxt(foldname + '/' + name + '_weight.txt').reshape((3, qubits))
        return theta, alpha, weight


def write_epochs_file(chi, problem, qubits, entanglement, layers, method, name):

    """
    This function creates a text file for saving data only in the SGD_step_by_step function
    INPUT:
        -chi: cost function, to choose between 'fidelity_chi' or 'weighted_fidelity_chi'
        -problem: name of the problem, to choose among
            ['circle', '3 circles', 'hypersphere', 'tricrown', 'non convex', 'crown', 'sphere', 'squares', 'wavy lines']
        -qubits: number of qubits, must be an integer
        -entanglement: whether there is entanglement or not in the Ansätze, just 'y'/'n'
        -layers: number of layers, must be an integer. If layers == 1, entanglement is not taken in account
        -method: minimization method, to choose among ['SGD', another valid for function scipy.optimize.minimize]
        -name: a name we want for our our files to be save with
    OUTPUT:
        -file_text: an object which is an open textfile ready to be used
    """
    foldname = name_folder(chi, problem, qubits, entanglement, layers, method)
    create_folder(foldname)
    filename = foldname + '/' + name + '_epochs.txt'
    file_text = open(filename,'w')
    return file_text

def write_epoch(file_text, epoch, theta, alpha, chi_value, acc_train, acc_test):
    """
    This function takes a text file and write information on it
    INPUT:
        -file_text: an object which is an open textfile ready to be used, output of write_epochs_file
        -epoch: the number of epoch providing this information
        -theta: set of parameters needed for the circuit. Must be an array with shape (qubits, layers, 3)
        -alpha: set of parameters needed for the circuit. Must be an array with shape (qubits, layers, dimension of data)
        -weight: set of parameters needed fot the circuit only if chi == 'weighted_fidelity_chi'. Must be an array with shape (classes, qubits)
        -chi_value: Value of the cost function after minimization
        -acc_train: accuracy for the training set
        -acc_test: accuracy for the test set
    OUTPUT:
        -file_text: with more information on it
    """
    file_text.write('\n Epoch = ' + str(epoch))
    file_text.write('\nTHETA = \n')
    file_text.write(str(theta))
    file_text.write('\nALPHA = \n')
    file_text.write(str(alpha))
    file_text.write('\n chi**2 = \n')
    file_text.write(str(chi_value))
    file_text.write('\nacc_train = \n')
    file_text.write(str(acc_train))
    file_text.write('\nacc_test = \n')
    file_text.write(str(acc_test))

def close_epochs_file(file_text, best_epoch):
    """
    This function takes a text file and closes it
    INPUT:
        -file_text: an object which is an open textfile ready to be used, output of write_epochs_file after write_epoch
        -best_epoch: the epoch with the best possible results
    OUTPUT:
        -file_text: closed
    """
    file_text.write('\n\n\nBest epoch = ' + str(best_epoch))
    file_text.close()

def write_epochs_error_rate(chi, problem, qubits, entanglement, layers, method, name,
                    accs_train, accs_test):
    """
    This function takes information from the SGD_step_by_step function and saves the accuracies for training and test sets. It is required for studying the overlearning
    INPUT:
        -chi: cost function, to choose between 'fidelity_chi' or 'weighted_fidelity_chi'
        -problem: name of the problem, to choose among
            ['circle', '3 circles', 'hypersphere', 'tricrown', 'non convex', 'crown', 'sphere', 'squares', 'wavy lines']
        -qubits: number of qubits, must be an integer
        -entanglement: whether there is entanglement or not in the Ansätze, just 'y'/'n'
        -layers: number of layers, must be an integer. If layers == 1, entanglement is not taken in account
        -method: minimization method, to choose among ['SGD', another valid for function scipy.optimize.minimize]
        -name: a name we want for our our files to be save with
        -accs_train: list or array with the accuracies of the training set for all epochs
        -accs_test: list or array with the accuracies of the test set for all epochs
    OUTPUT:
        Two files with the error rates in them
    """
    foldname = name_folder(chi, problem, qubits, entanglement, layers, method)
    create_folder(foldname)
    filename_train = foldname + '/' + name + '_train.txt'
    filename_test = foldname + '/' + name + '_test.txt'

    np.savetxt(filename_train, 1 - np.array(accs_train))
    np.savetxt(filename_test, 1 - np.array(accs_test))

def samples_paint(problem, settings, sol, foldname, filename, bw):
    """
    This function takes the problem and the points when they are already classified, and saves a picture of them
    INPUT:
        -problem: name of the problem, to choose among
            ['circle', '3 circles', 'hypersphere', 'tricrown', 'non convex', 'crown', 'sphere', 'squares', 'wavy lines']
        -settings: parameters the function needs for drawing. Provided by problem_gen.problem_gen
        -sol: solutions of the points alreafy classified
        -foldname : name of the folder where we store results
        -filename: name of the files we will produce
        -bw: black and white, True/False
    OUTPUT:
        a file with the points and their classes, and whether they are right or wrong
    """
    if bw == False:
        colors_classes = get_cmap('plasma')
        norm_class = Normalize(vmin=-.5,vmax=np.max(sol[:,-3]) + .5)

        colors_rightwrong = get_cmap('RdYlGn')
        norm_rightwrong = Normalize(vmin=-.1,vmax=1.1)

    if bw == True:
        colors_classes = get_cmap('Greys')
        norm_class = Normalize(vmin=-.1,vmax=np.max(sol[:,-3]) + .1)

        colors_rightwrong = get_cmap('Greys')
        norm_rightwrong = Normalize(vmin=-.1,vmax=1.1)

    fig, axs = plt.subplots(ncols = 2, figsize=(10,5))
    ax = axs[0]
    if problem in ['circle', '3 circles', 'crown', 'tricrown']:
        centers, radii = settings
        for c, r in zip(centers, radii):
            ca = plt.Circle(c, r, color='k', fill=False, linewidth=2)
            ax.add_artist(ca)
    elif problem == 'wavy circle':
        centers, radii, wave, freq = settings
        phi = np.linspace(0, 2*np.pi, 1000)
        for (c,r, w, f) in zip(centers, radii, wave, freq):
            ax.plot(c[0] + r*(1 + w * np.cos(f * phi)) * np.cos(phi),
                    c[1] + r*(1 + w * np.cos(f * phi)) * np.sin(phi),
                    'k-')
    elif problem == 'wavy lines':
        freq = settings
        s = np.linspace(-1,1,100)
        ax.plot(s, np.clip(s + np.sin(freq * np.pi * s), -1, 1), 'k-')
        ax.plot(s, -s + np.sin(freq * np.pi * s), 'k-')
    elif problem == 'squares':
        freq = settings
        s = np.linspace(-1,1,10)
        ax.plot(s, np.zeros(10), 'k-')
        ax.plot(np.zeros(10), s, 'k-')

    elif problem == 'non convex':
        freq, x_val, sin_val = settings
        s = np.linspace(-1,1,100)
        ax.plot(s, np.clip(-x_val * s + sin_val * np.sin(freq * np.pi * s), -1, 1), 'k-')

    ax.scatter(sol[:,0], sol[:,1], c=sol[:,-2], cmap = colors_classes, s=2, norm=norm_class)

    ax.set_xlabel('x', fontsize=16)
    ax.set_ylabel('y', fontsize=16)
    ax.tick_params(axis='both',labelsize=16)
    ax.set_xlim(-1, 1)
    ax.set_ylim(-1, 1)
    ax.margins(0)
    ax.axis('equal')

    bx = axs[1]
    bx.scatter(sol[:,0], sol[:,1], c=sol[:,-1], cmap = colors_rightwrong, s=2, norm=norm_rightwrong)
    if problem in ['circle', '3 circles', 'crown', 'tricrown']:
        centers, radii = settings
        for c, r in zip(centers, radii):
            ca = plt.Circle(c, r, color='k', fill=False, linewidth=2)
            bx.add_artist(ca)
    elif problem == 'wavy circle':
        centers, radii, wave, freq = settings
        phi = np.linspace(0, 2*np.pi, 1000)
        bx.plot(c[0] + r*(1 + wave * np.cos(freq * phi)) * np.cos(phi),
                c[1] + r*(1 + wave * np.cos(freq * phi)) * np.sin(phi),
                'k-')
    elif problem == 'wavy lines':
        freq = settings
        s = np.linspace(-1,1,100)
        bx.plot(s, np.clip(s + np.sin(freq * np.pi * s), -1, 1), 'k-')
        bx.plot(s, -s + np.sin(freq * np.pi * s), 'k-')

    elif problem == 'squares':
        freq = settings
        s = np.linspace(-1,1,10)
        bx.plot(s, np.zeros(10), 'k-')
        bx.plot(np.zeros(10), s, 'k-')

    elif problem == 'non convex':
        freq, x_val, sin_val = settings
        s = np.linspace(-1,1,100)
        bx.plot(s, np.clip(-x_val * s + sin_val * np.sin(freq * np.pi * s), -1, 1), 'k-')


    bx.set_xlabel('x', fontsize=16)
    bx.tick_params(axis='x', labelsize = 16)
    bx.tick_params(axis='y', labelsize=0)
    bx.set_xlim([-1, 1])
    bx.set_ylim([-1, 1])
    bx.margins(0)
    bx.axis('equal')

    fig.savefig(foldname + '/' + filename)
    plt.close('all')

def laea_x(lamb, phi):
    return 2*np.sqrt(2) * np.cos(phi)*np.sin(lamb / 2) / np.sqrt(1 + np.cos(phi)*np.cos(lamb/2))


def laea_y(lamb, phi):
    return np.sqrt(2) * np.sin(phi) / np.sqrt(1 + np.cos(phi)*np.cos(lamb/2))


def samples_paint_worldmap(problem, settings, sol, foldname, filename, bw):
    """
    This function takes the problem and the points when they are already classified, and saves a picture of them
    INPUT:
        -problem: name of the problem, to choose among
            ['circle', '3 circles', 'hypersphere', 'tricrown', 'non convex', 'crown', 'sphere', 'squares', 'wavy lines']
        -settings: parameters the function needs for drawing. Provided by problem_gen.problem_gen
        -sol: solutions of the points alreafy classified
        -foldname : name of the folder where we store results
        -filename: name of the files we will produce
        -bw: black and white, True/False
    OUTPUT:
        a file with the points and their classes, and whether they are right or wrong
    """
    if bw == False:
        colors_classes = get_cmap('plasma')
        norm_class = Normalize(vmin=-.5,vmax=np.max(sol[:,-3]) + .5)

        colors_rightwrong = get_cmap('RdYlGn')
        norm_rightwrong = Normalize(vmin=-.1,vmax=1.1)

    if bw == True:
        colors_classes = get_cmap('Greys')
        norm_class = Normalize(vmin=-.5,vmax=np.max(sol[:,-3]) + .5)

        colors_rightwrong = get_cmap('Greys')
        norm_rightwrong = Normalize(vmin=-.1,vmax=1.1)

    fig, axs = plt.subplots(nrows = 3, figsize=(5,15))

    line1 = _winkel_map((np.linspace(-np.pi,np.pi), np.zeros(50)))
    line2 = _winkel_map((np.linspace(-np.pi,np.pi), np.ones(50)))
    line3 = _winkel_map((np.linspace(-np.pi,np.pi), -np.ones(50)))
    line4 = _winkel_map((np.zeros(50), np.linspace(-np.pi/2,.5*np.pi)))
    line5 = _winkel_map((np.pi*np.ones(50), np.linspace(-np.pi/2,.5*np.pi)))
    line6 = _winkel_map((-np.pi*np.ones(50), np.linspace(-np.pi/2,.5*np.pi)))
    ax = axs[0]
    ax.plot(line1[0], line1[1], 'k')
    ax.plot(line2[0], line2[1], 'k')
    ax.plot(line3[0], line3[1], 'k')
    ax.plot(line4[0], line4[1], 'k')
    ax.plot(line5[0], line5[1], 'k')
    ax.plot(line6[0], line6[1], 'k')

    X = np.empty((len(sol), 2))
    for i,s in enumerate(sol):
        mapped = _winkel_map(s[:2])
        X[i] = mapped

    ax.scatter(X[:,0], X[:,1], c=sol[:,-3], cmap = colors_classes, s=2, norm=norm_class)

    #ax.set_xlabel('x', fontsize=16)
    #ax.set_ylabel('y', fontsize=16)
    #ax.tick_params(axis='both',labelsize=16)
    #ax.set_xlim(-1, 1)
    #ax.set_ylim(-1, 1)
    #ax.margins(0)
    #ax.axis('equal')

    bx = axs[1]
    bx.scatter(X[:,0], X[:,1], c=sol[:,-2], cmap = colors_classes, s=2, norm=norm_class)

    cx = axs[2]
    cx.scatter(X[:,0], X[:,1], c=sol[:,-1], cmap = colors_rightwrong, s=2, norm=norm_rightwrong)

    #bx.set_xlabel('x', fontsize=16)
    #bx.tick_params(axis='x', labelsize = 16)
    #bx.tick_params(axis='y', labelsize=0)
    #bx.set_xlim([-1, 1])
    #bx.set_ylim([-1, 1])
    #bx.margins(0)
    #bx.axis('equal')

    fig.savefig(foldname + '/' + filename + '_worldmap')
    plt.close('all')

def _winkel_map(angles):

    alpha = np.arccos(np.cos(angles[1])*np.cos(angles[0] / 2))
    x = .5 * (angles[0] * 180 / np.pi + 2 * np.cos(angles[1] * np.sin(.5 * angles[0])) / np.sinc(alpha / np.pi))
    y = .5 * (angles[1] * 180 / np.pi + np.sin(angles[1])/np.sinc(alpha/np.pi))

    return np.array([x,y])


def create_folder(directory):
    """
    Auxiliar function for creating directories with name directory

    """
    try:
        if not os.path.exists(directory):
            os.makedirs(directory)
    except OSError:
        print ('Error: Creating directory. ' + directory)

def name_folder(chi, problem, qubits, entanglement, layers, method):
    """
    This function takes information from the SGD_step_by_step function and saves the accuracies for training and test sets. It is required for studying the overlearning
    INPUT:
        -chi: cost function, to choose between 'fidelity_chi' or 'weighted_fidelity_chi'
        -problem: name of the problem, to choose among
            ['circle', '3 circles', 'hypersphere', 'tricrown', 'non convex', 'crown', 'sphere', 'squares', 'wavy lines']
        -qubits: number of qubits, must be an integer
        -entanglement: whether there is entanglement or not in the Ansätze, just 'y'/'n'
        -layers: number of layers, must be an integer. If layers == 1, entanglement is not taken in account
        -method: minimization method, to choose among ['SGD', another valid for function scipy.optimize.minimize]
        -name: a name we want for our our files to be save with
        -accs_train: list or array with the accuracies of the training set for all epochs
        -accs_test: list or array with the accuracies of the test set for all epochs
    OUTPUT:
        -foldname: A name for a folder
    """
    chi = chi.lower().replace(' ','_')
    if chi in ['fidelity', 'weighted_fidelity']: chi += '_chi'
    if chi not in ['fidelity_chi', 'weighted_fidelity_chi']:
        raise ValueError('Figure of merit is not valid')
    foldname = chi + '/'
    problem = problem.replace(' ', '_')
    foldname += problem + '/'
    foldname += str(qubits) + '_qubits/'
    if qubits != 1:
        if entanglement.lower()[0] == 'y':
            foldname += 'entangled/'
        if entanglement.lower()[0] == 'n':
            foldname += 'not_entangled/'

    foldname += str(layers) + '_layers/'
    foldname += method

    return foldname
```

Text Compare

```python
# coding=utf-8
##########################################################################
#Quantum classifier
#Sara Aminpour, Mike Banad, Sarah Sharif
#September 25th 2024

#School of Electrical and Computer Engineering/ Center for Quantum and Technology, University of Oklahoma, Norman, OK 73019 USA,
##########################################################################
#IMPORTANT_NOTE:
#The code on the left was developed by Sara Aminpour, while the code on the right serves as the reference implementation by Adrián Pérez-Salinas.
#The code on the left has been restructured to handle random data. So some certain sections has been deleted from the reference code.
#Additionally, our code on the left developed to analyze trace distance cost function and linear classification problem
#as well as necessary modification to apply COBYLA, L-BFGS-B, NELDER-MEAD, and SLSQP minimization methods.
##########################################################################


#This file provides useful tools checking how good our results are

from circuitery import code_coords, circuit
from fidelity_minimization import fidelity
from trace_minimization import trace_dis
from weighted_fidelity_minimization import mat_fidelities, w_fidelities
import numpy as np

def _claim(theta, alpha, weight, x, reprs, entanglement, chi):
    """
    This function takes the parameters of a solved problem and one data computes classification of this point

    INPUT:
        -theta: initial point for the theta parameters. The shape must be correct (qubits, layers, 3)

        -alpha: initial point for the alpha parameters. The shape must be correct (qubits, layers, dim)

        -weight: set of parameters needed fot the circuit. Must be an array with shape (classes, qubits)

        -x: coordinates of data for testing.
        -reprs: variable encoding the label states of the different classes
        -entanglement: whether there is entanglement or not in the Ansätze, just 'y'/'n'
        -chi: cost function, to choose between 'fidelity_chi' or 'weighted_fidelity_chi'
    OUTPUT:
        -y_: the class of x, according to the classifier
    """
    chi = chi.lower().replace(' ','_')
    if chi in ['fidelity', 'weighted_fidelity','trace']: chi += '_chi'
    if chi not in ['fidelity_chi', 'weighted_fidelity_chi','trace_chi']:
        raise ValueError('Figure of merit is not valid')

    if chi == 'fidelity_chi':
        y_ = _claim_fidelity(theta, alpha, x, reprs, entanglement)

    if chi == 'trace_chi':
        y_ = _claim_trace(theta, alpha, x, reprs, entanglement)

    if chi == 'weighted_fidelity_chi':
        y_ = _claim_weighted_fidelity(theta, alpha, weight, x, reprs, entanglement)

    return y_


def _claim_fidelity(theta, alpha, x, reprs, entanglement):
    """
    This function is inside _claim for fidelity_chi
    INPUT:
        -theta: initial point for the theta parameters. The shape must be correct (qubits, layers, 3)

        -alpha: initial point for the alpha parameters. The shape must be correct (qubits, layers, dim)

        -weight: set of parameters needed fot the circuit. Must be an array with shape (classes, qubits)

        -x: coordinates of data for testing.
        -reprs: variable encoding the label states of the different classes
        -entanglement: whether there is entanglement or not in the Ansätze, just 'y'/'n'
    OUTPUT:
        the class of x, according to the classifier
    """
    theta_aux = code_coords(theta, alpha, x)
    C = circuit(theta_aux, entanglement)
    Fidelities = [fidelity(r, C.psi) for r in reprs]
    return np.argmax(Fidelities)




#==========================================================================
#==========================================================================
#==========================================================================
#==========================================================================
#==========================================================================

def _claim_trace(theta, alpha, x, reprs, entanglement):
    """
    This function is inside _claim for fidelity_chi
    INPUT:
        -theta: initial point for the theta parameters. The shape must be correct (qubits, layers, 3)
        -alpha: initial point for the alpha parameters. The shape must be correct (qubits, layers, dim)
        -weight: set of parameters needed fot the circuit. Must be an array with shape (classes, qubits)
        -x: coordinates of data for testing.
        -reprs: variable encoding the label states of the different classes
        -entanglement: whether there is entanglement or not in the Ansätze, just 'y'/'n'
    OUTPUT:
        the class of x, according to the classifier
    """
    theta_aux = code_coords(theta, alpha, x)
    C = circuit(theta_aux, entanglement)
    #for r1 in reprs:
    #   Trace=trace_dis(r1, C.r)
    Trace = [trace_dis(r1, C.r) for r1 in reprs]
    #print('td=',Trace)
    #print('reprs[y]=',r1)
    #print('C.r=',C.r)
    #print('min=',np.argmin(Trace))
    return np.argmax(Trace)


#==========================================================================
#==========================================================================
#==========================================================================
#==========================================================================
#==========================================================================



def _claim_weighted_fidelity(theta, alpha, weight, x, reprs, entanglement):
    """
    This function is inside _claim for weighted_fidelity_chi
    INPUT:
        -theta: initial point for the theta parameters. The shape must be correct (qubits, layers, 3)

        -alpha: initial point for the alpha parameters. The shape must be correct (qubits, layers, dim)

        -weight: set of parameters needed fot the circuit. Must be an array with shape (classes, qubits)

        -x: coordinates of data for testing.
        -reprs: variable encoding the label states of the different classes
        -entanglement: whether there is entanglement or not in the Ansätze, just 'y'/'n'
    OUTPUT:
        the class of x, according to the classifier
    """
    theta_aux = code_coords(theta, alpha, x)
    fids = mat_fidelities(theta_aux, weight, reprs, entanglement)
    w_fid = w_fidelities(fids, weight)
    return np.argmax(w_fid)

def tester(theta, alpha, test_data, reprs, entanglement, chi, weights=None):
    """
    This function takes the parameters of a solved problem and one data computes how many points are correct

    INPUT:
        -theta: initial point for the theta parameters. The shape must be correct (qubits, layers, 3)

        -alpha: initial point for the alpha parameters. The shape must be correct (qubits, layers, dim)

        -weight: set of parameters needed fot the circuit. Must be an array with shape (classes, qubits)

        -test_data: set of data for testing
        -reprs: variable encoding the label states of the different classes
        -entanglement: whether there is entanglement or not in the Ansätze, just 'y'/'n'
        -chi: cost function, to choose between 'fidelity_chi' or 'weighted_fidelity_chi'
    OUTPUT:
        -success normalized
    """
    acc = 0
    for i, d in enumerate(test_data):
        x, y = d
        y_ = _claim(theta, alpha, weights, x, reprs, entanglement, chi)
        if y == y_:
            acc += 1
    return acc / len(test_data)


def Accuracy_test(theta, alpha, test_data, reprs, entanglement, chi, weights=None):
    """
    This function takes the parameters of a solved problem and one data computes how many points are correct

    INPUT:
        -theta: initial point for the theta parameters. The shape must be correct (qubits, layers, 3)

        -alpha: initial point for the alpha parameters. The shape must be correct (qubits, layers, dim)

        -weight: set of parameters needed fot the circuit. Must be an array with shape (classes, qubits)

        -test_data: set of data for testing
        -reprs: variable encoding the label states of the different classes
        -entanglement: whether there is entanglement or not in the Ansätze, just 'y'/'n'
        -chi: cost function, to choose between 'fidelity_chi' or 'weighted_fidelity_chi'
    OUTPUT:
        -solutions of the classification
        -success normalized
    """
    dim = len(test_data[0][0])
    solutions = np.zeros((len(test_data), dim + 3)) #data  #Esto se podrá mejorar en el futuro
    for i, d in enumerate(test_data):
        x, y = d
        y_ = _claim(theta, alpha, weights, x, reprs, entanglement, chi)
        solutions[i,:dim] = x
        solutions[i, -3] = y
        solutions[i, -2] = y_
        solutions[i, -1] = int(y == y_)

    acc = np.sum(solutions[:, -1]) / (i + 1)

    return solutions, acc

```

```python
##########################################################################
#Quantum classifier
#Adrián Pérez-Salinas, Alba Cervera-Lierta, Elies Gil, J. Ignacio Latorre
#Code by APS
#Code-checks by ACL
#June 3rd 2019


#Universitat de Barcelona / Barcelona Supercomputing Center/Institut de Ciències del Cosmos

##########################################################################


#This file provides useful tools checking how good our results are

from circuitery import code_coords, circuit
from fidelity_minimization import fidelity
from weighted_fidelity_minimization import mat_fidelities, w_fidelities
import numpy as np

def _claim(theta, alpha, weight, x, reprs, entanglement, chi):
    """
    This function takes the parameters of a solved problem and one data computes classification of this point
    INPUT:
        -theta: initial point for the theta parameters. The shape must be correct (qubits, layers, 3)
        -alpha: initial point for the alpha parameters. The shape must be correct (qubits, layers, dim)
        -weight: set of parameters needed fot the circuit. Must be an array with shape (classes, qubits)
        -x: coordinates of data for testing.
        -reprs: variable encoding the label states of the different classes
        -entanglement: whether there is entanglement or not in the Ansätze, just 'y'/'n'
        -chi: cost function, to choose between 'fidelity_chi' or 'weighted_fidelity_chi'
    OUTPUT:
        -y_: the class of x, according to the classifier
    """
    chi = chi.lower().replace(' ','_')
    if chi in ['fidelity', 'weighted_fidelity']: chi += '_chi'
    if chi not in ['fidelity_chi', 'weighted_fidelity_chi']:
        raise ValueError('Figure of merit is not valid')

    if chi == 'fidelity_chi':
        y_ = _claim_fidelity(theta, alpha, x, reprs, entanglement)

    if chi == 'weighted_fidelity_chi':
        y_ = _claim_weighted_fidelity(theta, alpha, weight, x, reprs, entanglement)

    return y_


def _claim_fidelity(theta, alpha, x, reprs, entanglement):
    """
    This function is inside _claim for fidelity_chi
    INPUT:
        -theta: initial point for the theta parameters. The shape must be correct (qubits, layers, 3)
        -alpha: initial point for the alpha parameters. The shape must be correct (qubits, layers, dim)
        -weight: set of parameters needed fot the circuit. Must be an array with shape (classes, qubits)
        -x: coordinates of data for testing.
        -reprs: variable encoding the label states of the different classes
        -entanglement: whether there is entanglement or not in the Ansätze, just 'y'/'n'
    OUTPUT:
        the class of x, according to the classifier
    """
    theta_aux = code_coords(theta, alpha, x)
    C = circuit(theta_aux, entanglement)
    Fidelities = [fidelity(r, C.psi) for r in reprs]

    return np.argmax(Fidelities)


def _claim_weighted_fidelity(theta, alpha, weight, x, reprs, entanglement):
    """
    This function is inside _claim for weighted_fidelity_chi
    INPUT:
        -theta: initial point for the theta parameters. The shape must be correct (qubits, layers, 3)
        -alpha: initial point for the alpha parameters. The shape must be correct (qubits, layers, dim)
        -weight: set of parameters needed fot the circuit. Must be an array with shape (classes, qubits)
        -x: coordinates of data for testing.
        -reprs: variable encoding the label states of the different classes
        -entanglement: whether there is entanglement or not in the Ansätze, just 'y'/'n'
    OUTPUT:
        the class of x, according to the classifier
    """
    theta_aux = code_coords(theta, alpha, x)
    fids = mat_fidelities(theta_aux, weight, reprs, entanglement)
    w_fid = w_fidelities(fids, weight)
    return np.argmax(w_fid)

def tester(theta, alpha, test_data, reprs, entanglement, chi, weights=None):
    """
    This function takes the parameters of a solved problem and one data computes how many points are correct
    INPUT:
        -theta: initial point for the theta parameters. The shape must be correct (qubits, layers, 3)
        -alpha: initial point for the alpha parameters. The shape must be correct (qubits, layers, dim)
        -weight: set of parameters needed fot the circuit. Must be an array with shape (classes, qubits)
        -test_data: set of data for testing
        -reprs: variable encoding the label states of the different classes
        -entanglement: whether there is entanglement or not in the Ansätze, just 'y'/'n'
        -chi: cost function, to choose between 'fidelity_chi' or 'weighted_fidelity_chi'
    OUTPUT:
        -success normalized
    """
    acc = 0
    for i, d in enumerate(test_data):
        x, y = d
        y_ = _claim(theta, alpha, weights, x, reprs, entanglement, chi)
        if y == y_:
            acc += 1

    return acc / len(test_data)


def Accuracy_test(theta, alpha, test_data, reprs, entanglement, chi, weights=None):
    """
    This function takes the parameters of a solved problem and one data computes how many points are correct
    INPUT:
        -theta: initial point for the theta parameters. The shape must be correct (qubits, layers, 3)
        -alpha: initial point for the alpha parameters. The shape must be correct (qubits, layers, dim)
        -weight: set of parameters needed fot the circuit. Must be an array with shape (classes, qubits)
        -test_data: set of data for testing
        -reprs: variable encoding the label states of the different classes
        -entanglement: whether there is entanglement or not in the Ansätze, just 'y'/'n'
        -chi: cost function, to choose between 'fidelity_chi' or 'weighted_fidelity_chi'
    OUTPUT:
        -solutions of the classification
        -success normalized
    """
    dim = len(test_data[0][0])
    solutions = np.zeros((len(test_data), dim + 3)) #data  #Esto se podrá mejorar en el futuro
    for i, d in enumerate(test_data):
        x, y = d
        y_ = _claim(theta, alpha, weights, x, reprs, entanglement, chi)
        solutions[i,:dim] = x
        solutions[i, -3] = y
        solutions[i, -2] = y_
        solutions[i, -1] = int(y == y_)

    acc = np.sum(solutions[:, -1]) / (i + 1)

    return solutions, acc

```

Text Compare

```
# coding=utf-8
####################################################################
#Quantum classifier
#Sara Aminpour, Mike Banad, Sarah Sharif
#September 25th 2024

#School of Electrical and Computer Engineering/ Center for Quantum and Technology, University of Oklahoma, Norman, OK 73019 USA,
####################################################################
#IMPORTANT_NOTE:
#The code on the left was developed by Sara Aminpour, while the code on the right serves as the reference implementation by Adrián Pérez-Salinas.
#The code on the left has been restructured to handle random data. So some certain sections has been deleted from the reference code.
#Additionally, our code on the left developed to analyze trace distance cost function and linear classification problem
#as well as necessary modification to apply COBYLA, L-BFGS-B, NELDER-MEAD, and SLSQP minimization methods.
####################################################################


#This file provides the minimization for the cheap chi square
from circuitery import code_coords, circuit
import numpy as np
import random
from scipy.optimize import minimize

def trace_minimization(theta, alpha, train_data, reprs,
                      entanglement, method,
                      batch_size, eta, epochs):
    """
    This function takes the parameters of a problem and computes the optimal parameters for it, using different functions. It uses the trace minimization
    INPUT:
        -theta: initial point for the theta parameters. The shape must be correct (qubits, layers, 3)
        -alpha: initial point for the alpha parameters. The shape must be correct (qubits, layers, dim)
        -train_data: set of data for training. There must be several entries (x,y)
        -reprs: variable encoding the label states of the different classes
        -entanglement: whether there is entanglement or not in the Ansätze, just 'y'/'n'
        -method: minimization method, to choose among ['SGD', another valid for function scipy.optimize.minimize]
        -batch_size: size of the batches for stochastic gradient descent, only for 'SGD' method
        -eta: learning rate, only for 'SGD' method
        -epochs: number of epochs , only for 'SGD' method
    OUTPUT:
        -theta: optimized point for the theta parameters. The shape is correct (qubits, layers, 3)
        -alpha: optimized point for the alpha parameters. The shape is correct (qubits, layers, dim)
        -chi: value of the minimization function
    """

    if method == 'SGD':
        thetas, alphas, chis = _sgd(theta, alpha, train_data, reprs,
                                    entanglement, eta, batch_size, epochs)
        i = chis.index(max(chis))
        return thetas[i], alphas[i], chis[i]

    else:
        params, hypars = _translate_to_scipy(theta, alpha)
        results = minimize(_scipy_minimizing, params,
                           args = (hypars, train_data, reprs, entanglement),
                           method=method)
        theta, alpha = _translate_from_scipy(results['x'], hypars)

        return theta, alpha, results['fun']


def _gradient(theta, alpha, data, reprs, entanglement):
    """
    This function computes a gradient step for the SGD minimization
    INPUT:
        -theta: initial point for the theta parameters. The shape must be correct (qubits, layers, 3)
        -alpha: initial point for the alpha parameters. The shape must be correct (qubits, layers, dim)
        -data: one data for training. It must be (x,y)
        -reprs: variable encoding the label states of the different classes
        -entanglement: whether there is entanglement or not in the Ansätze, just 'y'/'n'

    OUTPUT:
        -grad_theta: gradient for the theta parameters. The shape is correct (qubits, layers, 3)
        -grad_alpha: gradient for the alpha parameters. The shape is correct (qubits, layers, dim)
        -results['fun']: value of the minimization function
    """

    x,y = data
    theta_aux = code_coords(theta, alpha, x)
    C = circuit(theta_aux, entanglement)
    prod1 = np.dot(np.conj(reprs[y]), C.psi)
    prods2 = np.zeros(theta.shape, dtype='complex')
    (Q, L, I) = theta_aux.shape

    for q in range(Q):
        for l in range(L):
            for i in range(I):
                theta_aux[q, l, i] += np.pi
                der_c = circuit(theta_aux, entanglement)
                prods2[q, l, i] = np.dot(reprs[y], np.conj(der_c.psi))
                theta_aux[q, l, i] -= np.pi
    grad_theta = np.asfarray(np.real(prod1 * prods2))
    if len(x) <= 3:
        dim = len(x)
        grad_alpha = np.empty((theta.shape[0], theta.shape[1], dim))
        for q in range(Q):
            for l in range(L):
                for i in range(dim):
                    grad_alpha[q, l, i] = x[i] * grad_theta[q, l, i]

    if len(x) == 4:
        grad_alpha = np.empty((theta.shape[0], theta.shape[1], 4))
        for q in range(Q):
            grad_alpha[q, l, 0] = x[0] * grad_theta[q, l, 0]
            grad_alpha[q, l, 1] = x[1] * grad_theta[q, l, 1]
            grad_alpha[q, l, 2] = x[2] * grad_theta[q, l, 3]
            grad_alpha[q, l, 3] = x[3] * grad_theta[q, l, 4]

    return grad_theta, grad_alpha


def _train_batch(theta, alpha, batch, reprs, entanglement):
    """
    This function computes a gradient step for a complete batch for the SGD minimization
    INPUT:
        -theta: initial point for the theta parameters. The shape must be correct (qubits, layers, 3)
        -alpha: initial point for the alpha parameters. The shape must be correct (qubits, layers, dim)
        -batch: small set of data for training. It must be several (x,y)
        -reprs: variable encoding the label states of the different classes
        -entanglement: whether there is entanglement or not in the Ansätze, just 'y'/'n'

    OUTPUT:
        -grad_theta: gradient for the theta parameters averaged in batch. The shape is correct (qubits, layers, 3)
        -grad_alpha: gradient for the alpha parameters averaged in batch. The shape is correct (qubits, layers, dim)
    """
    gradient_theta = np.zeros(theta.shape)
    gradient_alpha = np.zeros(alpha.shape)
    for d in batch:
        g_t, g_a = _gradient(theta, alpha, d, reprs, entanglement)
        gradient_theta += g_t
        gradient_alpha += g_a

    return gradient_theta / len(batch), gradient_alpha / len(batch)


def _session_sgd(theta, alpha, train_data, reprs, entanglement, eta, batch_size):
    """
    This function computes a gradient descent step for all batches
    INPUT:
        -theta: initial point for the theta parameters. The shape must be correct (qubits, layers, 3)
        -alpha: initial point for the alpha parameters. The shape must be correct (qubits, layers, dim)
        -train_data: set of data for training. There must be several entries (x,y)
        -reprs: variable encoding the label states of the different classes
        -entanglement: whether there is entanglement or not in the Ansätze, just 'y'/'n'
        -eta: learning rate, only for 'SGD' method
        -batch_size: size of the batches for stochastic gradient descent, only for 'SGD' method

    OUTPUT:
        -theta: updated point for the theta parameters. The shape is correct (qubits, layers, 3)
        -alpha: updated point for the alpha parameters. The shape is correct (qubits, layers, dim)
        -Av_chi_square: value of the minimization function
    """
    batches = [train_data[k:k + batch_size] for k in range(0,
               len(train_data), batch_size)]
    for batch in batches:
        gradient_theta_batch, gradient_alpha_batch = _train_batch(
                theta, alpha, batch, reprs, entanglement)
        theta += eta * gradient_theta_batch #This sign is very important, it is the difference between maximizing or minimizing.
        alpha += eta * gradient_alpha_batch

    return theta, alpha, Av_Tr(theta, alpha, train_data, reprs, entanglement)



def _sgd(theta, alpha, train_data, reprs, entanglement, eta, batch_size, epochs):
    """
    This function completes the whole SGD strategy
    INPUT:
        -theta: initial point for the theta parameters. The shape must be correct (qubits, layers, 3)
        -alpha: initial point for the alpha parameters. The shape must be correct (qubits, layers, dim)
        -train_data: set of data for training. There must be several entries (x,y)
        -reprs: variable encoding the label states of the different classes
        -entanglement: whether there is entanglement or not in the Ansätze, just 'y'/'n'
        -method: minimization method, to choose among ['SGD', another valid for function scipy.optimize.minimize]
        -batch_size: size of the batches for stochastic gradient descent, only for 'SGD' method
        -eta: learning rate, only for 'SGD' method
        -epochs: number of epochs , only for 'SGD' method
    OUTPUT:
        -thetas: optimized points for the theta parameters for all epochs. The shape is correct (qubits, layers, 3)
        -alphas: optimized points for the theta parameters for all epochs. The shape is correct (qubits, layers, dim)
        -chis: value of the minimization function at every step
    """
    thetas = [np.empty(theta.shape)] * epochs
    alphas = [np.empty(alpha.shape)] * epochs
    chis = [0] * epochs
    for e in range(epochs):
        random.shuffle(train_data)
        theta_, alpha_, chi_ = _session_sgd(theta, alpha, train_data, reprs,
                                            entanglement, eta, batch_size)
        thetas[e] = theta_
        alphas[e] = alpha_
        chis[e] = chi_           #Storage for solution

        theta = theta_
        alpha = alpha_           #Next step initialization

    return thetas, alphas, chis



def _translate_to_scipy(theta, alpha):
    """
    This function is a intermediate step for translating theta and alpha to a single variable for scipy.optimize.minimize
    """
    qubits = theta.shape[0]
    layers = theta.shape[1]
    dim = alpha.shape[-1]

    return np.concatenate((theta.flatten(), alpha.flatten())), (qubits, layers, dim)


def _translate_from_scipy(params, hypars):
    """
    This function is a intermediate step for getting theta and alpha from a single variable for scipy.optimize.minimize
    """
    (qubits, layers, dim) = hypars
    if dim <= 3:
        theta = params[:qubits * layers * 3]. reshape(qubits, layers, 3)
        alpha = params[qubits * layers * 3: qubits * layers * 3 + qubits * layers * dim].reshape(qubits, layers, dim)

    if dim == 4:
        theta = params[:qubits * layers * 6]. reshape(qubits, layers, 6)
        alpha = params[qubits * layers * 6: qubits * layers * 6 + qubits * layers * dim].reshape(qubits, layers, dim)
    return theta, alpha



#==============================================================================
#Sara
def _scipy_minimizing(params, hypars, train_data, reprs, entanglement):
    """
    This function returns the chi^2 function for using scipy
    INPUT:
        -params: theta and alpha inside the same variable
        -hypars: hyperparameters needed to rebuild theta and alpha
        -train_data: training dataset for the classifier
        -reprs: variable encoding the label states of the different classes
        -entanglement: whether there is entanglement or not in the Ansätze, just 'y'/'n'
    OUTPUT:
        - -Av_Tr, which is the function we want to minimize
    """
    theta, alpha = _translate_from_scipy(params, hypars)
    return -Av_Tr(theta, alpha, train_data, reprs, entanglement)


#Sara
#==============================================================================


def trace_dis(r,s):
    """
    This function returns the trace distance of two pure states
    INPUT:
        -2 vectors of pure states of the same dimension
    OUTPUT:
        -trace distance
    """
    dist = np.linalg.norm(r - s)
    td=dist/2

    return td

#==============================================================================
def _Tr(theta, alpha, data, reprs, entanglement): #Chi for one point
    """
    This function compute chi^2 for only one point
    INPUT:
        -theta: set of parameters needed for the circuit. Must be an array with shape (qubits, layers, 3)
        -alpha: set of parameters needed for the circuit. Must be an array with shape (qubits, layers, dimension of data)
        -data: one data for training. It must be (x,y)
        -reprs: variable encoding the label states of the different classes
        -entanglement: whether there is entanglement or not in the Ansätze, just 'y'/'n'
    OUTPUT:
        -chi^2 for data
    """
    #
    x, y = data
    #print('data=',data)
    theta_aux = code_coords(theta, alpha, x)
    C = circuit(theta_aux, entanglement)


    '''if y==0:
        s=np.array([0,0,-1])

    elif y==1:
        s=np.array([0,0,1])

    elif y==2:
        s=np.array([1,0,0])

    elif y==3:
        s=np.array([-1,0,0])

    elif y==4:
        s=np.array([0,1,0])

    elif y==5:
        s=np.array([0,-1,0])'''

    ans = trace_dis(reprs[y], C.r)
    return ans

#==============================================================================
#Sara
def Av_Tr(theta, alpha, train_data, reprs, entanglement): #Chi in average
    """
    This function compute chi^2 for only one point
    INPUT:
        -theta: set of parameters needed for the circuit. Must be an array with shape (qubits, layers, 3)
        -alpha: set of parameters needed for the circuit. Must be an array with shape (qubits, layers, dimension of data)
        -data: one data for training. It must be (x,y)
        -reprs: variable encoding the label states of the different classes
        -entanglement: whether there is entanglement or not in the Ansätze, just 'y'/'n'
    OUTPUT:
        -Averaged chi^2 for data
    """
    Av_Tr = 0
    for d in train_data:
        Av_Tr += _Tr(theta, alpha, d, reprs, entanglement)

    return Av_Tr / len(train_data)
#Sara
#==============================================================================
```